\documentclass{fac}
\usepackage{graphicx}
\usepackage{amssymb}
\usepackage{amsfonts}
\usepackage{amsmath}
\usepackage{dsfont}
\usepackage{MnSymbol}
\usepackage{mathtools}
\usepackage{epsfig}
\usepackage{epstopdf}
\usepackage{tikz}
\usepackage{lmerge}
\ifprodtf \else \usepackage{latexsym}\fi

\newcommand\black{\ensuremath{\blacktriangleright}}
\newcommand\white{\ensuremath{\vartriangleright}}

\newif\ifamsfontsloaded
\ifprodtf
  \newcommand\whbl{\white\kern-.1em--\kern-.1em\black}
  \newcommand\blwh{\black\kern-.1em--\kern-.1em\white}
  \newcommand\blbl{\black\kern-.1em--\kern-.1em\black}
  \newcommand\whwh{\white\kern-.1em--\kern-.1em\white}
  \amsfontsloadedtrue
\else
  \checkfont{msam10}
  \iffontfound
    \IfFileExists{amssymb.sty}
      {\usepackage{amssymb}\amsfontsloadedtrue
       \newcommand\whbl{\white\kern-.125em--\kern-.125em\black}%
       \newcommand\blwh{\black\kern-.125em--\kern-.125em\white}%
       \newcommand\blbl{\black\kern-.125em--\kern-.125em\black}%
       \newcommand\whwh{\white\kern-.125em--\kern-.125em\white}}
      {}
  \fi
\fi

\newtheorem{theorem}{Theorem}[section]
\newtheorem{definition}[theorem]{Definition}
\newtheorem{proposition}[theorem]{Proposition}

\title[Probabilistic Process Algebra to Unifying Quantum and Classical Computing in Closed Systems]
      {Probabilistic Process Algebra to Unifying Quantum and Classical Computing in Closed Systems}

\author[Yong Wang]
    {Yong Wang\\
     College of Computer Science and Technology,\\
     Faculty of Information Technology,\\
     Beijing University of Technology, Beijing, China\\
     }

\correspond{Yong Wang, Pingleyuan 100, Chaoyang District, Beijing, China.
            e-mail: wangy@bjut.edu.cn}

\pubyear{2016}
\pagerange{\pageref{firstpage}--\pageref{lastpage}}

\begin{document}
\label{firstpage}

\makecorrespond

\maketitle

\begin{abstract}
We have unified quantum and classical computing in open quantum systems called qACP which is a quantum generalization of process algebra ACP. But, an axiomatization for quantum and classical processes with an assumption of closed quantum systems is still missing. For closed quantum systems, unitary operator, quantum measurement and quantum entanglement are three basic components for quantum computing. This leads to probability unavoidable. Along the solution of qACP to unify quantum and classical computing in open quantum systems, we unify quantum and classical computing with an assumption of closed systems under the framework of ACP-like probabilistic process algebra. This unification make it can be used widely in verification for quantum and classical computing mixed systems, such as most quantum communication protocols.
\end{abstract}


\section{Introduction}\label{Introduction}

Quantum process algebra provides formal tools for modeling, analysis and verification of quantum communication protocols, which combine quantum communications and quantum computing together. \cite{CQP} \cite{CQP2} defined a language called CQP (Communicating Quantum Processes) by adding primitives for quantum measurements and transformation of quantum states to $\pi$-calculus. An operational semantics and a type system for CQP were also presented to prove that the semantics preserves typing and typing guarantees that each qubit is owned by a unique process within a system.

\cite{QPA} \cite{QPA2} \cite{PAOS} \cite{BC} defined a language called QPAlg (Quantum Process Algebra), in which, based on CCS \cite{CCS}, primitives of unitary transformations and quantum measurements were added to CCS. An operational semantics based on probabilistic branching bisimulation was given in QPAlg.

qCCS \cite{qCCS} was introduced as a kind of algebra of pure quantum processes (no classical data involved) based on CCS. qCCS aimed at providing a suitable framework, in which the mechanism of quantum concurrent computation can be understood, and interactions and conjugation of computation and communication in quantum systems can be observed. In qCCS, quantum operations (super operators) were chosen to describe transformations of quantum states, and quantum variables and their substitutions were carefully treated. An operational semantics for qCCS based on exact (strong) bisimulation and an approximation version of bisimulation were presented for qCCS. Based on \cite{qCCS}, several kind of bisimulations were presented for qCCS, such as probabilistic bisimulation\cite{PSQP}, a kind of weak probabilistic bisimulation\cite{BQP}, open bisimulation \cite{OBQP} and symbolic bisimulation \cite{SB} \cite{SBQP}. These bisimulations provided qCSS with more semantic models. In some bisimulations, not only pure quantum data, but also classical data could be involved in qCCS.

Several years ago, we proposed an axiomatization of quantum processes called qACP \cite{qACP}, which is a quantum generalization of process algebra ACP \cite{ACP}. This work uses some results of the previous works, especially qCCS, in the following ways. (1) qACP still uses the concept of a quantum process configuration which is widely used in quantum process algebra area $\langle p,\varrho \rangle$ \cite{PSQP} \cite{QPA} \cite{QPA2} \cite{CQP} \cite{CQP2} \cite{qCCS} \cite{BQP} \cite{PSQP} \cite{SBQP}, which is usually consisted of a process term $p$ and quantum state information $\varrho$ of all (public) quantum information variables. (2) Like qCCS, quantum system is treated as an open system, and quantum operations are chosen to describe transformations of quantum states, and behave as the atomic actions of a pure quantum process. Quantum measurements are treated as quantum operations, so probabilistic bisimulations are avoided. In qACP, quantum and classical computing are unified with the same equational logic and the same structured operational semantics based on process configuration. Unlike CCS \cite{CCS} and CSP \cite{CSP}, ACP-like process algebras \cite{ACP} are logic on actions, and data are hidden behind actions \cite{PA}. For qACP, it is the duty of atomic actions -- quantum operations and atomic quantum communication actions, to obey the rules of quantum mechanics, such as the non-cloning theorem (it is also suitable for ACP-like probabilistic process algebra in this paper).

But, qACP relies on the assumption of an open quantum system and is not suitable for a closed quantum system. For a closed quantum system, the basic two quantum operations are unitary operator and quantum measurement, so the probability are unavoidable. Thanks for Andova's ACP-like probabilistic process algebra \cite{PPA}, we go long the solution of qACP for an open quantum system to unify quantum and classical computing for a closed quantum system by use of ACP-like probabilistic process algebra.

Fortunately, quantum and classical computing in closed systems are unified with Andova's ACP-like probabilistic process algebra, which has the same equational logic and the same structured operational semantics based on quantum configuration. There are several innovations in this paper, we enumerate them as follows. (1) With the assumption for a closed quantum system, we solve the modeling of the three main components in a closed quantum system: unitary operator, quantum measurement and quantum entanglement, with a full support for quantum and classical computing in closed systems. (2) We still use the framework of a quantum process configuration $\langle p, \varrho\rangle$, and let the quantum part $\varrho$ be the outcomes of execution of $p$ to examine and observe the function of the basic theory of quantum mechanics. We establish the relationship between probabilistic quantum bisimilarity and classical probabilistic bisimilarity, including strong probabilistic bisimilarity and weak probabilistic bisimilarity, which makes an axiomatization of quantum processes possible. With this framework, quantum and classical computing mixed processes are unified with the same structured operational semantics. (3) We establish a series of axiomatizations of quantum process algebra, including Fully Probabilistic Quantum Basic process algebra $fpqBPA$, Basic Quantum Process Algebra with Probabilistic Choice $pqBPA$, Probabilistic Basic Quantum Process Algebra with Projection $pqBPA+PR$, Probabilistic Quantum Variant of ACP $pqACP^+$, Fully Probabilistic Quantum Process Algebra with Abstraction $fpqBPA_{\tau}$, and some extensions, including an extension for quantum entanglement $pqACP^+_{QE}$, for renaming $pqACP^+_{RN}$, for priorities $pqACP^+_{\Theta}$. These process algebras support nearly all main computation properties. (4) In this paper, quantum and classical computing in closed quantum systems are unified with the same equational logic and the same structured operational semantics under the framework of ACP-like probabilistic process algebra. This unification means that our work can be used widely for verification for quantum and classical computing mixed systems, for example, most quantum communication protocols. (5) ACP-like axiomatization also inherits the advantages of ACP, for example, modularity makes it can be extended in an elegant way.

This paper is organized as follows. In section \ref{Pre}, we introduce some preliminaries, including basic quantum mechanics, and especially a brief introduction to Andova's ACP-like probabilistic process algebra. Respectively, we model unitary operator, quantum measurement and quantum entanglement under the framework of probabilistic process algebra in section \ref{PQRA}. Quantum and classical computing in closed systems are unified in section \ref{Unifying}. The applications of verification for quantum and classical computing mixed systems are shown in section \ref{App}, including verifications for quantum teleportation protocol, BB84 protocol and E91 protocol. In section \ref{Extensions}, we show the advantage of modularity of ACP-like process algebra, including extensions of renaming and priorities. Finally, in section \ref{Conclusions}, we conclude this paper. 

\section{Preliminaries}\label{Pre}

For convenience of the reader, we introduce some basic concepts about basic linear algebra, basic quantum mechanics (Please refer to \cite{QCQI} for details), and probabilistic process algebra (Please refer to \cite{PPA} for more details).

\subsection{Basic Quantum Mechanics}\label{BQM}

\definition[Hilbert space]{An isolated physical system is associated with a Hilbert space, which is called the state space of the system. A finite-dimensional Hilbert space is a complex vector space $\mathcal{H}$ together with an inner product, which is a mapping $\langle\cdot|\cdot\rangle:\mathcal{H}\times\mathcal{H}\rightarrow \mathbf{C}$ satisfying: (1)$\langle \varphi|\varphi\rangle\geq 0$ with equality if and only if $|\varphi\rangle=0$; (2)$\langle \varphi|\psi\rangle=\langle \psi|\varphi\rangle^*$; (3) $\langle \varphi|\lambda_1\psi_1+\lambda_2\psi_2\rangle=\lambda_1\langle\varphi|\psi_1\rangle+\lambda_2\langle\varphi|\psi_2\rangle$,
 where $\mathbf{C}$ is the set of complex numbers, and $\lambda^*$ denotes the conjugate of $\lambda$ ($\lambda\in\mathbf{C}$).}

\definition[Orthonormal basis]{For any vector $|\psi\rangle$ in $\mathcal{H}$, the length $||\psi||=\sqrt{\langle\psi|\psi\rangle}$. A vector $|\psi\rangle$ with $||\psi||=1$ is called a unit vector in its state space. An orthonormal basis of a Hilbert space $\mathcal{H}$ is a basis $\{|i\rangle\}$ with

$$\langle i|j\rangle=
\begin{cases}
1& \text{if i=j,}\\
0& \text{otherwise.}
\end{cases}$$
}

\definition[Trace of a linear operator]{The trace of a linear operator $A$ on $\mathcal{H}$ is defined as

$$tr(A)=\sum_{i}\langle i|A|i\rangle.$$
}

\definition{(Tensor products). The state space of a composite system is the tensor product of the state space of its components. Let $\mathcal{H}_1$ and $\mathcal{H}_2$ be two Hilbert spaces, then their tensor product $\mathcal{H}_1\otimes\mathcal{H}_2$ consists of linear vectors $|\psi_1\psi_2\rangle=|\psi_1\rangle\otimes|\psi_2\rangle$, where $\psi_1\in\mathcal{H}_1$ and $\psi_2\in\mathcal{H}_2$.

For two linear operator $A_1$ on Hilbert space $\mathcal{H}_1$, $A_2$ on Hilbert space $\mathcal{H}_2$, $A_1\otimes A_2$ is defined as

$$(A_1\otimes A_2)|\psi_1\psi_2\rangle=A_1|\psi_1\rangle\otimes A_2|\psi_2\rangle$$

where $|\psi_1\rangle\in\mathcal{H}_1$ and $|\psi_2\rangle\in\mathcal{H}_2$.

Let $|\varphi\rangle=\sum_{i}\alpha_i|\varphi_{1i}\varphi_{2i}\rangle\in\mathcal{H}_1\otimes\mathcal{H}_2$ and $|\psi\rangle=\sum_{j}\beta_i|\psi_{1j}\psi_{2j}\rangle\in\mathcal{H}_1\otimes\mathcal{H}_2$. Then the inner product of $|\varphi\rangle$ and $|\psi\rangle$ is defined as follows.

$$\langle\varphi|\psi\rangle=\sum_{i,j}\alpha_i^*\beta_j\langle\varphi_{1i}|\psi_{1j}\rangle\langle\varphi_{2i}|\psi_{2j}\rangle.$$
}

\definition[Density operator]{A mixed state of quantum system is represented by a density operator. A density operator in $\mathcal{H}$ is a linear operator $\varrho$ satisfying:(1) $\varrho$ is positive, that is, $\langle \psi|\varrho|\psi\rangle\geq 0$ for all $|\psi\rangle$; (2) $tr(\varrho)=1$. Let $\mathcal{D}(\mathcal{H})$ denote the set of all positive operators on $\mathcal{H}$.}

\definition[Unitary operator]{The evolution of a closed quantum system is described by a unitary operator on its state space. A unitary operator is a linear operator $U$ on a Hilbert space $\mathcal{H}$ with $U^{\dagger}U=\mathcal{I}_{\mathcal{H}}$, where $\mathcal{I}_{\mathcal{H}}$ is the identity operator on $\mathcal{H}$ and $U^{\dagger}$ is the adjoint of $U$. The unitary transformation $U$ applied on the closed system $\varrho$ has the effect $\varrho\rightarrow U\varrho U^{\dagger}$.}

\definition[Quantum measurement]{Observation of quantum system is a quantum measurement represented by a Hermitian Operator $M$ on the associated Hilbert space of the closed system. A quantum measurement consists of a collection of measurement operators $M=\sum_{m}mP_m$, where $m$ is the measurement outcomes and $P_m$ is the projector onto the eigenspace of $M$ with eigenvalue $m$. The probability of getting $m$ with an initial state $\varrho=|\psi\rangle$ and $p(m)=tr(P_m\varrho)=\langle\psi|P_m|\psi\rangle$, after measurement, the state of the system given the outcome $m$ becomes $P_m\varrho P_m/tr(P_m\varrho)=P_m|\psi\rangle/\sqrt{p(m)}$.}

\subsection{Probabilistic Process Algebra}\label{PPA}

In the following, the variables $x,x',y,y',z,z'$ range over the collection of process terms, $s,s',t,t',u,u'$ are closed items, $\tau$ is the special constant silent step, $\delta$ is the special constant deadlock, $A$ is the collection of atomic actions, atomic actions $a,b\in A$, $A_{\delta}=A\cup\{\delta\}$, $A_{\tau}=A\cup\{\tau\}$. And the predicate $\xrightarrow{a}\surd$ represents successful termination after execution of the action $a$.

Probabilistic process algebra \cite{PPA} distinguishes probabilistic choice and non-deterministic choice to introduce a probabilistic version of ACP. In \cite{PPA}, there are several probabilistic process algebra working in a modular way (we do not concern the part of probabilistic process algebra with discrete time):

\begin{enumerate}
  \item Fully Probabilistic Basic process algebra $fpBPA$: it consists of a (finite) set of constants, $A=\{a,b,c,...\}$ and two binary operators, sequential composition operator $\cdot$ and probabilistic choice operator $\boxplus_{\pi}$ ($\pi\in\langle 0,1\rangle$). The axiom system is as Table \ref{AxiomForfpBPA} shows.

      \begin{center}
        \begin{table}
            \begin{tabular}{@{}ll@{}}
            \hline No. &Axiom\\
                A5 & $(x\cdot y)\cdot z = x\cdot (y\cdot z)$\\
                PrAC1 & $x\boxplus_{\pi} y = y \boxplus_{1-\pi} x$\\
                PrAC2 & $x\boxplus_{\pi}(y\boxplus_{\rho} z) = (x\boxplus_{\frac{\pi}{\pi+\rho-\pi\rho}}y)\boxplus_{\pi+\rho-\pi\rho}z$\\
                PrAC3 & $x\boxplus_{\pi}x = x$\\
                PrAC4 & $(x\boxplus_{\pi} y)\cdot z = x\cdot z \boxplus_{\pi} y \cdot z$\\
            \end{tabular}
            \caption{Axioms for $fpBPA$}
            \label{AxiomForfpBPA}
        \end{table}
    \end{center}

  \item Basic Process Algebra with Probabilistic Choice (without deadlock) $pBPA_{-\delta}$: it consists of the signature of $fpBPA$ and the non-deterministic choice (alternative composition) operator $+$. The axiom system is shown in Table \ref{AxiomForpBPA-del}.

      \begin{center}
        \begin{table}
            \begin{tabular}{@{}ll@{}}
            \hline No. &Axiom\\
                A1 & $x + y = y + x$\\
                A2 & $(x + y) + z = x + (y + z)$\\
                AA3 & $a + a = a$\\
                A4 & $(x + y)\cdot z = x \cdot z + y \cdot z$\\
                PrAC5 & $(x\boxplus_{\pi} y) + z = (x + z)\boxplus_{\pi}(y + z)$\\
            \end{tabular}
            \caption{Axioms for $pBPA_{-\delta}$}
            \label{AxiomForpBPA-del}
        \end{table}
    \end{center}

  \item Basic Process Algebra with Probabilistic Choice $pBPA$: it is an extension of $pBPA_{-\delta}$ with the inaction process (deadlock process) by adding a new deadlock constant $\delta$. The axiom system of $pBPA$ is shown in Table \ref{AxiomForpBPA}.

      \begin{center}
        \begin{table}
            \begin{tabular}{@{}ll@{}}
            \hline No. &Axiom\\
                A6 & $x + \delta = x$\\
                A7 & $\delta\cdot x = \delta$\\
            \end{tabular}
            \caption{Axioms for $pBPA$}
            \label{AxiomForpBPA}
        \end{table}
    \end{center}

  \item Probabilistic Basic Process Algebra with Projection $pBPA+PR$: it is an extension of $pBPA$ with the projection operator $\Pi_n$, which is used to approximate infinite processes, that is, the $n$-th projection of a process $p$ is a process that behaves exactly like $p$ till at most $n$ steps are executed. The axiom system is shown in Table \ref{AxiomForpBPA+PR}.

      \begin{center}
        \begin{table}
            \begin{tabular}{@{}ll@{}}
            \hline No. &Axiom\\
                PR1 & $\Pi_n(a) = a$\\
                PR2 & $\Pi_1(a\cdot x) = a$\\
                PR3 & $\Pi_{n+1}(a\cdot x) = a\cdot\Pi_n(x)$\\
                PR4 & $\Pi_n(x + y) = \Pi_n(x) + \Pi_n(y)$\\
                prPR & $\Pi_n(x \boxplus_{\rho} y) = \Pi_n(x) \boxplus_{\rho} \Pi_n(y)$\\
            \end{tabular}
            \caption{Axioms for $pBPA+PR$}
            \label{AxiomForpBPA+PR}
        \end{table}
    \end{center}

  \item Probabilistic Variant of ACP $pACP^+$: it is an extension of $pBPA$ by adding a variant of asynchronous probabilistic parallel composition $\parallel$. The asynchronous probabilistic parallel composition $\parallel$ is defined by three auxiliary operators: merge with memory $][$, left merge $\leftmerge$, and communication merge $\mid$. The axiom system is shown in Table \ref{AxiomForpACP+}.

      \begin{center}
        \begin{table}
            \begin{tabular}{@{}ll@{}}
            \hline No. &Axiom\\
                PrMM1 & $x \parallel y = (x,x)][(y,y)$\\
                PrMM2 & $(x\boxplus_{\pi}x',z)][(y,w)=(x,z)][(y,w)\boxplus_{\pi} (x',z)][(y,w)$\\
                PrMM3 & $(x,z)][(y\boxplus_{\pi}y',w)=(x,z)][(y,w)\boxplus_{\pi} (x,z)][(y',w)$\\
                PrMM4 & $x=x+x,y=y+y\Rightarrow(x,z)][(y,w)=x\leftmerge w+y\leftmerge z+x\mid y$\\
                CF & $a\mid b = \gamma(a,b)$\\
                CM2 & $a\leftmerge x = a\cdot x$\\
                CM3 & $a\cdot x \leftmerge y = a\cdot(x\parallel y)$\\
                CM4 & $(x + y)\leftmerge z = x\leftmerge z + y\leftmerge z$\\
                PrCM1 & $(x\boxplus_{\pi} y)\leftmerge z = x\leftmerge z \boxplus_{\pi} y\leftmerge z$\\
                CM5 & $a\mid b\cdot x = (a\mid b)\cdot x$\\
                CM6 & $a\cdot x\mid b = (a\mid b)\cdot x$\\
                CM7 & $a\cdot x\mid b\cdot y = (a\mid b)\cdot (x\parallel y)$\\
                PrCM2 & $(x\boxplus_{\pi} y)\mid z = x\mid z \boxplus_{\pi} y\mid z$\\
                PrCM3 & $x\mid (y\boxplus_{\pi} z) = x\mid y \boxplus_{\pi} x\mid z$\\
                D1 & $\partial_H(a) = a$ if $a\notin H$\\
                D2 & $\partial_H(a) = \delta$ if $a\in H$\\
                D3 & $\partial_H(x+y) = \partial_H(x)+\partial_H(y)$\\
                D4 & $\partial_H(x\cdot y) = \partial_H(x)\cdot\partial_H(y)$\\
                PrD5 & $\partial_H(x\boxplus_{\pi} y) = \partial_H(x)\boxplus_{\pi}\partial_H(y)$\\
                PrCM4 & $z=z+z\Rightarrow (x+y)\mid z = x\mid z + y\mid z$\\
                PrCM5 & $z=z+z\Rightarrow z\mid(x+y) = z\mid x + z\mid y$\\
            \end{tabular}
            \caption{Axioms for $pACP^+$}
            \label{AxiomForpACP+}
        \end{table}
    \end{center}

  \item Fully Probabilistic Process Algebra with Abstraction $fpBPA_{\tau}$: it is an extension of $fpBPA$ by adding the new silent constant $\tau$ and the abstraction operator $\tau_I$ for $I\subseteq A$, where $A$ is the set of atomic action. The axiom system of $fpBPA_{\tau}$ is in Table \ref{AxiomForfpBPAtau}. Note that the presence of both probabilistic choice and non-deterministic choice, and also abstraction at the same time is still missing, and its corresponding axiomatization remains an open problem.

      \begin{center}
        \begin{table}
            \begin{tabular}{@{}ll@{}}
            \hline No. &Axiom\\
                T1 & $x\cdot\tau = x$\\
                TI0 & $\tau_I(\tau) = \tau$\\
                TI1 & $\tau_I(a) = a$ if $a\notin I$\\
                TI2 & $\tau_I(a) = \tau$ if $a\in I$\\
                TI4 & $\tau_I(x\cdot y) = \tau_I(x)\cdot \tau_I(y)$\\
                PrTI & $\tau_I(x\boxplus_{\pi} y) = \tau_I(x)\boxplus_{\pi}\tau_I(y)$\\
            \end{tabular}
            \caption{Axioms for $fpBPA_{\tau}$}
            \label{AxiomForfpBPAtau}
        \end{table}
    \end{center}

  \item To abstract away internal cycle (zero probability for infinite $\tau$ sequences), the following set of probabilistic verification rules $PVR_n$ with $n\geq 1$ for fairness are defined. $fpBPA_{\tau}$ extended with these rules for fairness is called $fpBPA_{\tau}+PVR_1+PVR_2+\cdots$.

    $$\frac{X_1=i\cdot X_1\boxplus_{\pi_1} Y_1,\tau\neq i\in I}{\tau\cdot\tau_I(X_1)=\tau\cdot\tau_I(Y_1)}\quad\quad(PVR_1)$$

    $$X_1=i_1\cdot X_2\boxplus_{\pi_1} Y_1$$
    $$X_2=i_2\cdot X_3\boxplus_{\pi_2} Y_2$$
    $$\cdot$$
    $$\cdot$$
    $$X_{n-1}=i_{n-1}\cdot X_n\boxplus_{\pi_{n-1}} Y_{n-1}\\$$
    $$\frac{X_n=i_n\cdot X_1\boxplus_{\pi_n} Y_n,\{\tau\}\neq \{i_1,i_2,...,i_n\}\subseteq I\cup\{\tau\}}{\tau\cdot\tau_I(X_1)=\tau\cdot(\tau_I(Y_1)\boxplus_{\alpha_1} \tau_I(Y_2)\boxplus_{\alpha_2}...\boxplus_{\alpha_{n-2}}\tau_I(Y_{n-1}) \boxplus_{\alpha_{n-1}}\tau_I(Y_n))}\quad\quad(PVR_n)$$

    where $\alpha_1=\frac{1-\pi_1}{1-\pi_1\cdot\pi_2\cdot...\pi_n}$, $\alpha_j=\frac{\pi_1\cdot...\cdot\pi_{j-1}(1-\pi_j)}{1-\pi_1\cdot\pi_2\cdot...\pi_n}$ for $1\leq j\leq n$ and $\pi_k\in\langle 0,1\rangle$ for $1\leq k\leq n$.

\end{enumerate}

$pACP^+$ is a theory for structured specification of probabilistic system. The operational semantics of $pACP^+$ is defined by a term deduction system, which consists of an extended set of constants (each atomic action $a$ has a dynamic counterpart $\breve{a}$) and its deduction rules include two kinds of transition type: probabilistic and action transition. Instead of labelling probabilistic transitions, the probability distribution function is defined, which gives a probability with which a probabilistic transition may occur. The concept of probabilistic bisimulation is used in the construction of the term models, and soundness and completeness of the term models with respect to the axiom system of $pACP^+$ is proven.

In the following, $\rightsquigarrow$ denotes probabilistic transition, and action transition labelled by an atomic action $a\in A$, $\xrightarrow{a}$ and $\xrightarrow{a}\surd$. $x\xrightarrow{a}p$ means that by performing action $a$ process $x$ evolves into $p$; while $x\xrightarrow{a}\surd$ means that $x$ performs an $a$ action and then terminates. $p\rightsquigarrow x$ denotes that process $p$ chooses to behave like process $x$ with a non-zero probability $\pi >0$.

Let $PRA$ denotes the above probabilistic algebra generally. $\mathcal{SP}(PRA)$ denotes the set of all closed terms over the signature $\Sigma_{PRA}$. Terms in $\mathcal{D}(PRA)$ represent processes with a trivial probability distribution--only one process is assigned the non-zero probability $1$.

The finite and infinite processes are denoted $\mathbb{PT}(PRA)$ and $\mathbb{PT}^{(\infty)}(PRA)$. Processes that make probabilistic transitions will be called static processes, denoted $\mathbb{SP}^{(\infty)}(PRA)$. The set of dynamic processes that make action transitions or deadlock, denoted $\mathbb{DP}^{(\infty)}(PRA)$. Processes that perform only one trivial probabilistic transition from processes with non-trivial probabilistic transitions is denoted $\mathbb{D}^{(\infty)}(PRA)$. The counter part of process $p$ is denoted $\breve{p}$ which represents a dynamic processes, and $p\rightsquigarrow\breve{p}$ with probability $1$. That is, for every $p\in\mathbb{D}^{(\infty)}(PRA)$, there is a $\breve{p}\in\mathbb{DP}^{(\infty)}(PRA)$ such that $p\rightsquigarrow\breve{p}$ with probability $1$. A probability distribution function (PDF) is a map $\mu:\mathbb{PT}^{(\infty)}(PRA)\times\mathbb{PT}^{(\infty)}(PRA)\rightarrow\langle 0,1\rangle$ and $\mu^*$ is the cumulative probability distribution function (cPDF). The deduction system $\mathcal{T}_{PRA}$ contains $\breve{\Sigma}_{PRA}=(A\cup\breve{A}\cup A_{\langle X|E\rangle},\textrm{ operators of } $PRA$)$, and (probabilistic and action) deductive transition rules for $PRA$.

We retype the main definitions and conclusions as follows. The (probabilistic and action) transition rules of $pBPA$, $pBPA+PR$ and $pACP^+$ are not retyped, and also the definition of PDF $\mu$ of $pBPA$, $pBPA+PR$ and $pACP^+$ are not retyped, for details, please refer to \cite{PPA}.

\definition[Basic terms]{The set of basic terms of $pBPA$, $\mathcal{B}(pBPA)$, is inductively defined with the help of an intermediate set $\mathcal{B}_+(pBPA)$ \cite{PPA}:
      \begin{enumerate}
        \item $A\cup\{\delta\}\subseteq\mathcal{B}_+(pBPA) \subset\mathcal{B}(pBPA)$;
        \item $a\in A, t\in\mathcal{B}(pBPA)\Rightarrow a\cdot t\in\mathcal{B}_+(pBPA)$;
        \item $t,s\in\mathcal{B}_+(pBPA)\Rightarrow t+s\in\mathcal{B}_+(pBPA)$;
        \item $t,s\in\mathcal{B}(pBPA)\Rightarrow t\boxplus_{\pi}s\in\mathcal{B}_+(pBPA)$ for $\pi\in\langle 0,1\rangle$.
      \end{enumerate}}

\theorem[Elimination theorem of $pBPA$]{Let $p$ be a closed $pBPA$ term, then there is a closed basic $pBPA$ term $q$ such that $pBPA\vdash p=q$.}

\theorem[Elimination theorem of $pACP^+$]{Let $p$ be a closed $pACP^+$ term, then there is a closed $pBPA$ term $q$ such that $pACP^+\vdash p=q$.}

\definition[(Strong) probabilistic bisimulation]{Let $R$ be an equivalence relation on $\mathbb{PT}^{(\infty)}(PRA)$. $R$ is a probabilistic bisimulation if:

\begin{enumerate}
  \item if $pRq$ and $p\rightsquigarrow s$ then there is a term $t$ such that $q\rightsquigarrow t$ and $sRt$;
  \item if $sRt$ and $s\xrightarrow{a}p$ for some $a\in A$, then there is a term $q$ such that $t\xrightarrow{a} q$ and $pRq$;
  \item if $sRt$ and $s\xrightarrow{a}\surd$, then $t\xrightarrow{a}\surd$;
  \item if $pRq$, then $\mu(p,M)=\mu(q,M)$ for each $M\in\mathbb{PT}^{(\infty)}(PRA)/R$.
\end{enumerate}

If there is a probabilistic bisimulation $R$ such that $pRq$, then $p$ is probabilistically bisimilar to $q$, denoted by $p\underline{\leftrightarrow} q$.
}

\theorem[Soundness of $pBPA+PR$]{Let $x$ and $y$ be $pBPA+PR$ terms. If $pBPA+PR\vdash x=y$ then $x\underline{\leftrightarrow} y$.}

\theorem[$AIP^-$ in $\mathbb{PT}^{(\infty)}(pBPA+PR)$]{If for all $n\geq 1$, $\Pi_n(p)\underline{\leftrightarrow}\Pi_n(q)$, then $p\underline{\leftrightarrow} q$.}

\theorem[Soundness of $pBPA$]{Let $x$ and $y$ be closed $pBPA$ terms. If $pBPA\vdash x=y$ then $x\underline{\leftrightarrow} y$.}

\theorem[Completeness of $pBPA$]{Let $z$ and $u$ are closed $pBPA$ terms, if $z\underline{\leftrightarrow} u$, then $pBPA\vdash z=u$.}

\theorem[Soundness of $pACP^+$]{Let $p$ and $q$ be closed $pACP^+$ terms. If $pACP^+\vdash p=q$ then $p\underline{\leftrightarrow} q$.}

\theorem[Completeness of $pACP^+$]{Let $z$ and $u$ are closed $pACP^+$ terms, if $z\underline{\leftrightarrow} u$, then $pACP^+\vdash z=u$.}

For the operational semantics of $fpBPA_{\tau}+PVR_1+PVR_2+\cdots$, a probabilistic branching bisimulation is defined to construct the term models.

An action transition may have the termination state, denoted $NIL$ as its incoming state. We also retype the main definitions and conclusions as follows.

\definition{A fully probabilistic graph $g$ is a tuple $(S_p\cup S_n\cup\{NIL\},\rightsquigarrow,\rightarrow,\mu,root)$} consisting of:

\begin{itemize}
  \item a countable set $S_p$ of probabilistic states,
  \item a countable set $S_n$ of action states such that $S_p\cap S_n=\emptyset$ and $NIL\notin S_p\cup S_n$,
  \item $root\in S_p$,
  \item a relation $\rightsquigarrow\subseteq S_p\times S_n$,
  \item a function $\rightarrow:S_n\rightarrow(S_p\cup\{NIL\})\times A_{\tau}$,
  \item a partial function $\mu:S_p\times S_n\rightarrow\langle 0,1]$ such that $\mu(p,n)$ is defined iff $(p,n)\in\rightsquigarrow$ for $(p,n)\in S_p\times S_n$ and $\sum_{n\in S_n}\mu(p,n)=1$ for any $p\in S_p$.
\end{itemize}

\definition[Probabilistic branching bisimulation]{Let $g$ and $h$ be fully probabilistic graphs. If $R$ is an equivalence relation on $S_g\cup S_h\cup\{NUL_g,NIL_h\}$ such that:

\begin{enumerate}
  \item $(root(g),root(h))\in R$;
  \item if $(p,q)\in R$ and $p\rightsquigarrow s$ then either
        \begin{enumerate}
          \item $(s,q)\in R$ or
          \item there are $v,t$ such that $(p,v),(s,t)\in R$ and $q\xRightarrow{\tau^*}v\rightsquigarrow t$ or $q\xrightarrow{\tau}\cdot\xRightarrow{\tau^*}v\rightsquigarrow t$;
        \end{enumerate}
  \item if $(p,q)\in R$ and $p\xrightarrow{a} s$ then either
        \begin{enumerate}
          \item $a=\tau$ and $(s,q)\in R$ or
          \item there are $v,t$ such that $(p,v),(s,t)\in R$ and $q\xRightarrow{\tau^*}\cdot\rightsquigarrow v\xrightarrow{a} t$ or $q(\xrightarrow{\tau}\cdot\rightsquigarrow)^* v\xrightarrow{a} t$;
        \end{enumerate}
  \item there is an equivalence relation $\tilde{R}$ on $Entry_R(g)\cup Entry_R(h)$ such that $\tilde{R}\subseteq R$ and
      \begin{enumerate}
        \item $(root(g),root(h))\in\tilde{R}$;
        \item if $(p,q)\in\tilde{R}$ then for any $C\in NextEntryC_{\tilde{R}}(p)\cup NextEntryC_{\tilde{R}}(q)$ and for any $a\in A$, $Prob_{[p]_R}(p,\tau^*,C)=Prob_{[q]_R}(q,\tau^*,C)$ and $Prob_{[p]_R}(p,\tau^*a,C)=Prob_{[q]_R}(q,\tau^*a,C)$;
      \end{enumerate}
\end{enumerate}

where $Prob$ is the probability measure, $\xRightarrow{\tau^*}$ is the transitive and reflexive closure of $\rightsquigarrow$, $Entry_R(g)$ is the set of entries of the fully probabilistic graph $g$ with a given equivalence relation $R$ on $g$, $[p]_R$ is the equivalence class of the probabilistic state $p$ in $g$. Then $(R,\tilde{R})$ is a probabilistic branching bisimulation relation between $g$ and $h$. We write $\underline{\leftrightarrow}_{pb}$ if there is a probabilistic bisimulation $R,\tilde{R}$ between $g$ and $h$.

For additional conditions of probabilistic root branching bisimulation $g \underline{\leftrightarrow}_{prb} h$, there is a probabilistic branching bisimulation $(R,\tilde{R})$ between $g$ and $h$ such that:

\begin{enumerate}
  \item if $root(g)\rightsquigarrow p$ then there is $q$ in $h$ such that $root(h)\rightsquigarrow q$ and $(p,q)\in R$;
  \item if $root(h)\rightsquigarrow q$ then there is $p$ in $g$ such that $root(g)\rightsquigarrow p$ and $(p,q)\in R$;
  \item if $root(g)\rightsquigarrow p\xrightarrow{a} s$ for $a\in A_{\tau}$ then there are $q$ and $t$ in $h$ such that $root(h)\rightsquigarrow q\xrightarrow{a} t$ and $(p,q)\in R$ and $(s,t)\in R$;
  \item if $root(h)\rightsquigarrow q\xrightarrow{a} t$ for $a\in A_{\tau}$ then there are $p$ and $s$ in $g$ such that $root(g)\rightsquigarrow p\xrightarrow{a} s$ and $(p,q)\in R$ and $(s,t)\in R$.
\end{enumerate}
}

\theorem[Soundness of $fpBPA_{\tau}+PVR_1+PVR_2+\cdots$]{G/$\underline{\leftrightarrow}_{prb}$ is a model of $fpBPA_{\tau}+PVR_1+PVR_2+\cdots$.}

\section{Probabilistic Process Algebra for Closed Quantum Systems}\label{PQRA}

Quantum operations (super operators) are used to describe the dynamics of an open quantum system. As done in quantum process algebras, qCCS \cite{qCCS} and qACP \cite{qACP}, quantum operations are deemed as the basic actions, so, a probabilistic algebra is avoided. But, in a closed quantum system, the basic action to describe the dynamics is unitary operator, and quantum measurement with a a probabilistic nature is unavoidable (see basic quantum mechanics in section \ref{BQM}).

In this section, we discuss the main ingredients in a closed quantum system: unitary operator, quantum measurement, and quantum entanglement. We try to model them under the framework of probabilistic process algebra (see in section \ref{PPA}, and for details in \cite{PPA}) and form a probabilistic quantum process algebra for pure quantum computing.

\subsection{Unitary Operator in Probabilistic Process Algebra}\label{UO}

In this pure probabilistic quantum process algebra, the basic atomic actions are all unitary operators, there is not classical actions involved (in section \ref{Unifying}, we will discuss actions mixed by unitary operations and classical actions). In section \ref{QM} and section \ref{QE}, we will discuss quantum measurement and quantum entanglement respectively, we will see that they do not affect the assumption about unitary operators as the basic atomic actions. And we use $\alpha, \beta$ to denote unitary operators in contrast to classical actions $a,b$, the variables $\upsilon,\omega$ range over the set $A$ of atomic unitary operators, $\alpha,\beta\in A$. And $A_{\delta}=A\cup\{\delta\}$, $A_{\tau}=A\cup\{\tau\}$.

The probabilistic quantum process algebra $PQRA$ is also defined by a term deduction system in a modular way, including $pqBPA$, $pqBPA+PR$, $pqACP^+$ and $fpqBPA_{\tau}$, which have the same operators and the same axiom systems shown in Table \ref{AxiomForfpBPA}--\ref{AxiomForfpBPAtau} as the corresponding $PRA$, including $pBPA$, $pBPA+PR$, $pACP^+$ and $fpBPA_{\tau}$, but different atomic actions (quantum unitary operators in contrast to classical atomic actions). The meaning of atomic actions does not affect the properties of probabilistic process algebra ($PQRA$ and $PRA$). The corresponding properties of $pqBPA$, $pqBPA+PR$, $pqACP^+$ and $fpqBPA_{\tau}$ are same as those of $pBPA$, $pBPA+PR$, $pACP^+$ and $fpBPA_{\tau}$, such as elimination property (For details, please refer to \cite{PPA}).

Indeed, The difference between $PQRA$ and $PRA$ is their operational semantics, which is caused by the distinction of a unitary operator in quantum processes and a classical atomic action in classical computing. In quantum processes, to avoid the abuse of quantum information which may violate the no-cloning theorem, a quantum process configuration $\langle p,\varrho \rangle$ \cite{PSQP} \cite{QPA} \cite{QPA2} \cite{CQP} \cite{CQP2} \cite{qCCS} \cite{BQP} \cite{PSQP} \cite{SBQP} \cite{qACP} is usually consisted of a process term $p$ and quantum state information $\varrho$ of all (public) quantum information variables. Since ACP-like process algebras are based on actions, quantum information variables are not explicitly defined and hidden behind atomic actions (unitary operators and atomic quantum communication actions), more importantly, the state information $\varrho$ is the effects of execution of a series of unitary operators on involved quantum systems, the execution of a series of atomic actions should not only obey the restrictions of the structure of the process terms, but also those of quantum mechanics principles, such as the no-cloning theorem. So, the operational semantics of quantum processes should be defined based on quantum process configuration $\langle p,\varrho\rangle$, in which $\varrho=\varsigma$ of two state information $\varrho$ and $\varsigma$ means equality under the framework of quantum computing, that is, these two quantum processes are in the same quantum state.

So, for an atomic unitary operator $\alpha$, without consideration of any probability, there should be an action transition rule as follows.

$$\frac{}{\langle\alpha,\varrho\rangle\xrightarrow{\alpha}\langle\surd,\varrho'\rangle}$$

where $\varrho'=\alpha\varrho\alpha^{\dagger}$.

Firstly, we introduce some concepts and conclusions about operational semantics for probabilistic quantum processes.

\definition[Quantum process configuration]{A quantum process configuration is defined to be a pair $\langle p,\varrho\rangle$, where $p$ is a (probabilistic) process (graph) called structural part of the configuration, and $\varrho\in\mathcal{D}(\mathcal{H})$ specifies the current state of the environment, which is called its quantum part.}

\definition[(Strong) probabilistic quantum bisimulation)]{Let $R$ be an equivalence relation on $\langle\mathbb{PT}^{(\infty)}(PQRA),\varrho\rangle$. $R$ is a (strong) probabilistic quantum bisimulation if:

\begin{enumerate}
  \item if $\langle p,\varrho\rangle R\langle q,\varrho\rangle$ and $\langle p,\varrho\rangle\rightsquigarrow \langle s,\varrho\rangle$ then there is a configuration $\langle t,\varrho\rangle$ such that $\langle q,\varrho\rangle\rightsquigarrow \langle t,\varrho\rangle$ and $\langle s,\varrho\rangle R\langle t,\varrho\rangle$;
  \item if $\langle s,\varrho\rangle R\langle t,\varrho\rangle$ and $\langle s,\varrho\rangle\xrightarrow{\alpha}\langle p,\varrho'\rangle$ for some $\alpha\in A$, then there is a configuration $\langle q,\varrho\rangle$ such that $\langle t,\varrho\rangle\xrightarrow{\alpha} \langle q,\varrho'\rangle$ and $\langle p,\varrho'\rangle R\langle q,\varrho'\rangle$;
  \item if $\langle s,\varrho\rangle R\langle t,\varrho\rangle$ and $\langle s,\varrho\rangle\xrightarrow{\alpha}\langle\surd,\varrho'\rangle$, then $\langle t,\varrho\rangle\xrightarrow{\alpha}\langle\surd,\varrho'\rangle$;
  \item if $\langle p,\varrho\rangle R\langle q,\varrho\rangle$, then $\mu(p,M)=\mu(q,M)$ for each $M\in\mathbb{PT}^{(\infty)}(PQRA)/R$.
\end{enumerate}

If there is a probabilistic quantum bisimulation $R$ such that $\langle p,\varrho\rangle R\langle q,\varrho\rangle$, then $\langle p,\varrho\rangle$ is probabilistically quantum bisimilar to $\langle q,\varrho\rangle$, denoted by $\langle p,\varrho\rangle\underline{\leftrightarrow}_q \langle q,\varrho\rangle$.
}\label{SPQB}

In the following, we give the transition rules based on quantum configurations for $pqBPA+PR$, $pqBPA$, $pqACP^+$, including sequential composition operator $\cdot$, probabilistic choice operator $\boxplus_{\pi}$, alternative composition operator $+$, encapsulation operator $\partial_H$, projection $\Pi_n$, parallel composition $\parallel$, merge with memory $][$, left merge $\leftmerge$ and communication merge $\mid$.

The probabilistic transition rules for $pqBPA$ based on quantum configuration are as follows:

$$\frac{}{\langle\alpha,\varrho\rangle\rightsquigarrow\langle\breve{\alpha},\varrho\rangle} \quad\quad\frac{}{\langle\delta,\varrho\rangle\rightsquigarrow\langle\breve{\delta},\varrho\rangle}$$

$$\frac{\langle x,\varrho\rangle\rightsquigarrow \langle x',\varrho\rangle}{\langle x\cdot y,\varrho\rangle\rightsquigarrow\langle x'\cdot y,\varrho\rangle}
\quad\quad\frac{\langle x,\varrho\rangle\rightsquigarrow \langle x',\varrho\rangle, \langle y,\varrho\rangle\rightsquigarrow \langle y',\varrho\rangle}{\langle x + y,\varrho\rangle\rightsquigarrow\langle x' + y',\varrho\rangle}
\quad\quad\frac{\langle x,\varrho\rangle\rightsquigarrow \langle z,\varrho\rangle}{\langle x\boxplus_{\pi} y,\varrho\rangle\rightsquigarrow \langle z,\varrho\rangle, \langle y\boxplus_{1-\pi} x,\varrho\rangle\rightsquigarrow\langle z,\varrho\rangle}$$

The action transition rules for $pqBPA$ based on quantum configuration are as follows.

$$\frac{}{\langle\breve{\alpha},\varrho\rangle\xrightarrow{\alpha}\langle\surd,\varrho'\rangle}
\quad\quad\frac{\langle x,\varrho\rangle\xrightarrow{\alpha} \langle x',\varrho'\rangle}{\langle x\cdot y,\varrho\rangle\xrightarrow{\alpha}\langle x'\cdot y,\varrho'\rangle}
\quad\quad\frac{\langle x,\varrho\rangle\xrightarrow{\alpha} \langle \surd,\varrho'\rangle}{\langle x\cdot y,\varrho\rangle\xrightarrow{\alpha}\langle y,\varrho'\rangle}$$

$$\frac{\langle x,\varrho\rangle\xrightarrow{\alpha} \langle x',\varrho'\rangle}{\langle x+ y,\varrho\rangle\xrightarrow{\alpha}\langle x',\varrho'\rangle,\langle y+x,\varrho\rangle\xrightarrow{\alpha}\langle x',\varrho'\rangle}
\quad\quad\frac{\langle x,\varrho\rangle\xrightarrow{\alpha} \langle \surd,\varrho'\rangle}{\langle x+ y,\varrho\rangle\xrightarrow{\alpha}\langle \surd,\varrho'\rangle,\langle y+x,\varrho\rangle\xrightarrow{\alpha}\langle \surd,\varrho'\rangle}$$

The probabilistic transition rules for projection in $pqBPA+PR$ are as follows.

$$\frac{\langle\langle t_X|E\rangle,\varrho\rangle\rightsquigarrow \langle u,\varrho\rangle}{\langle\langle X|E\rangle,\varrho\rangle\rightsquigarrow \langle u,\varrho\rangle}
\quad\quad\frac{\langle x,\varrho\rangle\rightsquigarrow\langle x',\varrho\rangle}{\langle\Pi_n(x),\varrho\rangle\rightsquigarrow\langle\Pi_n(x'),\varrho\rangle}$$

The action transition rules for projection in $pqBPA+PR$ are as follows.

$$\frac{\langle x,\varrho\rangle\xrightarrow{\alpha}\langle x',\varrho'\rangle}{\langle\Pi_{n+1}(x),\varrho\rangle\xrightarrow{\alpha}\langle\Pi_n(x'),\varrho'\rangle}
\quad\quad\frac{\langle x,\varrho\rangle\xrightarrow{\alpha}\langle \surd,\varrho'\rangle}{\langle\Pi_{n}(x),\varrho\rangle\xrightarrow{\alpha}\langle\surd,\varrho'\rangle}
\quad\quad\frac{\langle x,\varrho\rangle\xrightarrow{\alpha}\langle x',\varrho'\rangle}{\langle\Pi_{1}(x),\varrho\rangle\xrightarrow{\alpha}\langle\surd,\varrho'\rangle}$$

The probabilistic transition rules for $pqACP^+$ are as follows.

$$\frac{\langle x,\varrho\rangle\rightsquigarrow \langle x',\varrho\rangle, \langle y,\varrho\rangle\rightsquigarrow\langle y',\varrho\rangle}{\langle x\parallel y,\varrho\rangle\rightsquigarrow \langle x'\leftmerge y+y'\leftmerge x+x'\mid y',\varrho\rangle}$$

$$\frac{\langle x,\varrho\rangle\rightsquigarrow \langle x',\varrho\rangle, \langle y,\varrho\rangle\rightsquigarrow\langle y',\varrho\rangle}{\langle (x,z)][(y,w),\varrho\rangle\rightsquigarrow \langle x'\leftmerge w+y'\leftmerge z+x'\mid y',\varrho\rangle}$$

$$\frac{\langle x,\varrho\rangle\rightsquigarrow \langle x',\varrho\rangle}{\langle x\leftmerge y,\varrho\rangle\rightsquigarrow \langle x'\leftmerge y,\varrho\rangle}
\quad\quad\frac{\langle x,\varrho\rangle\rightsquigarrow \langle x',\varrho\rangle,\langle y,\varrho\rangle\rightsquigarrow \langle y',\varrho\rangle}{\langle x\mid y,\varrho\rangle\rightsquigarrow \langle x'\mid y',\varrho\rangle}
\quad\quad\frac{\langle x,\varrho\rangle\rightsquigarrow \langle x',\varrho\rangle}{\langle \partial_H(x),\varrho\rangle\rightsquigarrow \langle \partial_H(x'),\varrho\rangle}
$$

In quantum computing, the communication of quantum data should be processed carefully to obey the quantum no-cloning theorem. We assume that all quantum variables are utilized by reference, the communication actions are all classical actions, but not unitary operators. We define a collection of communication actions denoted $C$, and let communication actions $a,b,c\in C$. The action transition rules for $pqACP^+$ are as follows.

$$\frac{\langle x,\varrho\rangle\xrightarrow{\alpha}\langle x',\varrho'\rangle}{\langle x\leftmerge y,\varrho\rangle\xrightarrow{\alpha}\langle x'\parallel y,\varrho'\rangle}
\quad\quad\frac{\langle x,\varrho\rangle\xrightarrow{a}\langle x',\varrho\rangle}{\langle x\leftmerge y,\varrho\rangle\xrightarrow{a}\langle x'\parallel y,\varrho\rangle}
\quad\quad\frac{\langle x,\varrho\rangle\xrightarrow{\alpha}\langle \surd,\varrho'\rangle}{\langle x\leftmerge y,\varrho\rangle \xrightarrow{\alpha}\langle y,\varrho'\rangle}
\quad\quad\frac{\langle x,\varrho\rangle\xrightarrow{a}\langle \surd,\varrho\rangle}{\langle x\leftmerge y,\varrho\rangle\xrightarrow{a}\langle y,\varrho\rangle}$$

$$\frac{\langle x,\varrho\rangle\xrightarrow{a}\langle x',\varrho\rangle, \langle y,\varrho\rangle\xrightarrow{b}\langle y',\varrho\rangle, \gamma(a,b)=c}{\langle x\mid y,\varrho\rangle\xrightarrow{c}\langle x'\parallel y',\varrho\rangle}$$

$$\frac{\langle x,\varrho\rangle\xrightarrow{a}\langle x',\varrho\rangle, \langle y,\varrho\rangle\xrightarrow{b}\langle \surd,\varrho\rangle, \gamma(a,b)=c}{\langle x\mid y,\varrho\rangle\xrightarrow{c}\langle x',\varrho\rangle, \langle y\mid x,\varrho\rangle\xrightarrow{c}\langle x',\varrho\rangle}$$

$$\frac{\langle x,\varrho\rangle\xrightarrow{a}\langle \surd,\varrho\rangle, \langle y,\varrho\rangle\xrightarrow{b}\langle \surd,\varrho\rangle, \gamma(a,b)=c}{\langle x\mid y,\varrho\rangle\xrightarrow{c}\langle \surd,\varrho\rangle}$$

$$\frac{\langle x,\varrho\rangle\xrightarrow{\alpha}\langle x',\varrho'\rangle,\alpha\notin H}{\langle \partial_H(x),\varrho\rangle\xrightarrow{\alpha}\langle\partial_H(x'),\varrho'\rangle}
\quad\quad\frac{\langle x,\varrho\rangle\xrightarrow{a}\langle x',\varrho\rangle,a\notin H}{\langle \partial_H(x),\varrho\rangle\xrightarrow{a}\langle\partial_H(x'),\varrho\rangle}
\quad\quad\frac{\langle x,\varrho\rangle\xrightarrow{\alpha}\langle \surd,\varrho'\rangle,\alpha\notin H}{\langle \partial_H(x),\varrho\rangle\xrightarrow{\alpha}\langle\surd,\varrho'\rangle}
\quad\quad\frac{\langle x,\varrho\rangle\xrightarrow{a}\langle \surd,\varrho\rangle,a\notin H}{\langle \partial_H(x),\varrho\rangle\xrightarrow{a}\langle\surd,\varrho\rangle}$$

The definitions of PDF $\mu$ and cPDF $\mu^*$ in $pqBPA$, $pqBPA+PR$ and $pqACP^+$ are the same as those in probabilistic process algebra $pBPA$, $pBPA+PR$ and $pACP^+$, and the conclusions about PDF $\mu$ and cPDF $\mu^*$ in $pBPA$, $pBPA+PR$ and $pACP^+$ still hold in $pqBPA$, $pqBPA+PR$ and $pqACP^+$. For details, please refer to \cite{PPA}, and we do not retype here.

Then we can get the first obvious but important conclusion as follows.

\proposition{$\underline{\leftrightarrow}_q$ implies $\underline{\leftrightarrow}$ with respect to $pqBPA$, $pqBPA+PR$ and $pqACP^+$.}\label{PQBPB}

\begin{proof}
From the definition of $\underline{\leftrightarrow}_q$ (see in section \ref{UO}) and the definition of $\underline{\leftrightarrow}$ (see in section \ref{PPA}), it is obvious that $\underline{\leftrightarrow}_q$ adds additional conditions about quantum information state $\varrho$ into $\underline{\leftrightarrow}$, with respect to $pqBPA$, $pqBPA+PR$ and $pqACP^+$. So, let $x$ and $y$ be $pqBPA$, or $pqBPA+PR$, or $pqACP^+$ terms, $\langle x,\varrho\rangle\underline{\leftrightarrow}_q\langle y,\varrho\rangle\Rightarrow x\underline{\leftrightarrow} y$, as desired.
\end{proof}

\theorem[Congruence theorem of $pqBPA+PR$, $pqBPA$ and $pqACP^+$]{$\underline{\leftrightarrow}_q$ is a congruence relation on $\mathbb{PT}(pqBPA+PT)$, $\mathbb{PT}(pqBPA)$ and $\mathbb{PT}(pqACP^+)$ with respect to the $+$, $\cdot$, $\boxplus_{\pi}$, $\Pi_n$, $\leftmerge$, $\mid$, $\parallel$, $][$, and $\partial_H$ operators.}\label{congruence}

\begin{proof}
It is obvious based on the following facts:
\begin{enumerate}
  \item operators $+$, $\cdot$, $\boxplus_{\pi}$, $\Pi_n$, $\leftmerge$, $\mid$, $\parallel$, $][$, and $\partial_H$ are all defined on $\mathbb{PT}(pqBPA+PT)$, $\mathbb{PT}(pqBPA)$ and $\mathbb{PT}(pqACP^+)$, which are the same as $\mathbb{PT}(pBPA+PT)$, $\mathbb{PT}(pBPA)$ and $\mathbb{PT}(pACP^+)$ respectively;
  \item from Proposition \ref{PQBPB}, we know that $\underline{\leftrightarrow}_q$ implies $\underline{\leftrightarrow}$ with respect to $pqBPA$, $pqBPA+PR$ and $pqACP^+$.
  \item $\underline{\leftrightarrow}$ is a congruence relation on $\mathbb{PT}(pBPA+PT)$, $\mathbb{PT}(pBPA)$ and $\mathbb{PT}(pACP^+)$ with respect to the $+$, $\cdot$, $\boxplus_{\pi}$, $\Pi_n$, $\leftmerge$, $\mid$, $\parallel$, $][$, and $\partial_H$ operators;
  \item the quantum part $\varrho$ in a quantum configuration $\langle p,\varrho\rangle$ satisfies the congruence relation with respect to the $+$, $\cdot$, $\boxplus_{\pi}$, $\Pi_n$, $\leftmerge$, $\mid$, $\parallel$, $][$, and $\partial_H$ operators. It is trivial and we omit it.
\end{enumerate}
\end{proof}

\theorem[Soundness of $pqBPA+PR$]{Let $x$ and $y$ be $pqBPA+PR$ terms. If $pqBPA+PR\vdash x=y$ then $\langle x,\varrho\rangle\underline{\leftrightarrow}_q \langle y,\varrho\rangle$.}\label{SPQBPAPR}

\begin{proof}
It is already proven that $\underline{\leftrightarrow}_q$ is an equivalent and congruent relation on $\mathbb{PT}^{(\infty)}(pBPA+PR)$ (see details in \cite{PPA}), and $pBPA+PR\vdash x=y$ then $x\underline{\leftrightarrow} y$ (that is, the conditions on PDF $\mu$ are same in $\underline{\leftrightarrow}_q$ and $\underline{\leftrightarrow}$, see in section \ref{PPA} and \cite{PPA} for details), we only need to prove that the quantum information $\varrho$ related parts are also sound.

It is sufficient to examine every axiom in the axiom systems (in Table \ref{AxiomForfpBPA}, \ref{AxiomForpBPA-del}, \ref{AxiomForpBPA}, \ref{AxiomForpBPA+PR}) for $pqBPA+PR$ is sound.

\begin{itemize}
  \item \textbf{Axiom $A1$}. For a relation $R=R_1\cup R_2$, suppose that $Eq(\langle p+q,\varrho\rangle,\langle q+p,\varrho\rangle)= R_1$ with $p,q\in \mathbb{DP}^{(\infty)}(pqBPA+PR)$, and $Eq(\langle u+v,\varrho\rangle,\langle v+u,\varrho\rangle)= R_2$ with $u,v\in \mathbb{SP}^{(\infty)}(pqBPA+PR)$.

      By use of the probabilistic transition rules and the action transition rules for alternative operator $+$, we get:

      $$\frac{\langle u,\varrho\rangle\rightsquigarrow \langle u',\varrho\rangle, \langle v,\varrho\rangle\rightsquigarrow \langle v',\varrho\rangle}{\langle u + v,\varrho\rangle\rightsquigarrow\langle u' + v',\varrho\rangle}$$

      $$\frac{\langle u,\varrho\rangle\rightsquigarrow \langle u',\varrho\rangle, \langle v,\varrho\rangle\rightsquigarrow \langle v',\varrho\rangle}{\langle v + u,\varrho\rangle\rightsquigarrow\langle v' + u',\varrho\rangle}$$

      With the assumptions $u+v=v+u$ and $u'+v'=v'+u'$, we get $R_2$ satisfies condition 1 in the definition of $\underline{\leftrightarrow}_q$
      (see Definition \ref{SPQB}).

      $$\frac{\langle p,\varrho\rangle\xrightarrow{\alpha} \langle p',\varrho'\rangle}{\langle p+ q,\varrho\rangle\xrightarrow{\alpha}\langle p',\varrho'\rangle,\langle q+p,\varrho\rangle\xrightarrow{\alpha}\langle p',\varrho'\rangle}$$

      $$\frac{\langle p,\varrho\rangle\xrightarrow{\alpha} \langle \surd,\varrho'\rangle}{\langle p+ q,\varrho\rangle\xrightarrow{\alpha}\langle \surd,\varrho'\rangle,\langle q+p,\varrho\rangle\xrightarrow{\alpha}\langle \surd,\varrho'\rangle}$$

      With the assumption $p+q=q+p$, we get $R_1$ satisfies conditions 2 and 3 in the definition of $\underline{\leftrightarrow}_q$
      (see Definition \ref{SPQB}).

      So, $R=\underline{\leftrightarrow}_q$, as desired.

  \item \textbf{Axiom $A2$}. For a relation $R=R_1\cup R_2$, suppose that $Eq(\langle (p+q)+s,\varrho\rangle,\langle p+(q+s),\varrho\rangle)= R_1$ with $p,q,s\in \mathbb{DP}^{(\infty)}(pqBPA+PR)$, and $Eq(\langle (u+v)+w,\varrho\rangle,\langle u+(v+w),\varrho\rangle)= R_2$ with $u,v,w\in \mathbb{SP}^{(\infty)}(pqBPA+PR)$.

      By use of the probabilistic transition rules and the action transition rules for alternative operator $+$, we get:

      $$\frac{\langle u,\varrho\rangle\rightsquigarrow \langle u',\varrho\rangle, \langle v,\varrho\rangle\rightsquigarrow \langle v',\varrho\rangle, \langle w,\varrho\rangle\rightsquigarrow \langle w',\varrho\rangle}{\langle (u + v)+w,\varrho\rangle\rightsquigarrow\langle (u' + v')+w',\varrho\rangle}$$

      $$\frac{\langle u,\varrho\rangle\rightsquigarrow \langle u',\varrho\rangle, \langle v,\varrho\rangle\rightsquigarrow \langle v',\varrho\rangle,\langle w,\varrho\rangle\rightsquigarrow \langle w',\varrho\rangle}{\langle u+(v+w),\varrho\rangle\rightsquigarrow\langle u'+(v'+w'),\varrho\rangle}$$

      With the assumptions $(u+v)+w=u+(v+w)$ and $(u'+v')+w'=u'+(v'+w')$, we get $R_2$ satisfies condition 1 in the definition of $\underline{\leftrightarrow}_q$
      (see Definition \ref{SPQB}).

      $$\frac{\langle p,\varrho\rangle\xrightarrow{\alpha} \langle p',\varrho'\rangle}{\langle (p+ q)+s,\varrho\rangle\xrightarrow{\alpha}\langle p',\varrho'\rangle}$$

      $$\frac{\langle p,\varrho\rangle\xrightarrow{\alpha} \langle p',\varrho'\rangle}{\langle p+ (q+s),\varrho\rangle\xrightarrow{\alpha}\langle p',\varrho'\rangle}$$

      $$\frac{\langle p,\varrho\rangle\xrightarrow{\alpha} \langle \surd,\varrho'\rangle}{\langle (p+ q) +s,\varrho\rangle\xrightarrow{\alpha}\langle \surd,\varrho'\rangle}$$

      $$\frac{\langle p,\varrho\rangle\xrightarrow{\alpha} \langle \surd,\varrho'\rangle}{\langle p+ (q +s),\varrho\rangle\xrightarrow{\alpha}\langle \surd,\varrho'\rangle}$$

      With the assumption $(p+q)+s=p+(q+s)$, we get $R_1$ satisfies conditions 2 and 3 in the definition of $\underline{\leftrightarrow}_q$
      (see Definition \ref{SPQB}).

      So, $R=\underline{\leftrightarrow}_q$, as desired.

  \item \textbf{Axiom $AA3$}. For a relation $R=R_1\cup R_2$, suppose that $Eq(\langle \alpha+\alpha,\varrho\rangle,\langle \alpha,\varrho\rangle)= R_1$, and $Eq(\langle \breve{\alpha}+\breve{\alpha},\varrho\rangle,\langle \breve{\alpha},\varrho\rangle)= R_2$.

      By use of the probabilistic transition rules and the action transition rules for atomic unitary operator and alternative operator $+$, we get:

      $$\frac{\langle \alpha,\varrho\rangle\rightsquigarrow \langle \breve{\alpha},\varrho\rangle, \langle \alpha,\varrho\rangle\rightsquigarrow \langle \breve{\alpha},\varrho\rangle}{\langle \alpha+\alpha,\varrho\rangle\rightsquigarrow\langle \breve{\alpha}+\breve{\alpha},\varrho\rangle}$$

      $$\frac{\langle \alpha,\varrho\rangle\rightsquigarrow \langle \breve{\alpha},\varrho\rangle}{\langle \alpha,\varrho\rangle\rightsquigarrow\langle \breve{\alpha},\varrho\rangle}$$

      With the assumptions $\alpha+\alpha=\alpha$ and $\breve{\alpha}+\breve{\alpha}=\breve{\alpha}$, we get $R_1$ satisfies condition 1 in the definition of $\underline{\leftrightarrow}_q$
      (see Definition \ref{SPQB}).

      $$\frac{\langle \breve{\alpha},\varrho\rangle\xrightarrow{\alpha} \langle \surd,\varrho'\rangle}{\langle \breve{\alpha}+\breve{\alpha},\varrho\rangle\xrightarrow{\alpha}\langle \surd,\varrho'\rangle}$$

      $$\frac{\langle \breve{\alpha},\varrho\rangle\xrightarrow{\alpha} \langle \surd,\varrho'\rangle}{\langle \breve{\alpha},\varrho\rangle\xrightarrow{\alpha}\langle \surd,\varrho'\rangle}$$

      With the assumption $\breve{\alpha}+\breve{\alpha}=\breve{\alpha}$, we get $R_2$ satisfies conditions 2 and 3 in the definition of $\underline{\leftrightarrow}_q$
      (see Definition \ref{SPQB}).

      So, $R=\underline{\leftrightarrow}_q$, as desired.

  \item \textbf{Axiom $A4$}. For a relation $R=R_1\cup R_2$, suppose that $Eq(\langle (p+q)\cdot s,\varrho\rangle,\langle p\cdot s + q\cdot s,\varrho\rangle)= R_1$ with $p,q\in \mathbb{DP}^{(\infty)}(pqBPA+PR)$, $s\in \mathbb{SP}^{(\infty)}(pqBPA+PR)$, and $Eq(\langle (u+v)\cdot s,\varrho\rangle,\langle u\cdot s + v\cdot s,\varrho\rangle)= R_2$ with $u,v,s\in \mathbb{SP}^{(\infty)}(pqBPA+PR)$.

      By use of the probabilistic transition rules and the action transition rules for sequential composition $\cdot$, alternative operator $+$, we get:

      $$\frac{\langle u,\varrho\rangle\rightsquigarrow \langle u',\varrho\rangle, \langle v,\varrho\rangle\rightsquigarrow \langle v',\varrho\rangle}{\langle (u + v)\cdot s,\varrho\rangle\rightsquigarrow\langle (u' + v')\cdot s,\varrho\rangle}$$

      $$\frac{\langle u,\varrho\rangle\rightsquigarrow \langle u',\varrho\rangle, \langle v,\varrho\rangle\rightsquigarrow \langle v',\varrho\rangle}{\langle u\cdot s+v\cdot s,\varrho\rangle\rightsquigarrow\langle u'\cdot s+v'\cdot s,\varrho\rangle}$$

      With the assumptions $(u+v)\cdot s=u\cdot s+v\cdot s$ and $(u'+v')\cdot s=u'\cdot s+v'\cdot s$, we get $R_2$ satisfies condition 1 in the definition of $\underline{\leftrightarrow}_q$
      (see Definition \ref{SPQB}).

      $$\frac{\langle p,\varrho\rangle\xrightarrow{\alpha} \langle p',\varrho'\rangle}{\langle (p+ q)\cdot s,\varrho\rangle\xrightarrow{\alpha}\langle p'\cdot s,\varrho'\rangle}$$

      $$\frac{\langle p,\varrho\rangle\xrightarrow{\alpha} \langle p',\varrho'\rangle}{\langle p\cdot s+q\cdot s,\varrho\rangle\xrightarrow{\alpha}\langle p'\cdot s,\varrho'\rangle}$$

      $$\frac{\langle p,\varrho\rangle\xrightarrow{\alpha} \langle \surd,\varrho'\rangle}{\langle (p+ q)\cdot s,\varrho\rangle\xrightarrow{\alpha}\langle s,\varrho'\rangle}$$

      $$\frac{\langle p,\varrho\rangle\xrightarrow{\alpha} \langle \surd,\varrho'\rangle}{\langle p\cdot s+q\cdot s,\varrho\rangle\xrightarrow{\alpha}\langle s,\varrho'\rangle}$$

      With the assumptions $(p+q)\cdot s=p\cdot s + q\cdot s$, we get $R_1$ satisfies conditions 2 and 3 in the definition of $\underline{\leftrightarrow}_q$
      (see Definition \ref{SPQB}).

      So, $R=\underline{\leftrightarrow}_q$, as desired.

  \item \textbf{Axiom $A5$}. For a relation $R=R_1\cup R_2$, suppose that $Eq(\langle (p\cdot q)\cdot s,\varrho\rangle,\langle p\cdot(q\cdot s),\varrho\rangle)= R_1$ with $p\in \mathbb{DP}^{(\infty)}(pqBPA+PR)$, $q,s\in \mathbb{SP}^{(\infty)}(pqBPA+PR)$, and $Eq(\langle (u\cdot q)\cdot s,\varrho\rangle,\langle u\cdot(q\cdot s),\varrho\rangle)= R_2$ with $u,q,s\in \mathbb{SP}^{(\infty)}(pqBPA+PR)$.

      By use of the probabilistic transition rules and the action transition rules for sequential composition $\cdot$, we get:

      $$\frac{\langle u,\varrho\rangle\rightsquigarrow \langle u',\varrho\rangle}{\langle (u\cdot q)\cdot s,\varrho\rangle\rightsquigarrow\langle (u' \cdot q)\cdot s,\varrho\rangle}$$

      $$\frac{\langle u,\varrho\rangle\rightsquigarrow \langle u',\varrho\rangle}{\langle u\cdot (q\cdot s),\varrho\rangle\rightsquigarrow\langle u' \cdot (q\cdot s),\varrho\rangle}$$

      With the assumptions $(u\cdot q)\cdot s=u\cdot (q\cdot s)$ and $(u'\cdot q)\cdot s=u'\cdot (q\cdot s)$, we get $R_2$ satisfies condition 1 in the definition of $\underline{\leftrightarrow}_q$
      (see Definition \ref{SPQB}).

      $$\frac{\langle p,\varrho\rangle\xrightarrow{\alpha} \langle p',\varrho'\rangle}{\langle (p\cdot q)\cdot s,\varrho\rangle\xrightarrow{\alpha}\langle (p'\cdot q)\cdot s,\varrho'\rangle}$$

      $$\frac{\langle p,\varrho\rangle\xrightarrow{\alpha} \langle p',\varrho'\rangle}{\langle p\cdot (q\cdot s),\varrho\rangle\xrightarrow{\alpha}\langle p'\cdot (q\cdot s),\varrho'\rangle}$$

      $$\frac{\langle p,\varrho\rangle\xrightarrow{\alpha} \langle \surd,\varrho'\rangle}{\langle (p\cdot q)\cdot s,\varrho\rangle\xrightarrow{\alpha}\langle q\cdot s,\varrho'\rangle}$$

      $$\frac{\langle p,\varrho\rangle\xrightarrow{\alpha} \langle \surd,\varrho'\rangle}{\langle p\cdot(q\cdot s),\varrho\rangle\xrightarrow{\alpha}\langle q\cdot s,\varrho'\rangle}$$

      With the assumptions $(p\cdot q)\cdot s=p\cdot (q\cdot s)$ and $(p'\cdot q)\cdot s=p'\cdot (q\cdot s)$, we get $R_1$ satisfies conditions 2 and 3 in the definition of $\underline{\leftrightarrow}_q$
      (see Definition \ref{SPQB}).

      So, $R=\underline{\leftrightarrow}_q$, as desired.

  \item \textbf{Axiom $A6$}. For a relation $R=R_1\cup R_2$, suppose that $Eq(\langle p+\breve{\delta},\varrho\rangle,\langle p,\varrho\rangle)= R_1$ with $p\in \mathbb{DP}^{(\infty)}(pqBPA+PR)$, and $Eq(\langle u+\delta,\varrho\rangle,\langle u,\varrho\rangle)= R_2$ with $u\in \mathbb{SP}^{(\infty)}(pqBPA+PR)$.

      By use of the probabilistic transition rules and the action transition rules for alternative operator $+$ and the constant $\delta$, we get:

      $$\frac{\langle u,\varrho\rangle\rightsquigarrow \langle u',\varrho\rangle, \langle \delta,\varrho\rangle\rightsquigarrow \langle \breve{\delta},\varrho\rangle}{\langle u + \delta,\varrho\rangle\rightsquigarrow\langle u' + \breve{\delta},\varrho\rangle}$$

      $$\frac{\langle u,\varrho\rangle\rightsquigarrow \langle u',\varrho\rangle}{\langle u,\varrho\rangle\rightsquigarrow\langle u',\varrho\rangle}$$

      With the assumptions $u+\delta=u$ and $u'+\breve{\delta}=u'$, we get $R_2$ satisfies condition 1 in the definition of $\underline{\leftrightarrow}_q$
      (see Definition \ref{SPQB}).

      $$\frac{\langle p,\varrho\rangle\xrightarrow{\alpha} \langle p',\varrho'\rangle}{\langle p+ \breve{\delta},\varrho\rangle\xrightarrow{\alpha}\langle p'+\breve{\delta},\varrho'\rangle}$$

      $$\frac{\langle p,\varrho\rangle\xrightarrow{\alpha} \langle p',\varrho'\rangle}{\langle p,\varrho\rangle\xrightarrow{\alpha}\langle p',\varrho'\rangle}$$

      $$\frac{\langle p,\varrho\rangle\xrightarrow{\alpha} \langle \surd,\varrho'\rangle}{\langle p+\breve{\delta},\varrho\rangle\xrightarrow{\alpha}\langle \surd,\varrho'\rangle}$$

      $$\frac{\langle p,\varrho\rangle\xrightarrow{\alpha} \langle \surd,\varrho'\rangle}{\langle p,\varrho\rangle\xrightarrow{\alpha}\langle \surd,\varrho'\rangle}$$

      With the assumptions $p+\breve{\delta}=p$ and $p'+\breve{\delta}=p'$, we get $R_1$ satisfies conditions 2 and 3 in the definition of $\underline{\leftrightarrow}_q$
      (see Definition \ref{SPQB}).

      So, $R=\underline{\leftrightarrow}_q$, as desired.

  \item \textbf{Axiom $A7$}. For a relation $R=R_1\cup R_2$, suppose that $Eq(\langle \delta\cdot p,\varrho\rangle,\langle \delta,\varrho\rangle)= R_1$ with $p\in \mathbb{SP}^{(\infty)}(pqBPA+PR)$, and $Eq(\langle \breve{\delta}\cdot u,\varrho\rangle,\langle \breve{\delta},\varrho\rangle)= R_2$ with $u\in \mathbb{SP}^{(\infty)}(pqBPA+PR)$.

      By use of the probabilistic transition rules and the action transition rules for sequential operator $\cdot$ and the constant $\delta$, we get:

      $$\frac{\langle \delta,\varrho\rangle\rightsquigarrow \langle \breve{\delta},\varrho\rangle}{\langle \delta\cdot p,\varrho\rangle\rightsquigarrow\langle \breve{\delta}\cdot p,\varrho\rangle}$$

      $$\frac{\langle \delta,\varrho\rangle\rightsquigarrow \langle \breve{\delta},\varrho\rangle}{\langle \delta,\varrho\rangle\rightsquigarrow \langle \breve{\delta},\varrho\rangle}$$

      With the assumptions $\delta\cdot p=\delta$ and $\breve{\delta}\cdot p=\breve{\delta}$, we get $R_1$ satisfies condition 1 in the definition of $\underline{\leftrightarrow}_q$
      (see Definition \ref{SPQB}).

      There are not action transition rules for the constant $\breve{\delta}$, which mean that $\breve{\delta}$ leads to inaction processes, accompany with the action transition rules of sequential composition $\cdot$, $\breve{\delta}\cdot p\nrightarrow$ and $\breve{\delta}\nrightarrow$ all lead to inaction processes, with the assumption $\breve{\delta}\cdot p=\breve{\delta}$, we get $R_2$ satisfies conditions 2 and 3 in the definition of $\underline{\leftrightarrow}_q$
      (see Definition \ref{SPQB}).

      So, $R=\underline{\leftrightarrow}_q$, as desired.

  \item \textbf{Axiom $PrAC1$}. For a relation $R$, suppose that $Eq(\langle u\boxplus_{\pi}v,\varrho\rangle,\langle v\boxplus_{1-\pi}u,\varrho\rangle)= R$ with $u,v\in \mathbb{SP}^{(\infty)}(pqBPA+PR)$.

      By use of the probabilistic transition rules for probabilistic choice operator $\boxplus_{\pi}$, we get:

      $$\frac{\langle u,\varrho\rangle\rightsquigarrow \langle u',\varrho\rangle}{\langle u \boxplus_{\pi} v,\varrho\rangle\rightsquigarrow\langle u',\varrho\rangle,\langle v \boxplus_{1-\pi} u,\varrho\rangle\rightsquigarrow\langle u',\varrho\rangle}$$

      With the assumption $u\boxplus_{\pi}v=v\boxplus_{1-\pi}u$, we get $R$ satisfies condition 1 in the definition of $\underline{\leftrightarrow}_q$
      (see Definition \ref{SPQB}). Note that, for probabilistic choice operator $\boxplus_{\pi}$, there are not action transition rules.

      So, $R=\underline{\leftrightarrow}_q$, as desired.

  \item \textbf{Axiom $PrAC2$}. For a relation $R$, suppose that $Eq(\langle u\boxplus_{\pi}(v\boxplus_{\rho}w),\varrho\rangle,\langle (u\boxplus_{\frac{\pi}{\pi+\rho-\pi\rho}}v)\boxplus_{\pi+\rho-\pi\rho}w,\varrho\rangle)= R$ with $u,v,w\in \mathbb{SP}^{(\infty)}(pqBPA+PR)$.

      By use of the probabilistic transition rules for probabilistic choice operator $\boxplus_{\pi}$, we get:

      $$\frac{\langle u,\varrho\rangle\rightsquigarrow \langle u',\varrho\rangle}{\langle u\boxplus_{\pi}(v\boxplus_{\rho}w),\varrho\rangle\rightsquigarrow\langle u',\varrho\rangle,\langle (u\boxplus_{\frac{\pi}{\pi+\rho-\pi\rho}}v)\boxplus_{\pi+\rho-\pi\rho}w,\varrho\rangle\rightsquigarrow\langle u',\varrho\rangle}$$

      With the assumption $u\boxplus_{\pi}(v\boxplus_{\rho}w)=(u\boxplus_{\frac{\pi}{\pi+\rho-\pi\rho}}v)\boxplus_{\pi+\rho-\pi\rho}w$, we get $R$ satisfies condition 1 in the definition of $\underline{\leftrightarrow}_q$
      (see Definition \ref{SPQB}). Note that, for probabilistic choice operator $\boxplus_{\pi}$, there are not action transition rules.

      So, $R=\underline{\leftrightarrow}_q$, as desired.

  \item \textbf{Axiom $PrAC3$}. For a relation $R$, suppose that $Eq(\langle u\boxplus_{\pi}u,\varrho\rangle,\langle u,\varrho\rangle)= R$ with $u\in \mathbb{SP}^{(\infty)}(pqBPA+PR)$.

      By use of the probabilistic transition rules for probabilistic choice operator $\boxplus_{\pi}$, we get:

      $$\frac{\langle u,\varrho\rangle\rightsquigarrow \langle u',\varrho\rangle}{\langle u\boxplus_{\pi}u,\varrho\rangle\rightsquigarrow\langle u',\varrho\rangle}$$

      $$\frac{\langle u,\varrho\rangle\rightsquigarrow \langle u',\varrho\rangle}{\langle u,\varrho\rangle\rightsquigarrow\langle u',\varrho\rangle}$$

      With the assumption $u\boxplus_{\pi}u=u$, we get $R$ satisfies condition 1 in the definition of $\underline{\leftrightarrow}_q$
      (see Definition \ref{SPQB}). Note that, for probabilistic choice operator $\boxplus_{\pi}$, there are not action transition rules.

      So, $R=\underline{\leftrightarrow}_q$, as desired.

  \item \textbf{Axiom $PrAC4$}. For a relation $R$, suppose that $Eq(\langle (u\boxplus_{\pi}v)\cdot w,\varrho\rangle,\langle u\cdot w\boxplus_{\pi}v\cdot w,\varrho\rangle)= R$ with $u,v,w\in \mathbb{SP}^{(\infty)}(pqBPA+PR)$.

      By use of the probabilistic transition rules for probabilistic choice operator $\boxplus_{\pi}$ and sequential operator $\cdot$, we get:

      $$\frac{\langle u,\varrho\rangle\rightsquigarrow \langle u',\varrho\rangle}{\langle (u\boxplus_{\pi}v)\cdot w,\varrho\rangle\rightsquigarrow\langle u'\cdot w,\varrho\rangle}$$

      $$\frac{\langle u,\varrho\rangle\rightsquigarrow \langle u',\varrho\rangle}{\langle u\cdot w\boxplus_{\pi}v\cdot w,\varrho\rangle\rightsquigarrow\langle u'\cdot w,\varrho\rangle}$$

      With the assumption $(u\boxplus_{\pi}v)\cdot w=u\cdot w\boxplus_{\pi}v\cdot w$, we get $R$ satisfies condition 1 in the definition of $\underline{\leftrightarrow}_q$
      (see Definition \ref{SPQB}). Note that, for probabilistic choice operator $\boxplus_{\pi}$, there are not action transition rules.

      So, $R=\underline{\leftrightarrow}_q$, as desired.

  \item \textbf{Axiom $PrAC5$}. For a relation $R$, suppose that $Eq(\langle (u\boxplus_{\pi}v)+ w,\varrho\rangle,\langle u+ w\boxplus_{\pi}v+ w,\varrho\rangle)= R$ with $u,v,w\in \mathbb{SP}^{(\infty)}(pqBPA+PR)$.

      By use of the probabilistic transition rules for probabilistic choice operator $\boxplus_{\pi}$ and alternative operator $+$, we get:

      $$\frac{\langle u,\varrho\rangle\rightsquigarrow \langle u',\varrho\rangle}{\langle (u\boxplus_{\pi}v)+ w,\varrho\rangle\rightsquigarrow\langle u'+ w,\varrho\rangle}$$

      $$\frac{\langle u,\varrho\rangle\rightsquigarrow \langle u',\varrho\rangle}{\langle u+ w\boxplus_{\pi}v+ w,\varrho\rangle\rightsquigarrow\langle u'+ w,\varrho\rangle}$$

      With the assumption $(u\boxplus_{\pi}v)+ w=u+ w\boxplus_{\pi}v+ w$, we get $R$ satisfies condition 1 in the definition of $\underline{\leftrightarrow}_q$
      (see Definition \ref{SPQB}). Note that, for probabilistic choice operator $\boxplus_{\pi}$, there are not action transition rules.

      So, $R=\underline{\leftrightarrow}_q$, as desired.

  \item \textbf{Axiom $PR1$}. For a relation $R_n(n\geq 1)=Eq(\langle \Pi_n(\alpha),\varrho\rangle,\langle \alpha,\varrho\rangle)\cup Eq(\langle \Pi_n(\breve{\alpha}),\varrho\rangle,\langle \breve{\alpha},\varrho\rangle)$.

      By use of the probabilistic transition rules and the action transition rules for atomic unitary operator and projection operator $\Pi_n(n\geq 1)$, we get:

      $$\frac{\langle \alpha,\varrho\rangle\rightsquigarrow \langle \breve{\alpha},\varrho\rangle}{\langle \Pi_n(\alpha),\varrho\rangle\rightsquigarrow\langle \Pi_n(\breve{\alpha}),\varrho\rangle}$$

      $$\frac{\langle \alpha,\varrho\rangle\rightsquigarrow \langle \breve{\alpha},\varrho\rangle}{\langle \alpha,\varrho\rangle\rightsquigarrow\langle \breve{\alpha},\varrho\rangle}$$

      With the assumptions $\Pi_n(\alpha)=\alpha$ and $\Pi_n(\breve{\alpha})=\breve{\alpha}$, we get $R_n$ satisfies condition 1 in the definition of $\underline{\leftrightarrow}_q$
      (see Definition \ref{SPQB}).

      $$\frac{\langle \breve{\alpha},\varrho\rangle\xrightarrow{\alpha} \langle \surd,\varrho'\rangle}{\langle \Pi_n(\breve{\alpha}),\varrho\rangle\xrightarrow{\alpha}\langle \surd,\varrho'\rangle}$$

      $$\frac{\langle \breve{\alpha},\varrho\rangle\xrightarrow{\alpha} \langle \surd,\varrho'\rangle}{\langle \breve{\alpha},\varrho\rangle\xrightarrow{\alpha}\langle \surd,\varrho'\rangle}$$

      With the assumption $\Pi_n(\breve{\alpha})=\breve{\alpha}$, we get $R_n$ satisfies conditions 2 and 3 in the definition of $\underline{\leftrightarrow}_q$
      (see Definition \ref{SPQB}).

      So, $R_n=\underline{\leftrightarrow}_q$, as desired.

  \item \textbf{Axiom $PR2$}. For a relation $R_1=Eq(\langle \Pi_1(\alpha\cdot u),\varrho\rangle,\langle \alpha,\varrho\rangle)\cup Eq(\langle \Pi_1(\breve{\alpha}\cdot u),\varrho\rangle,\langle \breve{\alpha},\varrho\rangle)$, with $u\in\mathbb{SP}^{(\infty)}(pqBPA+PR)$.

      By use of the probabilistic transition rules and the action transition rules for sequential operator $\cdot$ and projection operator $\Pi_n(n\geq 1)$, we get:

      $$\frac{\langle \alpha,\varrho\rangle\rightsquigarrow \langle \breve{\alpha},\varrho\rangle}{\langle \Pi_1(\alpha\cdot u),\varrho\rangle\rightsquigarrow\langle \Pi_1(\breve{\alpha}\cdot u),\varrho\rangle}$$

      $$\frac{\langle \alpha,\varrho\rangle\rightsquigarrow \langle \breve{\alpha},\varrho\rangle}{\langle \alpha,\varrho\rangle\rightsquigarrow\langle \breve{\alpha},\varrho\rangle}$$

      With the assumptions $\Pi_1(\alpha\cdot u)=\alpha$ and $\Pi_1(\breve{\alpha}\cdot u)=\breve{\alpha}$, we get $R_1$ satisfies condition 1 in the definition of $\underline{\leftrightarrow}_q$
      (see Definition \ref{SPQB}).

      $$\frac{\langle \breve{\alpha},\varrho\rangle\xrightarrow{\alpha} \langle \surd,\varrho'\rangle}{\langle \Pi_1(\breve{\alpha}\cdot u),\varrho\rangle\xrightarrow{\alpha}\langle \surd,\varrho'\rangle}$$

      $$\frac{\langle \breve{\alpha},\varrho\rangle\xrightarrow{\alpha} \langle \surd,\varrho'\rangle}{\langle \breve{\alpha},\varrho\rangle\xrightarrow{\alpha}\langle \surd,\varrho'\rangle}$$

      With the assumption $\Pi_1(\breve{\alpha}\cdot u)=\breve{\alpha}$, we get $R_1$ satisfies conditions 2 and 3 in the definition of $\underline{\leftrightarrow}_q$
      (see Definition \ref{SPQB}).

      So, $R_1=\underline{\leftrightarrow}_q$, as desired.

  \item \textbf{Axiom $PR3$}. For a relation $R_n(n\geq 1)=Eq(\langle \Pi_{n+1}(\alpha\cdot u),\varrho\rangle,\langle \alpha\cdot\Pi_n(u),\varrho\rangle)\cup Eq(\langle \Pi_{n+1}(\breve{\alpha}\cdot u),\varrho\rangle,\langle \breve{\alpha}\cdot\Pi_n(u),\varrho\rangle)$, with $u\in\mathbb{SP}^{(\infty)}(pqBPA+PR)$.

      By use of the probabilistic transition rules and the action transition rules for atomic unitary operator, sequential operator $\cdot$ and projection operator $\Pi_n(n\geq 1)$, we get:

      $$\frac{\langle \alpha,\varrho\rangle\rightsquigarrow \langle \breve{\alpha},\varrho\rangle}{\langle \Pi_{n+1}(\alpha\cdot u),\varrho\rangle\rightsquigarrow\langle \Pi_{n+1}(\breve{\alpha}\cdot u),\varrho\rangle}$$

      $$\frac{\langle \alpha,\varrho\rangle\rightsquigarrow \langle \breve{\alpha},\varrho\rangle}{\langle \alpha\cdot\Pi_n(u),\varrho\rangle\rightsquigarrow\langle \breve{\alpha}\cdot\Pi_n(u),\varrho\rangle}$$

      With the assumptions $\Pi_{n+1}(\alpha\cdot u)=\alpha\cdot\Pi_n(u)$ and $\Pi_{n+1}(\breve{\alpha}\cdot u)=\breve{\alpha}\cdot\Pi_n(u)$, we get $R_n$ satisfies condition 1 in the definition of $\underline{\leftrightarrow}_q$
      (see Definition \ref{SPQB}).

      $$\frac{\langle \breve{\alpha},\varrho\rangle\xrightarrow{\alpha} \langle \surd,\varrho'\rangle}{\langle \Pi_{n+1}(\breve{\alpha}\cdot u),\varrho\rangle\xrightarrow{\alpha}\langle \Pi_n(u),\varrho'\rangle}$$

      $$\frac{\langle \breve{\alpha},\varrho\rangle\xrightarrow{\alpha} \langle \surd,\varrho'\rangle}{\langle \breve{\alpha}\cdot\Pi_n(u),\varrho\rangle\xrightarrow{\alpha}\langle \Pi_n(u),\varrho'\rangle}$$

      With the assumption $\Pi_{n+1}(\breve{\alpha}\cdot u)=\breve{\alpha}\cdot\Pi_n(u)$, we get $R_n$ satisfies conditions 2 and 3 in the definition of $\underline{\leftrightarrow}_q$
      (see Definition \ref{SPQB}).

      So, $R_n=\underline{\leftrightarrow}_q$, as desired.

  \item \textbf{Axiom $PR4$}. For a relation $R_n(n\geq 1)=Eq(\langle \Pi_n(u+v),\varrho\rangle,\langle \Pi_n(u)+\Pi_n(v),\varrho\rangle)\cup Eq(\langle \Pi_n(p+q),\varrho\rangle,\langle \Pi_n(p)+\Pi_n(q),\varrho\rangle)$, with $u,v\in\mathbb{SP}^{(\infty)}(pqBPA+PR)$ and $p,q\in\mathbb{DP}^{(\infty)}(pqBPA+PR)$.

      By use of the probabilistic transition rules and the action transition rules for alternative operator $+$ and projection operator $\Pi_n(n\geq 1)$, we get:

      $$\frac{\langle u,\varrho\rangle\rightsquigarrow \langle u',\varrho\rangle,\langle v,\varrho\rangle\rightsquigarrow \langle v',\varrho\rangle}{\langle \Pi_n(u+v),\varrho\rangle\rightsquigarrow\langle \Pi_n(u'+v'),\varrho\rangle}$$

      $$\frac{\langle u,\varrho\rangle\rightsquigarrow \langle u',\varrho\rangle,\langle v,\varrho\rangle\rightsquigarrow \langle v',\varrho\rangle}{\langle \Pi_n(u)+\Pi_n(v),\varrho\rangle\rightsquigarrow\langle \Pi_n(u')+\Pi_n(v'),\varrho\rangle}$$

      With the assumptions $\Pi_n(u+v)=\Pi_n(u)+\Pi_n(v)$ and $\Pi_n(u'+v')=\Pi_n(u')+\Pi_n(v')$, we get $R_n$ satisfies condition 1 in the definition of $\underline{\leftrightarrow}_q$
      (see Definition \ref{SPQB}).

      $$\frac{\langle p,\varrho\rangle\xrightarrow{\alpha} \langle p',\varrho'\rangle}{\langle \Pi_n(p+q),\varrho\rangle\xrightarrow{\alpha}\langle \Pi_{n-1}(p'),\varrho'\rangle}\quad n\geq 2$$

      $$\frac{\langle p,\varrho\rangle\xrightarrow{\alpha} \langle p',\varrho'\rangle}{\langle \Pi_n(p)+\Pi_n(q),\varrho\rangle\xrightarrow{\alpha}\langle \Pi_{n-1}(p'),\varrho'\rangle}\quad n\geq 2$$

      $$\frac{\langle p,\varrho\rangle\xrightarrow{\alpha} \langle \surd,\varrho'\rangle}{\langle \Pi_n(p+q),\varrho\rangle\xrightarrow{\alpha}\langle \surd,\varrho'\rangle}\quad n\geq 2$$

      $$\frac{\langle p,\varrho\rangle\xrightarrow{\alpha} \langle \surd,\varrho'\rangle}{\langle \Pi_n(p)+\Pi_n(q),\varrho\rangle\xrightarrow{\alpha}\langle \surd,\varrho'\rangle}\quad n\geq 2$$

      $$\frac{\langle p,\varrho\rangle\xrightarrow{\alpha} \langle p',\varrho'\rangle}{\langle \Pi_1(p+q),\varrho\rangle\xrightarrow{\alpha}\langle \surd,\varrho'\rangle}$$

      $$\frac{\langle p,\varrho\rangle\xrightarrow{\alpha} \langle p',\varrho'\rangle}{\langle \Pi_1(p)+\Pi_1(q),\varrho\rangle\xrightarrow{\alpha}\langle \surd,\varrho'\rangle}$$

      With the assumption $\Pi_n(p+q)=\Pi_n(p)+\Pi_n(q)$, we get $R_n$ satisfies conditions 2 and 3 in the definition of $\underline{\leftrightarrow}_q$
      (see Definition \ref{SPQB}).

      So, $R_n=\underline{\leftrightarrow}_q$, as desired.

  \item \textbf{Axiom $prPR$}. For a relation $R_n(n\geq 1)=Eq(\langle \Pi_n(u\boxplus_{\pi}v),\varrho\rangle,\langle \Pi_n(u)\boxplus_{\pi}\Pi_n(v),\varrho\rangle)$, with $u,v\in\mathbb{SP}^{(\infty)}(pqBPA+PR)$.

      By use of the probabilistic transition rules for probabilistic choice operator $\boxplus_{\pi}$ and projection operator $\Pi_n(n\geq 1)$, we get:

      $$\frac{\langle u,\varrho\rangle\rightsquigarrow \langle u',\varrho\rangle}{\langle \Pi_n(u\boxplus_{\pi}v),\varrho\rangle\rightsquigarrow\langle \Pi_n(u'),\varrho\rangle}$$

      $$\frac{\langle u,\varrho\rangle\rightsquigarrow \langle u',\varrho\rangle}{\langle \Pi_n(u)\boxplus_{\pi}\Pi_n(v),\varrho\rangle\rightsquigarrow\langle \Pi_n(u'),\varrho\rangle}$$

      With the assumption $\Pi_n(u\boxplus_{\pi}v)=\Pi_n(u)\boxplus_{\pi}\Pi_n(v)$, we get $R_n$ satisfies condition 1 in the definition of $\underline{\leftrightarrow}_q$
      (see Definition \ref{SPQB}).

      Note that, there are not action transition rules for probabilistic choice operator $\boxplus_{\pi}$.

      So, $R_n=\underline{\leftrightarrow}_q$, as desired.
\end{itemize}
\end{proof}

\theorem[$AIP^-$ in $\mathbb{PT}^{(\infty)}(pqBPA+PR)$]{If for all $n\geq 1$, $\langle\Pi_n(p),\varrho\rangle\underline{\leftrightarrow}_q \langle\Pi_n(q),\varrho\rangle$, then $\langle p,\varrho\rangle\underline{\leftrightarrow}_q \langle q,\varrho\rangle$.}

\begin{proof}
The conditions on PDF $\mu$ are same in $\underline{\leftrightarrow}_q$ and $\underline{\leftrightarrow}$, (see in section \ref{PPA} and \cite{PPA} for details), we only need to process the quantum information $\varrho$ related parts on projections.

The proof is similar to that of Theorem ($AIP^-$ in $\mathbb{PT}^{(\infty)}(pBPA+PR)$), please refer to \cite{PPA} for details. Next, we only give the skeleton of the proof.

For a relation $R=R_1\cup R_2$, $R_1=Eq(\langle u,\varrho\rangle,\langle v,\varrho\rangle)$, with $\langle \Pi_n(u),\varrho\rangle\underline{\leftrightarrow}_q\langle \Pi_n(v),\varrho\rangle(n\geq 1)$ and $u,v\in\mathbb{SP}^{(\infty)}(pqBPA+PR)$, $R_2=Eq(\langle p,\varrho\rangle,\langle q,\varrho\rangle)$, with $\langle \Pi_n(p),\varrho\rangle\underline{\leftrightarrow}_q\langle \Pi_n(q),\varrho\rangle(n\geq 1)$ and $p,q\in\mathbb{DP}^{(\infty)}(pqBPA+PR)$.

For probabilistic transitions, surely, $u$ and $v$ have head normal forms (HNFs) (see in \cite{PPA}). Suppose $\langle u,\varrho\rangle\rightsquigarrow \langle p,\varrho\rangle$, $p\equiv \breve{u_i}$ for some $i$, and will eventually lead to  $\langle v,\varrho\rangle\rightsquigarrow \langle q,\varrho\rangle$ for some $q$, and $\langle\Pi_n(p),\varrho\rangle\underline{\leftrightarrow}_q \langle\Pi_n(q),\varrho\rangle$, $(\langle p,\varrho\rangle,\langle q,\varrho\rangle)\in R$.

For action transitions, $p$ and $q$ have head normal forms (HNFs) (see in \cite{PPA}). Suppose $\langle p,\varrho\rangle\xrightarrow{\alpha} \langle u,\varrho'\rangle$ for some $u$, $u\equiv \breve{u_j}$ for some $j$, and will eventually lead to $\langle q,\varrho\rangle\xrightarrow{\alpha} \langle v,\varrho'\rangle$ for some $v$, and $\langle\Pi_n(u),\varrho\rangle\underline{\leftrightarrow}_q \langle\Pi_n(v),\varrho\rangle$, $(\langle u,\varrho'\rangle,\langle v,\varrho'\rangle)\in R$.

For action termination, $\langle p,\varrho\rangle\xrightarrow{\alpha}\langle \surd,\varrho'\rangle$, then $\langle\Pi_1(p),\varrho\rangle\xrightarrow{\alpha}\langle \surd,\varrho'\rangle$. Since $\langle\Pi_1(p),\varrho\rangle\underline{\leftrightarrow}_q \langle\Pi_1(q),\varrho\rangle$, so, we get $\langle\Pi_1(q),\varrho\rangle\xrightarrow{\alpha}\langle \surd,\varrho'\rangle$. So, $\langle q,\varrho\rangle\xrightarrow{\alpha}\langle \surd,\varrho'\rangle$.
\end{proof}

\theorem[Soundness of $pqBPA$]{Let $x$ and $y$ be closed $pqBPA$ terms. If $pqBPA\vdash x=y$ then $\langle x,\varrho\rangle\underline{\leftrightarrow}_q \langle y,\varrho\rangle$.}

\begin{proof}
By Theorem \ref{SPQBPAPR}.
\end{proof}

\theorem[Completeness of $pqBPA$]{Let $z$ and $u$ are closed $pqBPA$ terms, if $\langle z,\varrho\rangle\underline{\leftrightarrow}_q \langle u,\varrho\rangle$, then $pqBPA\vdash z=u$.}

\begin{proof}
It is based on the following three facts: let $z$ and $u$ are closed $pqBPA$ terms,

\begin{enumerate}
  \item $\langle z,\varrho\rangle\underline{\leftrightarrow}_q \langle u,\varrho\rangle \Rightarrow z\underline{\leftrightarrow} u$ by Proposition \ref{PQBPB};
  \item $z\underline{\leftrightarrow} u \Rightarrow pBPA\vdash z=u$ (see in section \ref{PPA}, and for details in \cite{PPA});
  \item the term systems of $pBPA$ and $pqBPA$ are same, including the same operators and the same axiom systems, but different atomic action meanings (semantics).
\end{enumerate}

So, we get $\langle z,\varrho\rangle\underline{\leftrightarrow}_q \langle u,\varrho\rangle \Rightarrow pqBPA\vdash z=u$, as desired.
\end{proof}

\theorem[Soundness of $pqACP^+$]{Let $p$ and $q$ be closed $pqACP^+$ terms. If $pqACP^+\vdash p=q$ then $\langle p,\varrho\rangle\underline{\leftrightarrow}_q \langle q,\varrho\rangle$.}

\begin{proof}
It is already proven that $\underline{\leftrightarrow}_q$ is an equivalent and congruent relation on $\mathbb{PT}^{(\infty)}(pACP^+)$ (see details in \cite{PPA}), and $pACP^+\vdash x=y$ then $x\underline{\leftrightarrow} y$ (that is, the conditions on PDF $\mu$ are same in $\underline{\leftrightarrow}_q$ and $\underline{\leftrightarrow}$, see in section \ref{PPA} and \cite{PPA} for details), we only need to prove that the quantum information $\varrho$ related parts are also sound.

It is sufficient to examine every axiom in the axiom systems which are added to $pqBPA$ to obtain $pqACP^+$ (in Table \ref{AxiomForpACP+}) is sound.

\begin{itemize}
  \item \textbf{Axiom $CF$}. For a relation $R=R_1\cup R_2$, suppose that $Eq(\langle a\mid b,\varrho\rangle,\langle c,\varrho\rangle)= R_1$ with $c=\gamma(a,b)$, and $Eq(\langle \breve{a}\mid\breve{b},\varrho\rangle,\langle \breve{c},\varrho\rangle)= R_2$ with $\breve{c}=\gamma(\breve{a},\breve{b})$.

      By use of the probabilistic transition rules and the action transition rules for atomic communication action and communication merge $\mid$, we get:

      $$\frac{\langle a,\varrho\rangle\rightsquigarrow \langle \breve{a},\varrho\rangle, \langle b,\varrho\rangle\rightsquigarrow \langle \breve{b},\varrho\rangle}{\langle a\mid b,\varrho\rangle\rightsquigarrow\langle \breve{a}\mid\breve{b},\varrho\rangle}$$

      $$\frac{\langle \breve{a},\varrho\rangle\xrightarrow{a} \langle \surd,\varrho\rangle,\langle \breve{b},\varrho\rangle\xrightarrow{b} \langle \surd,\varrho\rangle}{\langle \breve{a}\mid\breve{b},\varrho\rangle\xrightarrow{c}\langle \surd,\varrho\rangle}$$

      With the assumptions $a\mid b=c$ and $\breve{a}\mid\breve{b}=\breve{c}$, we get $R$ satisfies conditions 1, 2 and 3 in the definition of $\underline{\leftrightarrow}_q$
      (see Definition \ref{SPQB}).

      So, $R=\underline{\leftrightarrow}_q$, as desired.

  \item \textbf{Axiom $CM2$}. For a relation $R=R_1\cup R_2$, suppose that $Eq(\langle a\leftmerge u,\varrho\rangle,\langle a\cdot u,\varrho\rangle)= R_1$ with $u\in\mathbb{SP}(pqACP^+)$, and $Eq(\langle \breve{a}\leftmerge u,\varrho\rangle,\langle \breve{a}\cdot u,\varrho\rangle)= R_2$ with $u\in\mathbb{SP}(pqACP^+)$.

      By use of the probabilistic transition rules and the action transition rules for atomic communication action $a$ and left merge $\leftmerge$, we get:

      $$\frac{\langle a,\varrho\rangle\rightsquigarrow \langle \breve{a},\varrho\rangle}{\langle a\leftmerge u,\varrho\rangle\rightsquigarrow\langle \breve{a}\leftmerge u,\varrho\rangle}$$

      $$\frac{\langle a,\varrho\rangle\rightsquigarrow \langle \breve{a},\varrho\rangle}{\langle a\cdot u,\varrho\rangle\rightsquigarrow\langle \breve{a}\cdot u,\varrho\rangle}$$

      $$\frac{\langle \breve{a},\varrho\rangle\xrightarrow{a} \langle \surd,\varrho\rangle}{\langle \breve{a}\leftmerge u,\varrho\rangle\xrightarrow{a}\langle u,\varrho\rangle}$$

      $$\frac{\langle \breve{a},\varrho\rangle\xrightarrow{a} \langle \surd,\varrho\rangle}{\langle \breve{a}\cdot u,\varrho\rangle\xrightarrow{a}\langle u,\varrho\rangle}$$

      With the assumptions $a\leftmerge u=a\cdot u$ and $\breve{a}\leftmerge u=\breve{a}\cdot u$, we get $R$ satisfies conditions 1, 2 and 3 in the definition of $\underline{\leftrightarrow}_q$
      (see Definition \ref{SPQB}).

      So, $R=\underline{\leftrightarrow}_q$, as desired.

      For a relation $R'=R_1'\cup R_2'$, suppose that $Eq(\langle \alpha\leftmerge u,\varrho\rangle,\langle \alpha\cdot u,\varrho\rangle)= R_1'$ with $u\in\mathbb{SP}(pqACP^+)$, and $Eq(\langle \breve{\alpha}\leftmerge u,\varrho\rangle,\langle \breve{\alpha}\cdot u,\varrho\rangle)= R_2'$ with $u\in\mathbb{SP}(pqACP^+)$.

      By use of the probabilistic transition rules and the action transition rules for atomic unitary operator $\alpha$ and left merge $\leftmerge$, we get:

      $$\frac{\langle \alpha,\varrho\rangle\rightsquigarrow \langle \breve{\alpha},\varrho\rangle}{\langle \alpha\leftmerge u,\varrho\rangle\rightsquigarrow\langle \breve{\alpha}\leftmerge u,\varrho\rangle}$$

      $$\frac{\langle \alpha,\varrho\rangle\rightsquigarrow \langle \breve{\alpha},\varrho\rangle}{\langle \alpha\cdot u,\varrho\rangle\rightsquigarrow\langle \breve{\alpha}\cdot u,\varrho\rangle}$$

      $$\frac{\langle \breve{\alpha},\varrho\rangle\xrightarrow{\alpha} \langle \surd,\varrho'\rangle}{\langle \breve{\alpha}\leftmerge u,\varrho\rangle\xrightarrow{\alpha}\langle u,\varrho'\rangle}$$

      $$\frac{\langle \breve{\alpha},\varrho\rangle\xrightarrow{\alpha} \langle \surd,\varrho'\rangle}{\langle \breve{\alpha}\cdot u,\varrho\rangle\xrightarrow{\alpha}\langle u,\varrho'\rangle}$$

      With the assumptions $\alpha\leftmerge u=\alpha\cdot u$ and $\breve{\alpha}\leftmerge u=\breve{\alpha}\cdot u$, we get $R'$ satisfies conditions 1, 2 and 3 in the definition of $\underline{\leftrightarrow}_q$
      (see Definition \ref{SPQB}).

      So, $R'=\underline{\leftrightarrow}_q$, as desired.

  \item \textbf{Axiom $CM3$}. For a relation $R=R_1\cup R_2$, suppose that $Eq(\langle a\cdot u\leftmerge v,\varrho\rangle,\langle a\cdot (u\parallel v),\varrho\rangle)= R_1$ with $u,v\in\mathbb{SP}(pqACP^+)$, and $Eq(\langle \breve{a}\cdot u\leftmerge v,\varrho\rangle,\langle \breve{a}\cdot (u\parallel v),\varrho\rangle)= R_2$ with $u,v\in\mathbb{SP}(pqACP^+)$.

      By use of the probabilistic transition rules and the action transition rules for atomic communication action $a$, sequential composition $\cdot$ and left merge $\leftmerge$, we get:

      $$\frac{\langle a,\varrho\rangle\rightsquigarrow \langle \breve{a},\varrho\rangle}{\langle a\cdot u\leftmerge v,\varrho\rangle\rightsquigarrow\langle \breve{a}\cdot u\leftmerge v,\varrho\rangle}$$

      $$\frac{\langle a,\varrho\rangle\rightsquigarrow \langle \breve{a},\varrho\rangle}{\langle a\cdot (u\parallel v),\varrho\rangle\rightsquigarrow\langle \breve{a}\cdot (u\parallel v),\varrho\rangle}$$

      $$\frac{\langle \breve{a},\varrho\rangle\xrightarrow{a} \langle \surd,\varrho\rangle}{\langle \breve{a}\cdot u\leftmerge v,\varrho\rangle\xrightarrow{a}\langle u\parallel v,\varrho\rangle}$$

      $$\frac{\langle \breve{a},\varrho\rangle\xrightarrow{a} \langle \surd,\varrho\rangle}{\langle \breve{a}\cdot (u\parallel v),\varrho\rangle\xrightarrow{a}\langle u\parallel v,\varrho\rangle}$$

      With the assumptions $a\cdot u\leftmerge v=a\cdot (u\parallel v)$ and $\breve{a}\cdot u\leftmerge v=\breve{a}\cdot (u\parallel v)$, we get $R$ satisfies conditions 1, 2 and 3 in the definition of $\underline{\leftrightarrow}_q$
      (see Definition \ref{SPQB}).

      So, $R=\underline{\leftrightarrow}_q$, as desired.

      For a relation $R'=R_1'\cup R_2'$, suppose that $Eq(\langle \alpha\cdot u\leftmerge v,\varrho\rangle,\langle \alpha\cdot (u\parallel v),\varrho\rangle)= R_1'$ with $u,v\in\mathbb{SP}(pqACP^+)$, and $Eq(\langle \breve{\alpha}\cdot u\leftmerge v,\varrho\rangle,\langle \breve{\alpha}\cdot (u\parallel v),\varrho\rangle)= R_2'$ with $u,v\in\mathbb{SP}(pqACP^+)$.

      By use of the probabilistic transition rules and the action transition rules for atomic unitary operator $\alpha$, sequential composition $\cdot$ and left merge $\leftmerge$, we get:

      $$\frac{\langle \alpha,\varrho\rangle\rightsquigarrow \langle \breve{\alpha},\varrho\rangle}{\langle \alpha\cdot u\leftmerge v,\varrho\rangle\rightsquigarrow\langle \breve{\alpha}\cdot u\leftmerge v,\varrho\rangle}$$

      $$\frac{\langle \alpha,\varrho\rangle\rightsquigarrow \langle \breve{\alpha},\varrho\rangle}{\langle \alpha\cdot (u\parallel v),\varrho\rangle\rightsquigarrow\langle \breve{\alpha}\cdot (u\parallel v),\varrho\rangle}$$

      $$\frac{\langle \breve{\alpha},\varrho\rangle\xrightarrow{\alpha} \langle \surd,\varrho'\rangle}{\langle \breve{\alpha}\cdot u\leftmerge v,\varrho\rangle\xrightarrow{\alpha}\langle u\parallel v,\varrho'\rangle}$$

      $$\frac{\langle \breve{\alpha},\varrho\rangle\xrightarrow{\alpha} \langle \surd,\varrho'\rangle}{\langle \breve{\alpha}\cdot (u\parallel v),\varrho\rangle\xrightarrow{\alpha}\langle u\parallel v,\varrho'\rangle}$$

      With the assumptions $\alpha\cdot u\leftmerge v=\alpha\cdot (u\parallel v)$ and $\breve{\alpha}\cdot u\leftmerge v=\breve{\alpha}\cdot (u\parallel v)$, we get $R'$ satisfies conditions 1, 2 and 3 in the definition of $\underline{\leftrightarrow}_q$
      (see Definition \ref{SPQB}).

      So, $R'=\underline{\leftrightarrow}_q$, as desired.

  \item \textbf{Axiom $CM4$}. For a relation $R=R_1\cup R_2$, suppose that $Eq(\langle (p+q)\leftmerge s,\varrho\rangle,\langle p\leftmerge s + q\leftmerge s,\varrho\rangle)= R_1$ with $p,q\in \mathbb{DP}(pqACP^+)$, $s\in \mathbb{SP}(pqACP^+)$, and $Eq(\langle (u+v)\leftmerge s,\varrho\rangle,\langle u\leftmerge s + v\leftmerge s,\varrho\rangle)= R_2$ with $u,v,s\in \mathbb{SP}(pqACP^+)$.

      By use of the probabilistic transition rules and the action transition rules for left merge $\leftmerge$, alternative operator $+$, we get:

      $$\frac{\langle u,\varrho\rangle\rightsquigarrow \langle u',\varrho\rangle, \langle v,\varrho\rangle\rightsquigarrow \langle v',\varrho\rangle}{\langle (u + v)\leftmerge s,\varrho\rangle\rightsquigarrow\langle (u' + v')\leftmerge s,\varrho\rangle}$$

      $$\frac{\langle u,\varrho\rangle\rightsquigarrow \langle u',\varrho\rangle, \langle v,\varrho\rangle\rightsquigarrow \langle v',\varrho\rangle}{\langle u\leftmerge s+v\leftmerge s,\varrho\rangle\rightsquigarrow\langle u'\leftmerge s+v'\leftmerge s,\varrho\rangle}$$

      With the assumptions $(u+v)\leftmerge s=u\leftmerge s+v\leftmerge s$ and $(u'+v')\leftmerge s=u'\leftmerge s+v'\leftmerge s$, we get $R_2$ satisfies condition 1 in the definition of $\underline{\leftrightarrow}_q$
      (see Definition \ref{SPQB}).

      $$\frac{\langle p,\varrho\rangle\xrightarrow{\alpha} \langle p',\varrho'\rangle}{\langle (p+ q)\leftmerge s,\varrho\rangle\xrightarrow{\alpha}\langle p'\leftmerge s,\varrho'\rangle}$$

      $$\frac{\langle p,\varrho\rangle\xrightarrow{\alpha} \langle p',\varrho'\rangle}{\langle p\leftmerge s+q\leftmerge s,\varrho\rangle\xrightarrow{\alpha}\langle p'\leftmerge s,\varrho'\rangle}$$

      $$\frac{\langle p,\varrho\rangle\xrightarrow{\alpha} \langle \surd,\varrho'\rangle}{\langle (p+ q)\leftmerge s,\varrho\rangle\xrightarrow{\alpha}\langle s,\varrho'\rangle}$$

      $$\frac{\langle p,\varrho\rangle\xrightarrow{\alpha} \langle \surd,\varrho'\rangle}{\langle p\leftmerge s+q\leftmerge s,\varrho\rangle\xrightarrow{\alpha}\langle s,\varrho'\rangle}$$

      With the assumptions $(p+q)\leftmerge s=p\leftmerge s + q\leftmerge s$, we get $R_1$ satisfies conditions 2 and 3 in the definition of $\underline{\leftrightarrow}_q$
      (see Definition \ref{SPQB}).

      So, $R=\underline{\leftrightarrow}_q$, as desired.

  \item \textbf{Axiom $PrCM1$}. For a relation $R$, suppose that $Eq(\langle (u\boxplus_{\pi}v)\leftmerge w,\varrho\rangle,\langle u\leftmerge w\boxplus_{\pi}v\leftmerge w,\varrho\rangle)= R$ with $u,v,w\in \mathbb{SP}(pqACP^+)$.

      By use of the probabilistic transition rules for probabilistic choice operator $\boxplus_{\pi}$ and left merge $\leftmerge$, we get:

      $$\frac{\langle u,\varrho\rangle\rightsquigarrow \langle u',\varrho\rangle}{\langle (u\boxplus_{\pi}v)\leftmerge w,\varrho\rangle\rightsquigarrow\langle u'\leftmerge w,\varrho\rangle}$$

      $$\frac{\langle u,\varrho\rangle\rightsquigarrow \langle u',\varrho\rangle}{\langle u\leftmerge w\boxplus_{\pi}v\leftmerge w,\varrho\rangle\rightsquigarrow\langle u'\leftmerge w,\varrho\rangle}$$

      With the assumption $(u\boxplus_{\pi}v)\leftmerge w=u\leftmerge w\boxplus_{\pi}v\leftmerge w$, we get $R$ satisfies condition 1 in the definition of $\underline{\leftrightarrow}_q$
      (see Definition \ref{SPQB}). Note that, for probabilistic choice operator $\boxplus_{\pi}$, there are not action transition rules.

      So, $R=\underline{\leftrightarrow}_q$, as desired.

  \item \textbf{Axiom $CM5$}. For a relation $R=R_1\cup R_2$, suppose that $Eq(\langle a\cdot u\mid b,\varrho\rangle,\langle c\cdot u,\varrho\rangle)= R_1$ with $c=\gamma(a,b)$ and $u\in \mathbb{SP}(pqACP^+)$, and $Eq(\langle \breve{a}\cdot u\mid\breve{b},\varrho\rangle,\langle \breve{c}\cdot u,\varrho\rangle)= R_2$ with $\breve{c}=\gamma(\breve{a},\breve{b})$ and $u\in \mathbb{SP}(pqACP^+)$.

      By use of the probabilistic transition rules and the action transition rules for atomic communication action $a,b,c$, sequential composition $\cdot$ and communication merge $\mid$, we get:

      $$\frac{\langle a,\varrho\rangle\rightsquigarrow \langle \breve{a},\varrho\rangle, \langle b,\varrho\rangle\rightsquigarrow \langle \breve{b},\varrho\rangle}{\langle a\cdot u\mid b,\varrho\rangle\rightsquigarrow\langle \breve{a}\cdot u\mid\breve{b},\varrho\rangle}$$

      $$\frac{\langle c,\varrho\rangle\rightsquigarrow \langle \breve{c},\varrho\rangle}{\langle c\cdot u,\varrho\rangle\rightsquigarrow\langle \breve{c}\cdot u,\varrho\rangle}$$

      $$\frac{\langle \breve{a},\varrho\rangle\xrightarrow{a} \langle \surd,\varrho\rangle,\langle \breve{b},\varrho\rangle\xrightarrow{b} \langle \surd,\varrho\rangle}{\langle \breve{a}\cdot u\mid\breve{b},\varrho\rangle\xrightarrow{c}\langle u,\varrho\rangle}$$

      $$\frac{\langle \breve{c},\varrho\rangle\xrightarrow{c} \langle \surd,\varrho\rangle}{\langle \breve{c}\cdot u,\varrho\rangle\xrightarrow{c}\langle u,\varrho\rangle}$$

      With the assumptions $a\cdot u\mid b=c\cdot u$ and $\breve{a}\cdot u\mid\breve{b}=\breve{c}\cdot u$, we get $R$ satisfies conditions 1, 2 and 3 in the definition of $\underline{\leftrightarrow}_q$
      (see Definition \ref{SPQB}).

      So, $R=\underline{\leftrightarrow}_q$, as desired.

  \item \textbf{Axiom $CM6$}. For a relation $R=R_1\cup R_2$, suppose that $Eq(\langle a\mid b\cdot u,\varrho\rangle,\langle c\cdot u,\varrho\rangle)= R_1$ with $c=\gamma(a,b)$ and $u\in \mathbb{SP}(pqACP^+)$, and $Eq(\langle \breve{a}\mid\breve{b}\cdot u,\varrho\rangle,\langle \breve{c}\cdot u,\varrho\rangle)= R_2$ with $\breve{c}=\gamma(\breve{a},\breve{b})$ and $u\in \mathbb{SP}(pqACP^+)$.

      By use of the probabilistic transition rules and the action transition rules for atomic communication action $a,b,c$, sequential composition $\cdot$ and communication merge $\mid$, we get:

      $$\frac{\langle a,\varrho\rangle\rightsquigarrow \langle \breve{a},\varrho\rangle, \langle b,\varrho\rangle\rightsquigarrow \langle \breve{b},\varrho\rangle}{\langle a\mid b\cdot u,\varrho\rangle\rightsquigarrow\langle \breve{a}\mid\breve{b}\cdot u,\varrho\rangle}$$

      $$\frac{\langle c,\varrho\rangle\rightsquigarrow \langle \breve{c},\varrho\rangle}{\langle c\cdot u,\varrho\rangle\rightsquigarrow\langle \breve{c}\cdot u,\varrho\rangle}$$

      $$\frac{\langle \breve{a},\varrho\rangle\xrightarrow{a} \langle \surd,\varrho\rangle,\langle \breve{b},\varrho\rangle\xrightarrow{b} \langle \surd,\varrho\rangle}{\langle \breve{a}\mid\breve{b}\cdot u,\varrho\rangle\xrightarrow{c}\langle u,\varrho\rangle}$$

      $$\frac{\langle \breve{c},\varrho\rangle\xrightarrow{c} \langle \surd,\varrho\rangle}{\langle \breve{c}\cdot u,\varrho\rangle\xrightarrow{c}\langle u,\varrho\rangle}$$

      With the assumptions $a\mid b\cdot u=c\cdot u$ and $\breve{a}\mid\breve{b}\cdot u=\breve{c}\cdot u$, we get $R$ satisfies conditions 1, 2 and 3 in the definition of $\underline{\leftrightarrow}_q$
      (see Definition \ref{SPQB}).

      So, $R=\underline{\leftrightarrow}_q$, as desired.

  \item \textbf{Axiom $CM7$}. For a relation $R=R_1\cup R_2$, suppose that $Eq(\langle a\cdot u\mid b\cdot v,\varrho\rangle,\langle c\cdot (u\parallel v),\varrho\rangle)= R_1$ with $c=\gamma(a,b)$ and $u,v\in \mathbb{SP}(pqACP^+)$, and $Eq(\langle \breve{a}\cdot u\mid\breve{b}\cdot v,\varrho\rangle,\langle \breve{c}\cdot (u\parallel v),\varrho\rangle)= R_2$ with $\breve{c}=\gamma(\breve{a},\breve{b})$ and $u,v\in \mathbb{SP}(pqACP^+)$.

      By use of the probabilistic transition rules and the action transition rules for atomic communication action $a,b,c$, sequential composition $\cdot$ and communication merge $\mid$, we get:

      $$\frac{\langle a,\varrho\rangle\rightsquigarrow \langle \breve{a},\varrho\rangle, \langle b,\varrho\rangle\rightsquigarrow \langle \breve{b},\varrho\rangle}{\langle a\cdot u\mid b\cdot v,\varrho\rangle\rightsquigarrow\langle \breve{a}\cdot u\mid\breve{b}\cdot v,\varrho\rangle}$$

      $$\frac{\langle c,\varrho\rangle\rightsquigarrow \langle \breve{c},\varrho\rangle}{\langle c\cdot (u\parallel v),\varrho\rangle\rightsquigarrow\langle \breve{c}\cdot (u\parallel v),\varrho\rangle}$$

      $$\frac{\langle \breve{a},\varrho\rangle\xrightarrow{a} \langle \surd,\varrho\rangle,\langle \breve{b},\varrho\rangle\xrightarrow{b} \langle \surd,\varrho\rangle}{\langle \breve{a}\cdot u\mid\breve{b}\cdot v,\varrho\rangle\xrightarrow{c}\langle (u\parallel v),\varrho\rangle}$$

      $$\frac{\langle \breve{c},\varrho\rangle\xrightarrow{c} \langle \surd,\varrho\rangle}{\langle \breve{c}\cdot (u\parallel v),\varrho\rangle\xrightarrow{c}\langle (u\parallel v),\varrho\rangle}$$

      With the assumptions $a\cdot u\mid b\cdot v=c\cdot (u\parallel v)$ and $\breve{a}\cdot u\mid\breve{b}\cdot v=\breve{c}\cdot (u\parallel v)$, we get $R$ satisfies conditions 1, 2 and 3 in the definition of $\underline{\leftrightarrow}_q$
      (see Definition \ref{SPQB}).

      So, $R=\underline{\leftrightarrow}_q$, as desired.

  \item \textbf{Axiom $PrCM2$}. For a relation $R$, suppose that $Eq(\langle (u\boxplus_{\pi}v)\mid w,\varrho\rangle,\langle u\mid w\boxplus_{\pi}v\mid w,\varrho\rangle)= R$ with $u,v,w\in \mathbb{SP}(pqACP^+)$.

      By use of the probabilistic transition rules for probabilistic choice operator $\boxplus_{\pi}$ and communication merge $\mid$, we get:

      $$\frac{\langle u,\varrho\rangle\rightsquigarrow \langle u',\varrho\rangle}{\langle (u\boxplus_{\pi}v)\mid w,\varrho\rangle\rightsquigarrow\langle u'\mid w,\varrho\rangle}$$

      $$\frac{\langle u,\varrho\rangle\rightsquigarrow \langle u',\varrho\rangle}{\langle u\mid w\boxplus_{\pi}v\mid w,\varrho\rangle\rightsquigarrow\langle u'\mid w,\varrho\rangle}$$

      With the assumption $(u\boxplus_{\pi}v)\mid w=u\mid w\boxplus_{\pi}v\mid w$, we get $R$ satisfies condition 1 in the definition of $\underline{\leftrightarrow}_q$
      (see Definition \ref{SPQB}). Note that, for probabilistic choice operator $\boxplus_{\pi}$, there are not action transition rules.

      So, $R=\underline{\leftrightarrow}_q$, as desired.

  \item \textbf{Axiom $PrCM3$}. For a relation $R$, suppose that $Eq(\langle u\mid (v\boxplus_{\pi}w),\varrho\rangle,\langle u\mid v\boxplus_{\pi}u\mid w,\varrho\rangle)= R$ with $u,v,w\in \mathbb{SP}(pqACP^+)$.

      By use of the probabilistic transition rules for probabilistic choice operator $\boxplus_{\pi}$ and communication merge $\mid$, we get:

      $$\frac{\langle v,\varrho\rangle\rightsquigarrow \langle v',\varrho\rangle}{\langle u\mid (v\boxplus_{\pi}w),\varrho\rangle\rightsquigarrow\langle u\mid v',\varrho\rangle}$$

      $$\frac{\langle v,\varrho\rangle\rightsquigarrow \langle v',\varrho\rangle}{\langle u\mid v\boxplus_{\pi}u\mid w,\varrho\rangle\rightsquigarrow\langle u\mid v',\varrho\rangle}$$

      With the assumption $u\mid (v\boxplus_{\pi}w)=u\mid v\boxplus_{\pi}u\mid w$, we get $R$ satisfies condition 1 in the definition of $\underline{\leftrightarrow}_q$
      (see Definition \ref{SPQB}). Note that, for probabilistic choice operator $\boxplus_{\pi}$, there are not action transition rules.

      So, $R=\underline{\leftrightarrow}_q$, as desired.

  \item \textbf{Axiom $PrMM1$}. For a relation $R$, suppose that $Eq(\langle u\parallel v,\varrho\rangle,\langle (u,u)][(v,v),\varrho\rangle)= R$ with $u,v\in \mathbb{SP}(pqACP^+)$.

      By use of the probabilistic transition rules for operators $\parallel$ and $][$, we get:

      $$\frac{\langle u,\varrho\rangle\rightsquigarrow \langle u',\varrho\rangle, \langle v,\varrho\rangle\rightsquigarrow \langle v',\varrho\rangle}{\langle u \parallel v,\varrho\rangle\rightsquigarrow\langle u'\leftmerge v+v'\leftmerge u+u'\mid v',\varrho\rangle}$$

      $$\frac{\langle u,\varrho\rangle\rightsquigarrow \langle u',\varrho\rangle, \langle v,\varrho\rangle\rightsquigarrow \langle v',\varrho\rangle}{\langle (u,u)][(v,v),\varrho\rangle\rightsquigarrow\langle u'\leftmerge v+v'\leftmerge u+u'\mid v',\varrho\rangle}$$

      With the assumption $u\parallel v=(u,u)][(v,v)$, we get $R$ satisfies condition 1 in the definition of $\underline{\leftrightarrow}_q$
      (see Definition \ref{SPQB}). Note that, for operator $\parallel$ and $][$, there are not action transition rules.

      So, $R=\underline{\leftrightarrow}_q$, as desired.

  \item \textbf{Axiom $PrMM2$}. For a relation $R$, suppose that $Eq(\langle (u_1\boxplus_{\pi}u_2,z)][(v,w),\varrho\rangle, \langle (u_1,z)][(v,w) \boxplus_{\pi}(u_2,z)][(v,w),\varrho\rangle)= R$ with $u_1,u_2,v,z,w\in \mathbb{SP}(pqACP^+)$.

      By use of the probabilistic transition rules for probabilistic choice operator $\boxplus_{\pi}$, and $][$, we get:

      $$\frac{\langle u_1,\varrho\rangle\rightsquigarrow \langle u_1',\varrho\rangle, \langle v,\varrho\rangle\rightsquigarrow \langle v',\varrho\rangle}{\langle (u_1\boxplus_{\pi}u_2,z)][(v,w),\varrho\rangle\rightsquigarrow\langle u_1'\leftmerge w+v'\leftmerge z+u_1'\mid v',\varrho\rangle}$$

      $$\frac{\langle u_1,\varrho\rangle\rightsquigarrow \langle u_1',\varrho\rangle, \langle v,\varrho\rangle\rightsquigarrow \langle v',\varrho\rangle}{\langle (u_1,z)][(v,w)\boxplus_{\pi}(u_2,z)][(v,w),\varrho\rangle\rightsquigarrow\langle u_1'\leftmerge w+v'\leftmerge z+u_1'\mid v',\varrho\rangle}$$

      With the assumption $(u_1\boxplus_{\pi}u_2,z)][(v,w)= (u_1,z)][(v,w)\boxplus_{\pi}(u_2,z)][(v,w)$, we get $R$ satisfies condition 1 in the definition of $\underline{\leftrightarrow}_q$
      (see Definition \ref{SPQB}). Note that, for operator $\boxplus_{\pi}$ and $][$, there are not action transition rules.

      So, $R=\underline{\leftrightarrow}_q$, as desired.

  \item \textbf{Axiom $PrMM3$}. For a relation $R$, suppose that $Eq(\langle (u,z)][(v_1\boxplus_{\pi}v_2,w),\varrho\rangle,\langle (u,z)][(v_1,w)\boxplus_{\pi}(u,z)][(v_1,w),\varrho\rangle)= R$ with $u_1,u_2,v,z,w\in \mathbb{SP}(pqACP^+)$.

      By use of the probabilistic transition rules for probabilistic choice operator $\boxplus_{\pi}$, and $][$, we get:

      $$\frac{\langle u,\varrho\rangle\rightsquigarrow \langle u',\varrho\rangle, \langle v_1,\varrho\rangle\rightsquigarrow \langle v_1',\varrho\rangle}{\langle (u,z)][(v_1\boxplus_{\pi}v_2,w),\varrho\rangle\rightsquigarrow\langle u'\leftmerge w+v_1'\leftmerge z+u'\mid v_1',\varrho\rangle}$$

      $$\frac{\langle u,\varrho\rangle\rightsquigarrow \langle u',\varrho\rangle, \langle v_1,\varrho\rangle\rightsquigarrow \langle v_1',\varrho\rangle}{\langle (u,z)][(v_1,w)\boxplus_{\pi}(u,z)][(v_1,w),\varrho\rangle\rightsquigarrow\langle u'\leftmerge w+v_1'\leftmerge z+u'\mid v_1',\varrho\rangle}$$

      With the assumption $(u,z)][(v_1\boxplus_{\pi}v_2,w)=(u,z)][(v_1,w)\boxplus_{\pi}(u,z)][(v_1,w)$, we get $R$ satisfies condition 1 in the definition of $\underline{\leftrightarrow}_q$
      (see Definition \ref{SPQB}). Note that, for operator $\boxplus_{\pi}$ and $][$, there are not action transition rules.

      So, $R=\underline{\leftrightarrow}_q$, as desired.

  \item \textbf{Axiom $D1$}. For a relation $R=R_1\cup R_2$, suppose that $Eq(\langle \partial_H(\alpha),\varrho\rangle,\langle \alpha,\varrho\rangle)= R_1$ with $\alpha\notin H$, and $Eq(\langle \partial_H(\breve{\alpha}),\varrho\rangle,\langle \breve{\alpha},\varrho\rangle)= R_2$ with $\breve{\alpha}\notin H$.

      By use of the probabilistic transition rules and the action transition rules for atomic unitary operator and encapsulation operator $\partial_H$, we get:

      $$\frac{\langle \alpha,\varrho\rangle\rightsquigarrow \langle \breve{\alpha},\varrho\rangle}{\langle \partial_H(\alpha),\varrho\rangle\rightsquigarrow\langle \partial_H(\breve{\alpha}),\varrho\rangle}$$

      $$\frac{\langle \alpha,\varrho\rangle\rightsquigarrow \langle \breve{\alpha},\varrho\rangle}{\langle \alpha,\varrho\rangle\rightsquigarrow\langle \breve{\alpha},\varrho\rangle}$$

      With the assumptions $\partial_H(\alpha)=\alpha$ and $\partial_H(\breve{\alpha})=\breve{\alpha}$ with $\alpha,\breve{\alpha}\notin H$, we get $R_1$ satisfies condition 1 in the definition of $\underline{\leftrightarrow}_q$
      (see Definition \ref{SPQB}).

      $$\frac{\langle \breve{\alpha},\varrho\rangle\xrightarrow{\alpha} \langle \surd,\varrho'\rangle}{\langle \partial_H(\breve{\alpha}),\varrho\rangle\xrightarrow{\alpha}\langle \surd,\varrho'\rangle}$$

      $$\frac{\langle \breve{\alpha},\varrho\rangle\xrightarrow{\alpha} \langle \surd,\varrho'\rangle}{\langle \breve{\alpha},\varrho\rangle\xrightarrow{\alpha}\langle \surd,\varrho'\rangle}$$

      With the assumption $\partial_H(\breve{\alpha})=\breve{\alpha}$, we get $R_2$ satisfies conditions 2 and 3 in the definition of $\underline{\leftrightarrow}_q$
      (see Definition \ref{SPQB}).

      So, $R=\underline{\leftrightarrow}_q$, as desired.

  \item \textbf{Axiom $D2$}. For a relation $R=R_1\cup R_2$, suppose that $Eq(\langle \partial_H(\alpha),\varrho\rangle,\langle \delta,\varrho\rangle)= R_1$ with $\alpha\in H$, and $Eq(\langle \partial_H(\breve{\alpha}),\varrho\rangle,\langle \breve{\delta},\varrho\rangle)= R_2$ with $\breve{\alpha}\in H$.

      By use of the probabilistic transition rules and the action transition rules for atomic unitary operator and encapsulation operator $\partial_H$, we get:

      $$\frac{\langle \alpha,\varrho\rangle\rightsquigarrow \langle \breve{\delta},\varrho\rangle}{\langle \partial_H(\alpha),\varrho\rangle\rightsquigarrow\langle \partial_H(\breve{\delta}),\varrho\rangle}$$

      There are not action transition rules for the constant $\breve{\delta}$, which mean that $\breve{\delta}$ leads to inaction processes, accompany with the action transition rules of operator $\partial_H$, $\partial_H(\breve{\delta})\nrightarrow$ and $\breve{\delta}\nrightarrow$ all lead to inaction processes.

      With the assumptions $\partial_H(\alpha)=\delta$ and $\partial_H(\breve{\alpha})=\breve{\delta}$ with $\alpha,\breve{\alpha}\in H$, we get $R$ satisfies conditions 1, 2 and 3 in the definition of $\underline{\leftrightarrow}_q$
      (see Definition \ref{SPQB}).

      So, $R=\underline{\leftrightarrow}_q$, as desired.

  \item \textbf{Axiom $D3$}. For a relation $R=Eq(\langle \partial_H(u+v),\varrho\rangle,\langle \partial_H(u)+\partial_H(v),\varrho\rangle)\cup Eq(\langle \partial_H(p+q),\varrho\rangle,\langle \partial_H(p)+\partial_H(q),\varrho\rangle)$, with $u,v\in\mathbb{SP}(pqACP^+)$ and $p,q\in\mathbb{DP}(pqACP^+)$.

      By use of the probabilistic transition rules and the action transition rules for alternative operator $+$ and encapsulation operator $\partial_H$, we get:

      $$\frac{\langle u,\varrho\rangle\rightsquigarrow \langle u',\varrho\rangle,\langle v,\varrho\rangle\rightsquigarrow \langle v',\varrho\rangle}{\langle \partial_H(u+v),\varrho\rangle\rightsquigarrow\langle \partial_H(u'+v'),\varrho\rangle}$$

      $$\frac{\langle u,\varrho\rangle\rightsquigarrow \langle u',\varrho\rangle,\langle v,\varrho\rangle\rightsquigarrow \langle v',\varrho\rangle}{\langle \partial_H(u)+\partial_H(v),\varrho\rangle\rightsquigarrow\langle \partial_H(u')+\partial_H(v'),\varrho\rangle}$$

      With the assumptions $\partial_H(u+v)=\partial_H(u)+\partial_H(v)$ and $\partial_H(u'+v')=\partial_H(u')+\partial_H(v')$, we get $R$ satisfies condition 1 in the definition of $\underline{\leftrightarrow}_q$
      (see Definition \ref{SPQB}).

      $$\frac{\langle p,\varrho\rangle\xrightarrow{\alpha} \langle p',\varrho'\rangle}{\langle \partial_H(p+q),\varrho\rangle\xrightarrow{\alpha}\langle \partial_H(p'),\varrho'\rangle}$$

      $$\frac{\langle p,\varrho\rangle\xrightarrow{\alpha} \langle p',\varrho'\rangle}{\langle \partial_H(p)+\partial_H(q),\varrho\rangle\xrightarrow{\alpha}\langle \partial_H(p'),\varrho'\rangle}$$

      $$\frac{\langle p,\varrho\rangle\xrightarrow{\alpha} \langle \surd,\varrho'\rangle}{\langle \partial_H(p+q),\varrho\rangle\xrightarrow{\alpha}\langle \surd,\varrho'\rangle}$$

      $$\frac{\langle p,\varrho\rangle\xrightarrow{\alpha} \langle \surd,\varrho'\rangle}{\langle \partial_H(p)+\partial_H(q),\varrho\rangle\xrightarrow{\alpha}\langle \surd,\varrho'\rangle}$$

      With the assumption $\partial_H(p+q)=\partial_H(p)+\partial_H(q)$, we get $R$ satisfies conditions 2 and 3 in the definition of $\underline{\leftrightarrow}_q$
      (see Definition \ref{SPQB}).

      So, $R=\underline{\leftrightarrow}_q$, as desired.

      For the case of the atomic communication action $a$ instead of atomic unitary operator $\alpha$, the proof is similar to those of axioms $CM2$ and $CM3$ and is omitted.

  \item \textbf{Axiom $D4$}. For a relation $R=Eq(\langle \partial_H(u\cdot v),\varrho\rangle,\langle \partial_H(u)\cdot\partial_H(v),\varrho\rangle)\cup Eq(\langle \partial_H(p\cdot v),\varrho\rangle,\langle \partial_H(p)\cdot\partial_H(v),\varrho\rangle)$, with $u,v\in\mathbb{SP}(pqACP^+)$ and $p\in\mathbb{DP}(pqACP^+)$.

      By use of the probabilistic transition rules and the action transition rules for sequential composition $\cdot$ and encapsulation operator $\partial_H$, we get:

      $$\frac{\langle u,\varrho\rangle\rightsquigarrow \langle u',\varrho\rangle}{\langle \partial_H(u\cdot v),\varrho\rangle\rightsquigarrow\langle \partial_H(u'\cdot v),\varrho\rangle}$$

      $$\frac{\langle u,\varrho\rangle\rightsquigarrow \langle u',\varrho\rangle}{\langle \partial_H(u)\cdot\partial_H(v),\varrho\rangle\rightsquigarrow\langle \partial_H(u')\cdot\partial_H(v),\varrho\rangle}$$

      With the assumptions $\partial_H(u\cdot v)=\partial_H(u)\cdot\partial_H(v)$ and $\partial_H(u'\cdot v)=\partial_H(u')\cdot\partial_H(v)$, we get $R$ satisfies condition 1 in the definition of $\underline{\leftrightarrow}_q$
      (see Definition \ref{SPQB}).

      $$\frac{\langle p,\varrho\rangle\xrightarrow{\alpha} \langle p',\varrho'\rangle}{\langle \partial_H(p\cdot v),\varrho\rangle\xrightarrow{\alpha}\langle \partial_H(p'\cdot v),\varrho'\rangle}$$

      $$\frac{\langle p,\varrho\rangle\xrightarrow{\alpha} \langle p',\varrho'\rangle}{\langle \partial_H(p)\cdot\partial_H(v),\varrho\rangle\xrightarrow{\alpha}\langle \partial_H(p')\cdot\partial_H(v),\varrho'\rangle}$$

      $$\frac{\langle p,\varrho\rangle\xrightarrow{\alpha} \langle \surd,\varrho'\rangle}{\langle \partial_H(p\cdot v),\varrho\rangle\xrightarrow{\alpha}\langle \partial_H(v),\varrho'\rangle}$$

      $$\frac{\langle p,\varrho\rangle\xrightarrow{\alpha} \langle \surd,\varrho'\rangle}{\langle \partial_H(p)\cdot\partial_H(v),\varrho\rangle\xrightarrow{\alpha}\langle \partial_H(v),\varrho'\rangle}$$

      With the assumption $\partial_H(p\cdot v)=\partial_H(p)\cdot\partial_H(v)$ and $\partial_H(p'\cdot v)=\partial_H(p')\cdot\partial_H(v)$, we get $R$ satisfies conditions 2 and 3 in the definition of $\underline{\leftrightarrow}_q$
      (see Definition \ref{SPQB}).

      So, $R=\underline{\leftrightarrow}_q$, as desired.

      For the case of the atomic communication action $a$ instead of atomic unitary operator $\alpha$, the proof is similar to those of axioms $CM2$ and $CM3$ and is omitted.

  \item \textbf{Axiom $PrD5$}. For a relation $R=Eq(\langle \partial_H(u\boxplus_{\pi}v),\varrho\rangle,\langle \partial_H(u)\boxplus_{\pi}\partial_H(v),\varrho\rangle)$, with $u,v\in\mathbb{SP}(pqACP^+)$.

      By use of the probabilistic transition rules for probabilistic choice operator $\boxplus_{\pi}$ and encapsulation operator $\partial_H$, we get:

      $$\frac{\langle u,\varrho\rangle\rightsquigarrow \langle u',\varrho\rangle}{\langle \partial_H(u\boxplus_{\pi}v),\varrho\rangle\rightsquigarrow\langle \partial_H(u'),\varrho\rangle}$$

      $$\frac{\langle u,\varrho\rangle\rightsquigarrow \langle u',\varrho\rangle}{\langle \partial_H(u)\boxplus_{\pi}\partial_H(v),\varrho\rangle\rightsquigarrow\langle \partial_H(u'),\varrho\rangle}$$

      With the assumption $\partial_H(u\boxplus_{\pi}v)=\partial_H(u)\boxplus_{\pi}\partial_H(v)$, we get $R$ satisfies condition 1 in the definition of $\underline{\leftrightarrow}_q$
      (see Definition \ref{SPQB}).

      Note that, there are not action transition rules for probabilistic choice operator $\boxplus_{\pi}$.

      So, $R=\underline{\leftrightarrow}_q$, as desired.

  \item \textbf{Axiom $PrMM4$}. For a relation $R$, suppose that $Eq(\langle (u,z)][(v,w),\varrho\rangle,\langle u\leftmerge w+v\leftmerge z+u\mid v,\varrho\rangle)= R$ with $u=u+u$, $v=v+v$, and $u,v,z,w\in \mathbb{SP}(pqACP^+)$.

      By use of the probabilistic transition rules for operators $\leftmerge$, $\mid$ and $][$, we get:

      $$\frac{\langle u,\varrho\rangle\rightsquigarrow \langle u',\varrho\rangle, \langle v,\varrho\rangle\rightsquigarrow \langle v',\varrho\rangle}{\langle (u,z)][(v,w),\varrho\rangle\rightsquigarrow\langle u'\leftmerge w+v'\leftmerge z+u'\mid v',\varrho\rangle}$$

      $$\frac{\langle u,\varrho\rangle\rightsquigarrow \langle u',\varrho\rangle, \langle v,\varrho\rangle\rightsquigarrow \langle v',\varrho\rangle}{\langle u\leftmerge w+v\leftmerge z+u\mid v,\varrho\rangle\rightsquigarrow\langle u'\leftmerge w+v'\leftmerge z+u'\mid v',\varrho\rangle}$$

      In the above two probabilistic transition rules, the condition $u=u+u$ ensures that $\langle u,\varrho\rangle\rightsquigarrow \langle u',\varrho\rangle$, $\langle u,\varrho\rangle\rightsquigarrow \langle u'',\varrho\rangle$, $\langle u',\varrho\rangle\underline{\leftrightarrow}_q \langle u'',\varrho\rangle$. $v=v+v$ ensures the similar things.

      With the assumption $(u,z)][(v,w)=u\leftmerge w+v\leftmerge z+u\mid v$, we get $R$ satisfies condition 1 in the definition of $\underline{\leftrightarrow}_q$
      (see Definition \ref{SPQB}). Note that, for operator $][$, there are not action transition rules.

      So, $R=\underline{\leftrightarrow}_q$, as desired.

  \item \textbf{Axiom $PrCM4$}. For a relation $R=R_1\cup R_2$, suppose that $Eq(\langle (p+q)\mid s,\varrho\rangle,\langle p\mid s + q\mid s,\varrho\rangle)= R_1$ with $p,q,s\in \mathbb{DP}(pqACP^+)$, and $Eq(\langle (u+v)\mid w,\varrho\rangle,\langle u\mid w + v\mid w,\varrho\rangle)= R_2$ with $w=w+w$ and $u,v,w\in \mathbb{SP}(pqACP^+)$.

      By use of the probabilistic transition rules and the action transition rules for communication merge $\mid$, alternative operator $+$, we get:

      $$\frac{\langle u,\varrho\rangle\rightsquigarrow \langle u',\varrho\rangle, \langle v,\varrho\rangle\rightsquigarrow \langle v',\varrho\rangle,\langle w,\varrho\rangle\rightsquigarrow \langle w',\varrho\rangle}{\langle (u + v)\mid w,\varrho\rangle\rightsquigarrow\langle (u' + v')\mid w',\varrho\rangle}$$

      $$\frac{\langle u,\varrho\rangle\rightsquigarrow \langle u',\varrho\rangle, \langle v,\varrho\rangle\rightsquigarrow \langle v',\varrho\rangle,\langle w,\varrho\rangle\rightsquigarrow \langle w',\varrho\rangle}{\langle u\mid w+v\mid w,\varrho\rangle\rightsquigarrow\langle u'\mid w'+v'\mid w',\varrho\rangle}$$

      In the above two probabilistic transition rules, the condition $w=w+w$ ensures that $\langle w,\varrho\rangle\rightsquigarrow \langle w',\varrho\rangle$, $\langle w,\varrho\rangle\rightsquigarrow \langle w'',\varrho\rangle$, $\langle w',\varrho\rangle\underline{\leftrightarrow}_q \langle w'',\varrho\rangle$.

      With the assumptions $(u+v)\mid w=u\mid w+v\mid w$ and $(u'+v')\mid w'=u'\mid w'+v'\mid w'$, we get $R_2$ satisfies condition 1 in the definition of $\underline{\leftrightarrow}_q$
      (see Definition \ref{SPQB}).

      $$\frac{\langle p,\varrho\rangle\xrightarrow{a} \langle p',\varrho\rangle,\langle s,\varrho\rangle\xrightarrow{b} \langle s',\varrho\rangle}{\langle (p+ q)\mid s,\varrho\rangle\xrightarrow{\gamma(a,b)}\langle p'\parallel s',\varrho\rangle}$$

      $$\frac{\langle p,\varrho\rangle\xrightarrow{a} \langle p',\varrho\rangle,\langle s,\varrho\rangle\xrightarrow{b} \langle s',\varrho\rangle}{\langle p\mid s+q\mid s,\varrho\rangle\xrightarrow{\gamma(a,b)}\langle p'\parallel s',\varrho\rangle}$$

      $$\frac{\langle p,\varrho\rangle\xrightarrow{a} \langle p',\varrho\rangle,\langle s,\varrho\rangle\xrightarrow{b} \langle \surd,\varrho\rangle}{\langle (p+ q)\mid s,\varrho\rangle\xrightarrow{\gamma(a,b)}\langle p',\varrho\rangle}$$

      $$\frac{\langle p,\varrho\rangle\xrightarrow{a} \langle p',\varrho\rangle,\langle s,\varrho\rangle\xrightarrow{b} \langle \surd,\varrho\rangle}{\langle p\mid s+q\mid s,\varrho\rangle\xrightarrow{\gamma(a,b)}\langle p',\varrho\rangle}$$

      $$\frac{\langle p,\varrho\rangle\xrightarrow{a} \langle \surd,\varrho\rangle,\langle s,\varrho\rangle\xrightarrow{b} \langle \surd,\varrho\rangle}{\langle (p+ q)\mid s,\varrho\rangle\xrightarrow{\gamma(a,b)}\langle \surd,\varrho\rangle}$$

      $$\frac{\langle p,\varrho\rangle\xrightarrow{a} \langle \surd,\varrho\rangle,\langle s,\varrho\rangle\xrightarrow{b} \langle \surd,\varrho\rangle}{\langle p\mid s+q\mid s,\varrho\rangle\xrightarrow{\gamma(a,b)}\langle \surd,\varrho\rangle}$$

      With the assumptions $(p+q)\mid s=p\mid s + q\mid s$, we get $R_1$ satisfies conditions 2 and 3 in the definition of $\underline{\leftrightarrow}_q$
      (see Definition \ref{SPQB}).

      So, $R=\underline{\leftrightarrow}_q$, as desired.

  \item \textbf{Axiom $PrCM5$}. For a relation $R=R_1\cup R_2$, suppose that $Eq(\langle s\mid(p+q),\varrho\rangle,\langle s\mid p + s\mid q,\varrho\rangle)= R_1$ with $p,q,s\in \mathbb{DP}(pqACP^+)$, and $Eq(\langle w\mid(u+v),\varrho\rangle,\langle w\mid u + w\mid v,\varrho\rangle)= R_2$ with $w=w+w$ and $u,v,w\in \mathbb{SP}(pqACP^+)$.

      By use of the probabilistic transition rules and the action transition rules for communication merge $\mid$, alternative operator $+$, we get:

      $$\frac{\langle u,\varrho\rangle\rightsquigarrow \langle u',\varrho\rangle, \langle v,\varrho\rangle\rightsquigarrow \langle v',\varrho\rangle,\langle w,\varrho\rangle\rightsquigarrow \langle w',\varrho\rangle}{\langle w\mid(u + v),\varrho\rangle\rightsquigarrow\langle w'\mid(u' + v'),\varrho\rangle}$$

      $$\frac{\langle u,\varrho\rangle\rightsquigarrow \langle u',\varrho\rangle, \langle v,\varrho\rangle\rightsquigarrow \langle v',\varrho\rangle,\langle w,\varrho\rangle\rightsquigarrow \langle w',\varrho\rangle}{\langle w\mid u+w\mid v,\varrho\rangle\rightsquigarrow\langle w'\mid u'+w'\mid v',\varrho\rangle}$$

      In the above two probabilistic transition rules, the condition $w=w+w$ ensures that $\langle w,\varrho\rangle\rightsquigarrow \langle w',\varrho\rangle$, $\langle w,\varrho\rangle\rightsquigarrow \langle w'',\varrho\rangle$, $\langle w',\varrho\rangle\underline{\leftrightarrow}_q \langle w'',\varrho\rangle$.

      With the assumptions $w\mid(u+v)=w\mid u+w\mid v$ and $w'\mid(u'+v')=w'\mid u'+w'\mid v'$, we get $R_2$ satisfies condition 1 in the definition of $\underline{\leftrightarrow}_q$
      (see Definition \ref{SPQB}).

      $$\frac{\langle p,\varrho\rangle\xrightarrow{a} \langle p',\varrho\rangle,\langle s,\varrho\rangle\xrightarrow{b} \langle s',\varrho\rangle}{\langle s\mid(p+ q),\varrho\rangle\xrightarrow{\gamma(a,b)}\langle s'\parallel p',\varrho\rangle}$$

      $$\frac{\langle p,\varrho\rangle\xrightarrow{a} \langle p',\varrho\rangle,\langle s,\varrho\rangle\xrightarrow{b} \langle s',\varrho\rangle}{\langle s\mid p+s\mid q,\varrho\rangle\xrightarrow{\gamma(a,b)}\langle s'\parallel p',\varrho\rangle}$$

      $$\frac{\langle p,\varrho\rangle\xrightarrow{a} \langle p',\varrho\rangle,\langle s,\varrho\rangle\xrightarrow{b} \langle \surd,\varrho\rangle}{\langle s\mid(p+ q),\varrho\rangle\xrightarrow{\gamma(a,b)}\langle p',\varrho\rangle}$$

      $$\frac{\langle p,\varrho\rangle\xrightarrow{a} \langle p',\varrho\rangle,\langle s,\varrho\rangle\xrightarrow{b} \langle \surd,\varrho\rangle}{\langle s\mid p+s\mid q,\varrho\rangle\xrightarrow{\gamma(a,b)}\langle p',\varrho\rangle}$$

      $$\frac{\langle p,\varrho\rangle\xrightarrow{a} \langle \surd,\varrho\rangle,\langle s,\varrho\rangle\xrightarrow{b} \langle \surd,\varrho\rangle}{\langle s\mid(p+ q),\varrho\rangle\xrightarrow{\gamma(a,b)}\langle \surd,\varrho\rangle}$$

      $$\frac{\langle p,\varrho\rangle\xrightarrow{a} \langle \surd,\varrho\rangle,\langle s,\varrho\rangle\xrightarrow{b} \langle \surd,\varrho\rangle}{\langle s\mid p+s\mid q,\varrho\rangle\xrightarrow{\gamma(a,b)}\langle \surd,\varrho\rangle}$$

      With the assumptions $s\mid(p+q)=s\mid p + s\mid q$, we get $R_1$ satisfies conditions 2 and 3 in the definition of $\underline{\leftrightarrow}_q$
      (see Definition \ref{SPQB}).

      So, $R=\underline{\leftrightarrow}_q$, as desired.

\end{itemize}
\end{proof}

\theorem[Completeness of $pqACP^+$]{Let $z$ and $u$ are closed $pqACP^+$ terms, if $\langle z,\varrho\rangle\underline{\leftrightarrow}_q \langle u,\varrho\rangle$, then $pqACP^+\vdash z=u$.}\label{CompletePQACP}

\begin{proof}
It is based on the following three facts: let $z$ and $u$ are closed $pqACP^+$ terms,

\begin{enumerate}
  \item $\langle z,\varrho\rangle\underline{\leftrightarrow}_q \langle u,\varrho\rangle \Rightarrow z\underline{\leftrightarrow} u$ by Proposition \ref{PQBPB};
  \item $z\underline{\leftrightarrow} u \Rightarrow pACP^+\vdash z=u$ (see in section \ref{PPA}, and for details in \cite{PPA});
  \item the term systems of $pACP^+$ and $pqACP^+$ are same, including the same operators and the same axiom systems, but different atomic action meanings (semantics).
\end{enumerate}

So, we get $\langle z,\varrho\rangle\underline{\leftrightarrow}_q \langle u,\varrho\rangle \Rightarrow pqACP^+\vdash z=u$, as desired.
\end{proof}

In the following, for the existence of $\tau$ action, the quantum state $\varrho$ is adjust to the state of all public quantum variables. Because of quantum entanglement, the entangled state that $\tau$ action affects are all in internal variables. This treatment of quantum entanglement is implicit, in section \ref{QE}, we will introduce an explicit treatment of quantum entanglement. With the adjustment of $\varrho$, there is $\tau(\varrho)=\varrho$. So, for silent action $\tau$, without consideration of any probability, there should be an action transition rule as follows.

$$\frac{}{\langle\tau,\varrho\rangle\xrightarrow{\tau}\langle\surd,\tau(\varrho)\rangle}$$

where $\tau(\varrho)=\varrho$ with the states of all public quantum variables $\varrho$.

\definition{A fully probabilistic quantum graph $g$ is a tuple $(S_p\cup S_n\cup\{NIL\},\rightsquigarrow,\rightarrow,\mu,\varrho,root)$} consisting of:

\begin{itemize}
  \item a countable set $S_p$ of probabilistic states,
  \item a countable set $S_n$ of action states such that $S_p\cap S_n=\emptyset$ and $NIL\notin S_p\cup S_n$,
  \item $root\in S_p$,
  \item a relation $\rightsquigarrow\subseteq S_p\times S_n$,
  \item a function $\rightarrow:S_n\rightarrow(S_p\cup\{NIL\})\times A_{\tau}$,
  \item a partial function $\mu:S_p\times S_n\rightarrow\langle 0,1]$ such that $\mu(p,n)$ is defined iff $(p,n)\in\rightsquigarrow$ for $(p,n)\in S_p\times S_n$ and $\sum_{n\in S_n}\mu(p,n)=1$ for any $p\in S_p$.
  \item a public quantum state $\varrho\in\mathcal{D}(\mathcal{H})$, and an implicit function $\mathcal{D}(\mathcal{H})\times A_{\tau}\rightarrow \mathcal{D}(\mathcal{H})$.
\end{itemize}

\definition[Probabilistic quantum branching bisimulation]{Let $g$ and $h$ be fully probabilistic quantum graphs. If $R$ is an equivalence relation on $S_g\cup S_h\cup\{NUL_g,NIL_h\}$ such that:

\begin{enumerate}
  \item $(root(g),root(h))\in R$;
  \item if $(p,q)\in R$ and $p\rightsquigarrow s$ then either
        \begin{enumerate}
          \item $(s,q)\in R$ or
          \item there are $v,t$ such that $(p,v),(s,t)\in R$ and $q\xRightarrow{\tau^*}v\rightsquigarrow t$ or $q\xrightarrow{\tau}\cdot\xRightarrow{\tau^*}v\rightsquigarrow t$;
        \end{enumerate}
  \item if $(p,q)\in R$ and $p\xrightarrow{\alpha} s$ with $\varrho_g\rightarrow\varrho_g'=\alpha\varrho_g\alpha^{\dagger}$, then either
        \begin{enumerate}
          \item $\alpha=\tau$ and $(s,q)\in R$ or
          \item there are $v,t$ such that $(p,v),(s,t)\in R$ and $q\xRightarrow{\tau^*}\cdot\rightsquigarrow v\xrightarrow{\alpha} t$ with $\varrho_h\rightarrow\varrho_h'=\alpha\varrho_h\alpha^{\dagger}$, $\varrho_g'=\varrho_h'$ iff $\varrho_g=\varrho_h$, or $q(\xrightarrow{\tau}\cdot\rightsquigarrow)^* v\xrightarrow{\alpha} t$ with $\varrho_h\rightarrow\varrho_h'=\alpha\varrho_h\alpha^{\dagger}$, $\varrho_g'=\varrho_h'$ iff $\varrho_g=\varrho_h$;
        \end{enumerate}
  \item there is an equivalence relation $\tilde{R}$ on $Entry_R(g)\cup Entry_R(h)$ such that $\tilde{R}\subseteq R$ and
      \begin{enumerate}
        \item $(root(g),root(h))\in\tilde{R}$;
        \item if $(p,q)\in\tilde{R}$ then for any $C\in NextEntryC_{\tilde{R}}(p)\cup NextEntryC_{\tilde{R}}(q)$ and for any $\alpha\in A$, $Prob_{[p]_R}(p,\tau^*,C)=Prob_{[q]_R}(q,\tau^*,C)$ and $Prob_{[p]_R}(p,\tau^*\alpha,C)=Prob_{[q]_R}(q,\tau^*\alpha,C)$;
      \end{enumerate}
\end{enumerate}

where $Prob$, $\xRightarrow{\tau^*}$, $Entry_R(g)$, $[p]_R$ have the same meanings as those in probabilistic branching bisimulation in section \ref{PPA}. Then $(R,\tilde{R})$ is a probabilistic quantum branching bisimulation relation between $g$ and $h$. We write $\underline{\leftrightarrow}_{pqb}$ if there is a probabilistic quantum bisimulation $R,\tilde{R}$ between $g$ and $h$.

For additional conditions of probabilistic quantum root branching bisimulation $g \underline{\leftrightarrow}_{pqrb} h$, there is a probabilistic quantum branching bisimulation $(R,\tilde{R})$ between $g$ and $h$ such that:
\begin{enumerate}
  \item if $root(g)\rightsquigarrow p$ then there is $q$ in $h$ such that $root(h)\rightsquigarrow q$ and $(p,q)\in R$;
  \item if $root(h)\rightsquigarrow q$ then there is $p$ in $g$ such that $root(g)\rightsquigarrow p$ and $(p,q)\in R$;
  \item if $root(g)\rightsquigarrow p\xrightarrow{\alpha} s$ for $\alpha\in A_{\tau}$ with $\varrho_g\rightarrow\varrho_g'=\alpha\varrho_g\alpha^{\dagger}$, then there are $q$ and $t$ in $h$ such that $root(h)\rightsquigarrow q\xrightarrow{\alpha} t$ with $\varrho_h\rightarrow\varrho_h'=\alpha\varrho_h\alpha^{\dagger}$, $\varrho_g'=\varrho_h'$ iff $\varrho_g=\varrho_h$, and $(p,q)\in R$ and $(s,t)\in R$;
  \item if $root(h)\rightsquigarrow q\xrightarrow{\alpha} t$ for $\alpha\in A_{\tau}$ with $\varrho_h\rightarrow\varrho_h'=\alpha\varrho_h\alpha^{\dagger}$, then there are $p$ and $s$ in $g$ such that $root(g)\rightsquigarrow p\xrightarrow{\alpha} s$ with $\varrho_g\rightarrow\varrho_g'=\alpha\varrho_g\alpha^{\dagger}$, $\varrho_g'=\varrho_h'$ iff $\varrho_g=\varrho_h$, and $(p,q)\in R$ and $(s,t)\in R$.
\end{enumerate}
}

We obtain the following obvious but important conclusion.

\proposition{$\underline{\leftrightarrow}_{pqrb}$ implies $\underline{\leftrightarrow}_{prb}$ with respect to $fpqBPA_{\tau}+PVR_1+PVR_2+\cdots$.}\label{PQRB}

\begin{proof}
From the definition of $\underline{\leftrightarrow}_{pqrb}$ and the definition of $\underline{\leftrightarrow}_{prb}$ (see in section \ref{PPA}), it is obvious that $\underline{\leftrightarrow}_{pqrb}$ adds additional conditions about quantum information state $\varrho$ into $\underline{\leftrightarrow}_{prb}$, with respect to $fpqBPA_{\tau}+PVR_1+PVR_2+\cdots$. So, let $x$ and $y$ be $fpqBPA_{\tau}+PVR_1+PVR_2+\cdots$ terms, $\langle x,\varrho\rangle\underline{\leftrightarrow}_{pqrb}\langle y,\varrho\rangle\Rightarrow x\underline{\leftrightarrow}_{prb} y$, as desired.
\end{proof}

\theorem[Soundness of $fpqBPA_{\tau}+PVR_1+PVR_2+\cdots$]{G/$\underline{\leftrightarrow}_{pqrb}$ is a model of $fpqBPA_{\tau}+PVR_1+PVR_2+\cdots$.}

\begin{proof}
It is already proven that $\underline{\leftrightarrow}_{pqrb}$ is an equivalent and congruent relation on $\mathbb{PT}(fpqBPA_{\tau}+PVR_1+PVR_2+\cdots)$ (see details in \cite{PPA}), and $fpBPA_{\tau}+PVR_1+PVR_2+\cdots\vdash x=y$ then $x\underline{\leftrightarrow}_{prb} y$ (see in section \ref{PPA} and \cite{PPA} for details), we only need to prove that the quantum information $\varrho$ related parts are also sound.

It is sufficient to examine every axiom in the axiom systems which are added to $fpqBPA$ to obtain $fpqBPA_{\tau}+PVR_1+PVR_2+\cdots$ (in Table \ref{AxiomForfpBPAtau}) and $PVR_1+PVR_2+\cdots$ are sound.

\begin{enumerate}
  \item \textbf{Axioms $TI0,TI1,TI2$}. The soundness of axioms $TI0,TI1,TI2$ are straightforward.
  \item \textbf{Axiom $TI4$}. For axiom $TI4$, $g$ represents the graph of $\tau_I(p+q)$ and $h$ represents the graph of $\tau_I(p)+\tau_I(q)$, the identity (diagonal) relation pair on $g$ and $h$, $(\triangle_S,\triangle_S)$, is still the desired probabilistic quantum root branching bisimulation relating $g$ and $h$, based on the probabilistic and action transition rules of alternative composition $+$.
  \item \textbf{Axiom $PrTI$} For axiom $PrTI$, $g$ represents the graph of $\tau_I(p\boxplus_{\pi}q)$ and $h$ represents the graph of $\tau_I(p)\boxplus_{\pi}\tau_I(q)$, the identity (diagonal) relation pair on $g$ and $h$, $(\triangle_S,\triangle_S)$, is still the desired probabilistic quantum root branching bisimulation relating $g$ and $h$, based on the probabilistic and action transition rules of probabilistic choice operator $\boxplus_{\pi}$.
  \item \textbf{Axiom $T1$}. For two graphs $g$ and $g\cdot \tau$, $\tau$ does not affect the (public) quantum state $\varrho$ for $\tau(\varrho)=\varrho$.
  \item \textbf{$PVR_1+PVR_2+\cdots$}. The way a probabilistic root branching bisimulation is constructed for $PVR_n$ in \cite{PPA}, is still suitable for constructing a probabilistic root quantum branching bisimulation for $PVR_n$.
\end{enumerate}
\end{proof}

\subsection{Quantum Measurement in Probabilistic Process Algebra}\label{QM}

In the above section, we introduce unitary operators as the atomic actions into probabilistic process algebra. Based on quantum configuration, we establish a pure quantum probabilistic process algebra called $PQRA$, which has the same equational logic as classical probabilistic process algebra $PRA$, and quantum configuration-based structured operational semantics.

But, in closed quantum systems, there is another basic quantum operation -- quantum measurement, besides the unitary operator. Quantum measurements have a probabilistic nature, so, if we treat quantum measurements as another kinds of atomic actions, there should be such as the following action transition rule.

$$\frac{}{\langle\beta,\varrho\rangle\xrightarrow{\beta}\langle\surd,\varrho_1\rangle \boxplus_{\pi_1}\langle\surd,\varrho_2\rangle\boxplus_{\pi_2}\cdots\boxplus_{\pi_{i-1}} \langle\surd,\varrho_i\rangle}$$

where $\beta$ denotes a quantum measurement, and $\beta=\sum_i\lambda_i \beta_i$, $\beta_i$ denotes the projection performed on the quantum system $\varrho$, $\pi_i=Tr(\beta_i\varrho)$, $\varrho_i=\beta_i\varrho \beta_i/\pi_i$.

Ideally, we extend the collection of atomic actions to contain two kinds of actions: the unitary operator $\alpha$ and the quantum measurement $\beta$, add the above action transition rule into semantic deductive system of $PQRA$, and also design new action transition rules of quantum measurement acting on operators of $PQRA$, to obtain an almost perfect theory on closed quantum systems.

But, the above idea is unrealistic, because $i$ in the above action transition rule is unknown and is dependent on the concrete quantum measurement performed on some concrete quantum system $\varrho$. It leads to unknown transitions, so, it is not positive after deduction \cite{SOS} and leads to probabilistic quantum bisimulation that the deduction system induces is not a congruence. Therefor, the soundness and the completeness of $PQRA$ are lost.

In fact, quantum measurement $\beta$ can not be captured explicitly in $PQRA$ and can only be processed during modeling phase in application for verifying behaviors of closed quantum systems. The behaviors of quantum measurement $\beta$ can be fully modeled by the probabilistic choice operator $\boxplus_{\pi}$ in $PQRA$, the reasons are following:

\begin{enumerate}
  \item by Elimination Theorem of $pqACP^+$, let $p$ be a closed $pqACP^+$ term, then there is a closed $pqBPA$ term $q$ such that $pqACP^+\vdash p=q$;
  \item by Elimination Theorem of $pqBPA$, let $p$ be a closed $pqBPA$ term, then there is a closed basic $pqBPA$ term $q$ such that $pqBPA\vdash p=q$;
  \item the basic terms are exactly the terms of the form

        $$x\equiv x_1\quad (x\in\mathcal{B}_+(pBPA))\textrm{ or }$$

        $$x\equiv x_1\boxplus_{\pi_1}x_2\boxplus_{\pi_2}\cdots\boxplus_{\pi_{n-1}}x_n\quad(n\geq 1)$$

        where $x_i\equiv\sum_{j<l_i}\alpha_{ij}\cdot t_{ij}+\sum_{k<m_i}\beta_{ik}$ for certain $\alpha_{ij},\beta_{ik}\in A_\delta$, basic $pqBPA$ terms $t_{ij}$ and $n,m_i,l_i\in\mathbb{N}$.

  \item informally, but in general, a process in quantum computing for closed systems with quantum measurement has the following form which can be captured by the above normal form of basic $pqBPA$ terms:

        $$t_1\boxplus_{\pi_1}t_2\boxplus_{\pi_2}\cdots\boxplus_{\pi_{i-1}}t_i$$

        where $\sum_i \pi_i=1$, and $t_i\in\mathcal{B}(pBPA)$.
\end{enumerate}

There is a concrete but non-trivial problem in modeling quantum measurement, which may make the whole theory fail. In a pure quantum probabilistic process algebra, it is reasonable that we assume that all probabilistic choices are caused by quantum measurements. If we model quantum measurement only by probabilistic choice operator $\boxplus_{\pi}$, remember that quantum measurement will make the quantum state $\varrho$ evolve into $\varrho'$, this means that probabilistic action $\rightsquigarrow$ will make $\varrho$ change, that is, there will be such probabilistic transition rule $\frac{p\rightsquigarrow p'}{\langle p,\varrho\rangle\rightsquigarrow\langle p',\varrho'\rangle}$. We will have to make changes to the whole theory to let probabilistic choice change the quantum state $\varrho$. We consider the probabilistic transition rule for alternative composition $+$, it will change to the following form.

$$\frac{\langle x,\varrho\rangle\rightsquigarrow \langle x',\varrho'\rangle, \langle y,\varrho\rangle\rightsquigarrow \langle y',\varrho''\rangle}{\langle x + y,\varrho\rangle\rightsquigarrow\langle x' + y',\varrho'+\varrho''\rangle}$$ or,

$$\frac{\langle x,\varrho\rangle\rightsquigarrow \langle x',\varrho'\rangle, \langle y,\varrho\rangle\rightsquigarrow \langle y',\varrho''\rangle}{\langle x + y,\varrho\rangle\rightsquigarrow\langle x',\varrho'\rangle+\langle y',\varrho''\rangle}$$

Now, alternative composition $+$ is acting on quantum state $\varrho$ or quantum configuration $\langle p,\varrho\rangle$! The signature $\Sigma$ of $PQRA$ in section \ref{UO} must contain $\varrho$ or $\langle p,\varrho\rangle$, which leads to the whole theory of $PQRA$ fails.

Let the following process term represent quantum measurement during modeling phase,

$$\beta_1\cdot t_1\boxplus_{\pi_1}\beta_2\cdot t_2\boxplus_{\pi_2}\cdots\boxplus_{\pi_{i-1}}\beta_i\cdot t_i$$

where $\sum_i \pi_i=1$, $t_i\in\mathcal{B}(pBPA)$, $\beta$ denotes a quantum measurement, and $\beta=\sum_i\lambda_i \beta_i$, $\beta_i$ denotes the projection performed on the quantum system $\varrho$, $\pi_i=Tr(\beta_i\varrho)$, $\varrho_i=\beta_i\varrho \beta_i/\pi_i$.

The above term means that, firstly, we choose a projection $\beta_i$ in a quantum measurement $\beta=\sum_i\lambda_i\beta_i$ probabilistically, then, we execute (perform) the projection $\beta_i$ on the closed quantum system. This also adheres to the intuition on quantum mechanics.

We define $B$ as the collection of all projections of all quantum measurements, and make the collection of atomic actions in $PQRA$ be $A\cup B\cup C$. We see that a projection $\beta_i\in B$ has the almost same semantics as a unitary operator $\alpha\in A$. So, we add the following (probabilistic and action)
transition rules into those of $PQRA$.

$$\frac{}{\langle\beta_i,\varrho\rangle\rightsquigarrow\langle\breve{\beta_i},\varrho\rangle}$$

$$\frac{}{\langle\breve{\beta_i},\varrho\rangle\xrightarrow{\beta_i}\langle\surd,\varrho'\rangle}$$

Until now, $PQRA$ works again. The two main quantum operations in a closed quantum system -- the unitary operator and the quantum measurement, are fully modeled in probabilistic process algebra.

\subsection{Quantum Entanglement in Probabilistic Process Algebra}\label{QE}

In section \ref{UO}, when we introduce the silent step $\tau$, we re-define the quantum state of all variables $\varrho$ to the quantum state of all public quantum variables. We say that if two quantum variables are entangled, then a quantum operation performed on one variable, then state of the other quantum variable is also changed. So, the entangled states must be all the inner variables or all the public variables. Since there is exactly only one quantum operation executed, so it is an implicit manner to model quantum entanglement.

In \cite{QE}, we introduced a mechanism to explicitly define quantum entanglement in open quantum systems. This mechanism is still suitable for defining quantum entanglement in closed systems. A new constant called shadow constant denoted $\circledS_\alpha$ corresponding to a specific unitary operator or a projection, $\alpha$, and also a new kind of entanglement merge $\between$, are added to $pqACP^+$ to obtain $pqACP^+_{QE}$. If there are $n$ quantum variables entangled, they maybe be distributed in different quantum systems, with a unitary operator or a projection $\alpha\in A\cup B$ performed on one variable, there should be one $\circledS_\alpha$ executed on each variable in the other $n-1$ variables. Thus, distributed variables are all hidden behind actions. Entanglement merge happens with the interaction between $\alpha$ and $\circledS_\alpha$, and it is modeled as a new kind parallelism mechanism. In the following, we let $\circledS\in A\cup B$.

The axiom system for the shadow constant $\circledS$ and the entanglement merge $\between$ is shown in Table \ref{AxiomsForQE}. Note that the axiom $PrMM4$ in Table \ref{AxiomForpACP+} is changed into the new $PrMM4$ in Table \ref{AxiomsForQE}, and the rest axioms in Table \ref{AxiomForpACP+} still hold (we will prove the Elimination Theorem and the Conservativity Theorem of $pqACP^+_{QE}$ with respect to $pqACP^+$).

\begin{center}
\begin{table}
  \begin{tabular}{@{}ll@{}}
\hline No. &Axiom\\
  $PrMM4$ & $x=x+x,y=y+y\Rightarrow (x,z)][(y,w) = x\leftmerge w + y\leftmerge z + x\mid y + x\between y$\\
  $EM1$ & $\alpha\between\circledS_{\alpha}=\alpha$\\
  $EM2$ & $\circledS_{\alpha}\between\alpha=\alpha$\\
  $EM3$ & $\alpha\between(\circledS_{\alpha}\cdot y) = \alpha\cdot y$\\
  $EM4$ & $\circledS_{\alpha}\between(\alpha\cdot y) = \alpha\cdot y$\\
  $EM5$ & $(\alpha\cdot x)\between\circledS_{\alpha} = \alpha\cdot x$\\
  $EM6$ & $(\circledS_{\alpha}\cdot x)\between\alpha = \alpha\cdot x$\\
  $EM7$ & $(\alpha\cdot x)\between(\circledS_{\alpha}\cdot y) = \alpha\cdot (x\parallel y)$\\
  $EM8$ & $(\circledS_{\alpha}\cdot x)\between(\alpha\cdot y) = \alpha\cdot (x\parallel y)$\\
  $PrEM1$ & $(x\boxplus_{\pi} y)\between z = x\between z \boxplus_{\pi} y\between z$\\
  $PrEM2$ & $x\between (y\boxplus_{\pi} z) = x\between y \boxplus_{\pi} x\between z$\\
  $PrEM3$ & $z=z+z\Rightarrow (x + y)\between z = x\between z + y\between z$\\
  $PrEM4$ & $z=z+z\Rightarrow z\between (x+y) = z\between x + z\between y$\\
\end{tabular}
\caption{Axioms for quantum entanglement}
\label{AxiomsForQE}
\end{table}
\end{center}

The (probabilistic and action) transition rules for constant $\circledS$ and entanglement merge $\between$ are as follows. Note that the probabilistic transition rules for operators $\parallel$ and $][$ are changed into the following new ones.

$$\frac{}{\langle\circledS,\varrho\rangle\rightsquigarrow\langle\breve{\circledS},\varrho\rangle}$$

$$\frac{\langle x,\varrho\rangle\rightsquigarrow \langle x',\varrho\rangle, \langle y,\varrho\rangle\rightsquigarrow\langle y',\varrho\rangle}{\langle x\parallel y,\varrho\rangle\rightsquigarrow \langle x'\leftmerge y+y'\leftmerge x+x'\mid y'+x'\between y',\varrho\rangle}$$

$$\frac{\langle x,\varrho\rangle\rightsquigarrow \langle x',\varrho\rangle, \langle y,\varrho\rangle\rightsquigarrow\langle y',\varrho\rangle}{\langle (x,z)][(y,w),\varrho\rangle\rightsquigarrow \langle x'\leftmerge w+y'\leftmerge z+x'\mid y'+x'\between y',\varrho\rangle}$$

$$\frac{\langle x,\varrho\rangle\rightsquigarrow \langle x',\varrho\rangle,\langle y,\varrho\rangle\rightsquigarrow \langle y',\varrho\rangle}{\langle x\between y,\varrho\rangle\rightsquigarrow \langle x'\between y',\varrho\rangle}$$

$$\frac{}{\langle\breve{\circledS},\varrho\rangle\rightarrow\langle\surd,\varrho\rangle}$$

$$\frac{\langle x, \varrho\rangle\xrightarrow{\alpha}\langle x',\varrho'\rangle\quad \langle y, \varrho'\rangle\xrightarrow{\circledS_{\alpha}}\langle y',\varrho'\rangle}{\langle x\between y,\varrho\rangle\xrightarrow{\alpha}\langle x'\parallel y', \varrho'\rangle}$$

$$\frac{\langle x, \varrho\rangle\xrightarrow{\alpha}\langle\surd,\varrho'\rangle\quad \langle y, \varrho'\rangle\xrightarrow{\circledS_{\alpha}}\langle y',\varrho'\rangle}{\langle x\between y,\varrho\rangle\xrightarrow{\alpha}\langle y', \varrho'\rangle}$$

$$\frac{\langle x, \varrho'\rangle\xrightarrow{\circledS_{\alpha}}\langle\surd,\varrho'\rangle\quad \langle y, \varrho\rangle\xrightarrow{\alpha}\langle y',\varrho'\rangle}{\langle x\between y,\varrho\rangle\xrightarrow{\alpha}\langle y', \varrho'\rangle}$$

$$\frac{\langle x, \varrho\rangle\xrightarrow{\alpha}\langle\surd,\varrho'\rangle\quad \langle y, \varrho'\rangle\xrightarrow{\circledS_{\alpha}}\langle\surd,\varrho'\rangle}{\langle x\between y,\varrho\rangle\xrightarrow{\alpha}\langle \surd, \varrho'\rangle}$$

\definition[PDF $\mu$ for new operators]{We define PDF $\mu$ for operators $\parallel$, $][$, and $\between$ as follows, and the definitions of $\mu$ for operators $\leftmerge$, $\mid$ and $\partial_H$ are same as those in $pqACP^+$.

$$\mu(x\parallel y, x'\leftmerge y+y'\leftmerge x+x'\mid y'+x'\between y')=\mu(x,x')\cdot\mu(y,y')$$

$$\mu((x,z)][(y,w),x'\leftmerge w+y'\leftmerge z+x'\mid y'+x'\between y')=\mu(x,x')\cdot\mu(y,y')$$

$$\mu(x\between y,x'\between y')=\mu(x,x')\cdot\mu(y,y')$$
}

\proposition{$\mu$ and $\mu^*$ are well-defined on $\mathbb{PT}(pqACP^+_{QE})$}

\begin{proof}
It follows the two cases:

\begin{enumerate}
  \item Case $t\in\mathbb{SP}(pqACP^+_{RN})$ processes.
    \begin{itemize}
        \item for $t\equiv s\between r$, $\mu(s\between r,u)=\mu(s,v)\cdot\mu(r,w)$, if $u\equiv v\between w$; otherwise, $\mu(s\between r,u)=0$. $\mu(s,v)$ and $\mu(r,w)$ are defined by the inductive hypothesis, so $\mu(s\between r,u)$ is defined as well;
        \item for $t\equiv s\parallel r$, $\mu(s\parallel r,u)=\mu(s,v)\cdot\mu(r,w)$, if $u\equiv v\leftmerge r + w\leftmerge s +v\mid w +v\between w$; otherwise, $\mu(s\parallel r,u)=0$. $\mu(s,v)$ and $\mu(r,w)$ are defined by the inductive hypothesis, so $\mu(s\between r,u)$ is defined as well;
        \item for $t\equiv (s,z)][(r,w)$, $\mu((s,z)][(r,w),u)=\mu(s,x)\cdot\mu(r,y)$, if $u\equiv x\leftmerge w + y\leftmerge z +x\mid y +x\between y$; otherwise, $\mu((s,z)][(r,w),u)=0$. $\mu(s,x)$ and $\mu(r,y)$ are defined by the inductive hypothesis, so $\mu((s,z)][(r,w),u)$ is defined as well.
    \end{itemize}

  \item Case $t\in\mathbb{DP}(pqACP^+_{RN})$ processes. For $t\equiv s\between r$, $\mu(s\between r,u)=\mu(s,v)\cdot\mu(r,w)$, if $u\equiv v\between w$; otherwise, $\mu(s\between r,u)=0$. $\mu(s,v)=0$ and $\mu(r,w)=0$ are defined by the inductive hypothesis, so $\mu(s\between r,u)=0$ is defined as well;
\end{enumerate}

It is easy to check cPDF $\mu^*$ is also well-defined on $\mathbb{PT}(pqACP^+_{QE})$, we omit it.
\end{proof}

\theorem[Elimination theorem of $pqACP^+_{QE}$]{Let $p$ be a closed $pqACP^+_{QE}$ term. Then there is a closed $pqBPA$ term such that $pqACP^+_{QE}\vdash p=q$.}\label{ElimiQE}

\begin{proof}
The equational logic of $pqACP^+_{QE}$ ($pqBPA$ and $pqACP^+$) is same as that of $pACP^+_{QE}$ ($pBPA$ and $pACP^+$), and the same elimination properties. We only need to treat the new case of entanglement merge $\between$. The case of $\parallel$ and $][$ can be proved similarly as in elimination theorem of $pqACP^+$ just by adding the new case of $\between$, please refer to \cite{PPA}.

For the new case $p=p_1\between p_2$, $p_1$ and $p_2$ are closed $pqACP^+_{QE}$ terms. By the induction, there are closed $pqBPA$ terms $q_1$ and $q_2$ such that $pqACP^+_{QE}\vdash p_1=q_1$ and $pqACP^+_{QE}\vdash p_2=q_2$. By the elimination theorem of $pqBPA$, there are basic term $r_1$ and $r_2$ such that $pqBPA\vdash q_1=r_1$ and $pqBPA\vdash q_2=r_2$. Then also, $pqACP^+_{QE}\vdash p_1=r_1$, $pqACP^+_{QE}\vdash p_2=r_2$, and $pqACP^+_{QE}\vdash p_1\between p_2=r_1\between r_2$. In the following, we prove that there is a closed $pqBPA$ term $s$ such that $pqACP^+_{QE}\vdash r_1\between r_2=s$.

It is sufficient to induct on the structures of basic terms $r_1$ and $r_2$:

\begin{itemize}
  \item Case $r_1\equiv\alpha\in \{A\cup B\}_\delta$ and $r_2\equiv\circledS_\alpha\in \{A\cup B\}_\delta$. Then $pqACP^+_{QE}\vdash r_1\between r_2=\alpha\between\circledS_\alpha=\alpha$, and $\alpha$ is a closed $pqBPA$ term;
  \item Case $r_1\equiv\circledS_\alpha\in \{A\cup B\}_\delta$ and $r_2\equiv\alpha\in \{A\cup B\}_\delta$. Then $pqACP^+_{QE}\vdash r_1\between r_2=\circledS_\alpha\between\alpha=\alpha$, and $\alpha$ is a closed $pqBPA$ term;
  \item Case $r_1\equiv\alpha\in \{A\cup B\}_\delta$ and $r_2\equiv\circledS_\alpha\cdot r_2', \circledS_\alpha\in \{A\cup B\}_\delta$. Then $pqACP^+_{QE}\vdash r_1\between r_2=\alpha\between\circledS_\alpha\cdot r_2'=\alpha\cdot r_2'$, by induction $r_2'$ is a closed basic term, $\alpha\cdot r_2'$ is a closed $pqBPA$ term;
  \item Case $r_1\equiv\circledS_\alpha\cdot r_1', \circledS_\alpha\in \{A\cup B\}_\delta$ and $r_2\equiv\alpha\in \{A\cup B\}_\delta$. Then $pqACP^+_{QE}\vdash r_1\between r_2=\circledS_\alpha\cdot r_1'\between\alpha=\alpha\cdot r_1'$, by induction $r_1'$ is a closed $pqBPA$ term, $\alpha\cdot r_1'$ is a closed $pqBPA$ term;
  \item Case $r_1\equiv\circledS_\alpha\in \{A\cup B\}_\delta$ and $r_2\equiv\alpha\cdot r_2', \alpha\in \{A\cup B\}_\delta$. Then $pqACP^+_{QE}\vdash r_1\between r_2=\circledS_\alpha\between\alpha\cdot r_2'=\alpha\cdot r_2'$, by induction $r_2'$ is a closed $pqBPA$ term, $\alpha\cdot r_2'$ is a closed $pqBPA$ term;
  \item Case $r_1\equiv\alpha\cdot r_1', \alpha\in \{A\cup B\}_\delta$ and $r_2\equiv\circledS_\alpha\in \{A\cup B\}_\delta$. Then $pqACP^+_{QE}\vdash r_1\between r_2=\alpha\cdot r_1'\between\circledS_\alpha=\alpha\cdot r_1'$, by induction $r_1'$ is a closed $pqBPA$ term, $\alpha\cdot r_1'$ is a closed $pqBPA$ term;
  \item Case $r_1\equiv\circledS_\alpha\cdot r_1',\circledS_\alpha\in \{A\cup B\}_\delta$ and $r_2\equiv\alpha\cdot r_2', \alpha\in \{A\cup B\}_\delta$. Then $pqACP^+_{QE}\vdash r_1\between r_2=\circledS_\alpha\cdot r_1'\between\alpha\cdot r_2'=\alpha\cdot (r_1'\between r_2')$, by induction there is a closed $pqBPA$ term $s$ such that $pqACP^+_{QE}\vdash r_1'\between r_2'=s$,  and $pqACP^+_{QE}\vdash r_1'\between r_2'=\alpha\cdot s$ is a closed $pqBPA$ term;
  \item Case $r_1\equiv\alpha\cdot r_1', \alpha\in \{A\cup B\}_\delta$ and $r_2\equiv\circledS_\alpha\cdot r_2',\circledS_\alpha\in \{A\cup B\}_\delta$. Then $pqACP^+_{QE}\vdash r_1\between r_2=\alpha\cdot r_1'\between\circledS_\alpha\cdot r_2'=\alpha\cdot (r_1'\between r_2')$, by induction there is a closed $pBPA$ term $s$ such that $pqACP^+_{QE}\vdash r_1'\between r_2'=s$, and $pqACP^+_{QE}\vdash r_1'\between r_2'=\alpha\cdot s$ is a closed $pqBPA$ term;
  \item Case $r_1\equiv\alpha\in \{A\cup B\}_\delta$ and $r_2\equiv r_2'+r_2''$. Then $pqACP^+_{QE}\vdash r_1\between r_2=\alpha\between (r_2'+r_2'')=\alpha\between r_2'+\alpha\between r_2''$, by induction, there are two closed $pqBPA$ terms $s_1$ and $s_2$ such that $pqACP^+_{QE}\vdash \alpha\between r_2'=s_1$, and $pqACP^+_{QE}\vdash \alpha\between r_2''=s_2$, then $pqACP^+_{QE}\vdash r_1\between r_2=s_1+s_2$ which is a closed $pqBPA$ term;
  \item Case $r_1\equiv r_1'+r_1''$ and $r_2\equiv \alpha\in \{A\cup B\}_\delta$. Then $pqACP^+_{QE}\vdash r_1\between r_2=(r_1'+r_1'')\between\alpha =r_1'\between\alpha +r_1''\between\alpha $, by induction, there are two closed $pqBPA$ terms $s_1$ and $s_2$ such that $pqACP^+_{QE}\vdash r_1' \between\alpha=s_1$, and $pqACP^+_{QE}\vdash r_1''\between \alpha=s_2$, then $pqACP^+_{QE}\vdash r_1\between r_2=s_1+s_2$ which is a closed $pqBPA$ term;
  \item Case $r_1\equiv\alpha\cdot r_1',\alpha\in \{A\cup B\}_\delta$ and $r_2\equiv r_2'+r_2''$. Then $pqACP^+_{QE}\vdash r_1\between r_2=\alpha\cdot r_1'\between (r_2'+r_2'')=\alpha\cdot r_1'\between r_2'+\alpha\cdot r_1'\between r_2''$, by induction, there are two closed $pqBPA$ terms $s_1$ and $s_2$ such that $pqACP^+_{QE}\vdash \alpha\cdot r_1'\between r_2'=s_1$, and $pqACP^+_{QE}\vdash \alpha\cdot r_1'\between r_2''=s_2$, then $pqACP^+_{QE}\vdash r_1\between r_2=s_1+s_2$ which is a closed $pqBPA$ term;
  \item Case $r_1\equiv r_1'+r_1''$ and $r_2\equiv \alpha\cdot r_2',\alpha\in \{A\cup B\}_\delta$. Then $pqACP^+_{QE}\vdash r_1\between r_2=(r_1'+r_1'')\between\alpha \cdot r_2' =r_1'\between\alpha\cdot r_2' +r_1''\between\alpha\cdot r_2' $, by induction, there are two closed $pqBPA$ terms $s_1$ and $s_2$ such that $pqACP^+_{QE}\vdash r_1' \between\alpha\cdot r_2'=s_1$, and $pqACP^+_{QE}\vdash r_1''\between \alpha\cdot r_2'=s_2$, then $pqACP^+_{QE}\vdash r_1\between r_2=s_1+s_2$ which is a closed $pqBPA$ term;
  \item Case $r_1\equiv r_1'+r_1''$ and $r_2\equiv r_2' + r_2''$. Then $pqACP^+_{QE}\vdash r_1\between r_2=(r_1'+r_1'')\between(r_2'+r_2'') =r_1'\between r_2' +r_1''\between r_2' + r_1'\between r_2'' +r_1''\between r_2''$, by induction, there are four closed $pqBPA$ terms $s_1$, $s_2$, $s_3$ and $s_4$ such that $pqACP^+_{QE}\vdash r_1' \between r_2'=s_1$, $pqACP^+_{QE}\vdash r_1''\between r_2'=s_2$, $pqACP^+_{QE}\vdash r_1'\between r_2''=s_3$, $pqACP^+_{QE}\vdash r_1''\between r_2''=s_4$, then $pqACP^+_{QE}\vdash r_1\between r_2=s_1+s_2+s_3+s_4$ which is a closed $pqBPA$ term;
  \item Case $r_1\in\mathcal{B}_+$ and $r_2\in\mathcal{B}\setminus\mathcal{B}_+$. For some $m\in\mathbb{N}$, $m\geq 2$, $r_{2i}\in\mathcal{B}_+$ and $\pi_i\in\langle 0,1\rangle$ with $1\leq i\leq m$, $r_2\equiv r_{21}\boxplus_{\pi_1}r_{22}\boxplus_{\pi_2}\cdots\boxplus_{\pi_{m-1}}r_{2m}$. Then $pqACP^+_{QE}\vdash r_1\between r_2=r_1\between (r_{21}\boxplus_{\pi_1}r_{22}\boxplus_{\pi_2}\cdots\boxplus_{\pi_{m-1}}r_{2m})= (r_1\between r_{21})\boxplus_{\pi_1}(r_1\between r_{22})\boxplus_{\pi_2}\cdots\boxplus_{\pi_{m-1}}(r_1\between r_{2m})$, by induction hypothesis, there are $m$ closed $pBPA$ terms $s_1,\cdots,s_m$, such that $pqACP^+_{QE}\vdash r_1\between r_{21}=s_1$, $\cdots$, $pqACP^+_{QE}\vdash r_1\between r_{2m}=s_m$, then $pqACP^+_{QE}\vdash s_1\boxplus_{\pi_1}\cdots\boxplus_{\pi_{m-1}}s_m$ which is a closed $pqBPA$ term;
  \item Case $r_1\in\mathcal{B}\setminus\mathcal{B}_+$ and $r_2\in\mathcal{B}_+$. For some $m\in\mathbb{N}$, $m\geq 2$, $r_{1i}\in\mathcal{B}_+$ and $\pi_i\in\langle 0,1\rangle$ with $1\leq i\leq m$, $r_1\equiv r_{11}\boxplus_{\pi_1}r_{12}\boxplus_{\pi_2}\cdots\boxplus_{\pi_{m-1}}r_{1m}$. Then $pqACP^+_{QE}\vdash r_1\between r_2=(r_{11}\boxplus_{\pi_1}r_{12}\boxplus_{\pi_2}\cdots\boxplus_{\pi_{m-1}}r_{1m})\between r_2= (r_{11}\between r_2)\boxplus_{\pi_1}(r_{12}\between r_2)\boxplus_{\pi_2}\cdots\boxplus_{\pi_{m-1}}(r_{1m}\between r_2)$, by induction hypothesis, there are $m$ closed $pBPA$ terms $s_1,\cdots,s_m$, such that $pqACP^+_{QE}\vdash r_{11}\between r_2=s_1$, $\cdots$, $pqACP^+_{QE}\vdash r_{1m}\between r_2=s_m$, then $pqACP^+_{QE}\vdash s_1\boxplus_{\pi_1}\cdots\boxplus_{\pi_{m-1}}s_m$ which is a closed $pqBPA$ term;
  \item Case $r_1,r_2\in\mathcal{B}\setminus\mathcal{B}_+$. For some $m\in\mathbb{N}$, $m\geq 2$, $r_{1i}\in\mathcal{B}_+$ and $\pi_i\in\langle 0,1\rangle$ with $1\leq i\leq m$, $r_1\equiv r_{11}\boxplus_{\pi_1}r_{12}\boxplus_{\pi_2}\cdots\boxplus_{\pi_{m-1}}r_{1m}$. For some $n\in\mathbb{N}$, $n\geq 2$, $r_{2j}\in\mathcal{B}_+$ and $\rho_j\in\langle 0,1\rangle$ with $1\leq j\leq n$, $r_2\equiv r_{21}\boxplus_{\rho_1}r_{22}\boxplus_{\rho_2}\cdots\boxplus_{\rho_{n-1}}r_{2n}$. $pqACP^+_{QE}\vdash r_1\between r_2=\boxplus_{\pi_i\cdot\rho_j}(r_{1i}\between r_{2j})$. By the induction, there are closed $pBPA$ terms $s_{ij}$ such that $pqACP^+_{QE}\vdash r_{1i}\between r_{2j}=s_{ij}$, then $pqACP^+_{QE}\vdash \boxplus_{\pi_i\cdot\rho_j}s_{ij}$ which is a closed $pqBPA$ term.
\end{itemize}

Note that the mismatch of $\alpha\in \{A\cup B\}_\delta$ and $\circledS_\alpha\neq\beta\in \{A\cup B\}_\delta$ in $\alpha\between\beta=\delta$, and $\delta$ is also a closed basic $pqBPA$ term.
\end{proof}

\theorem[Congruence theorem of $pqACP^+_{QE}$]{$\underline{\leftrightarrow}_q$ is a congruence relation on $\mathbb{PT}(pqACP^+_{QE})$ with respect to the $+$, $\cdot$, $\boxplus_{\pi}$, $\leftmerge$, $\mid$, $\parallel$, $][$, $\partial_H$ and $\between$ operators.}

\begin{proof}
The part of the proof for the operators of $pqACP^+$ is the same as the proof of the congruence of $pqACP^+$. Here we only give the rest of the proof which concerns the operator $\between$. For the operators $][$ and $\parallel$, there are some changes related operator $\between$, the proof is similarly to that of $pqACP^+$ (please refer to \cite{PPA}).

Let $x$, $y$, $z$ and $w$ be $\mathbb{PT}(pqACP^+_{QE})$ processes such that $\langle x,\varrho\rangle\underline{\leftrightarrow}_q \langle y,\varrho\rangle$ and $\langle z,\varrho\rangle\underline{\leftrightarrow}_q \langle w,\varrho\rangle$. So, there exist probabilistic bisimulations $R_1$ and $R_2$ such that $(\langle x,\varrho\rangle,\langle y,\varrho\rangle)\in R_1$ and $(\langle z,\varrho\rangle,\langle w,\varrho\rangle)\in R_2$. We define a relation $R$ in the following way:

$$R=Eq(S'\cup D'\cup S\cup D\cup R_1\cup R_2)$$

where $S=\{(\langle p\parallel q,\varrho\rangle,\langle s\parallel t,\varrho\rangle):p,q,s,t\in\mathbb{SP}(pqACP^+_{QE}),(\langle p,\varrho\rangle,\langle s,\varrho\rangle)\in R_1,(\langle q,\varrho\rangle,\langle t,\varrho\rangle)\in R_2\}$,

$D=\{(\langle u\leftmerge q+v\leftmerge p+u\mid v+u\between v,\varrho\rangle,\langle l\leftmerge t+l\leftmerge t+l\mid k+l\between k,\varrho\rangle):p,q,s,t\in\mathbb{SP}(pqACP^+_{QE}),u,v,l,k\in\mathbb{DP}(pqACP^+_{QE}),(\langle p,\varrho\rangle,\langle s,\varrho\rangle), (\langle u,\varrho\rangle,\langle l,\varrho\rangle)\in R_1,(\langle q,\varrho\rangle,\langle t,\varrho\rangle),(\langle v,\varrho\rangle,\langle k,\varrho\rangle)\in R_2\}$,

$S'=\{(\langle p\between q,\varrho\rangle,\langle s\between t,\varrho\rangle):p,q,s,t\in\mathbb{SP}(pqACP^+_{QE}),(\langle p,\varrho\rangle,\langle s,\varrho\rangle)\in R_1,(\langle q,\varrho\rangle,\langle t,\varrho\rangle)\in R_2\}$,

$D'=\{(\langle u\between v,\varrho\rangle,\langle l\between k,\varrho\rangle):u,v,l,k\in\mathbb{DP}(pqACP^+_{QE}),(\langle u,\varrho\rangle,\langle l,\varrho\rangle)\in R_1,(\langle v,\varrho\rangle,\langle k,\varrho\rangle)\in R_2\}$.

\begin{itemize}
  \item \textbf{EM1}: $S'$ and $D'$ are equivalence relation, and $S'$, $S$, $R_1$, $R_2$ contain pairs of static processes relevant to $R$;
  \item \textbf{EM2}: if $(\langle p\between q,\varrho\rangle,\langle s\between t,\varrho\rangle)\in S'$ and $K\in \mathbb{DP}(pqACP^+_{QE})/D'$, then $\langle p\between q,\varrho\rangle\rightsquigarrow\langle K,\varrho\rangle$ iff $\langle s\between t,\varrho\rangle\rightsquigarrow\langle K,\varrho\rangle$;
  \item \textbf{EM3}: if $\langle p\between q,\varrho\rangle\rightsquigarrow\langle K,\varrho\rangle$ for $K\in \mathbb{DP}(pqACP^+_{QE})/D'$, then $K=[u\between v]_{D'}$ for some $u,v$ such that $\langle p,\varrho\rangle\rightsquigarrow \langle u,\varrho\rangle$ and $\langle u,\varrho\rangle\rightsquigarrow \langle v,\varrho\rangle$. From the definition of $D'$, $K=[u]_{R_1}\between [v]_{R_2}$;
  \item \textbf{EM4}: since $R_1$, $R_2$, $D$, $D'$ are all subsets of $R$ and they are equivalence relations themselves, if $M\in\mathbb{DP}(pqACP^+_{QE})/R$, then $M=\bigcup_{i_1\in I_1}M^1_{i_1}$, $M=\bigcup_{i_2\in I_2}M^2_{i_2}$, $M=\bigcup_{j\in J}K_j$ and $M=\bigcup_{n\in N}M_n$ for some non-empty index set $I_1,I_2,J,N$ and some equivalence classes $M^1_{i_1}(i_1\in I_1)$, $M^2_{i_2}(i_2\in I_2)$, $K_j(j\in J)$ and $M_n(n\in N)$ of $R_1$, $R_2$, $D'$ and $D$ respectively.
\end{itemize}

Now suppose that $(\langle r,\varrho\rangle,\langle r_1,\varrho\rangle)\in R$ for some $r,r_1\in\mathbb{SP}(pqACP^+_{QE})$ and $M\in\mathbb{DP}(pqACP^+_{QE})/R$, then:
\begin{enumerate}
  \item if $(\langle r,\varrho\rangle,\langle r_1,\varrho\rangle)\in R_k,k=1,2$, then the result follows from \textbf{EM4};
  \item if $(\langle r,\varrho\rangle,\langle r_1,\varrho\rangle)\in S$, then $r\equiv p\parallel q$ and $r_1\equiv s\parallel t$ for some $p,q,s,t\in \mathbb{SP}(pqACP^+_{QE})$ such that $(\langle p,\varrho\rangle,\langle s,\varrho\rangle)\in R_1$ and $(\langle q,\varrho\rangle,\langle t,\varrho\rangle)\in R_2$. Then $K_j=[u_j]_{R_1}^{[p]_{R_1}}\parallel^{[q]_{R_2}}[v_j]_{R_2}$ and $\langle p,\varrho\rangle\rightsquigarrow\langle u_j,\varrho\rangle$ and $\langle q,\varrho\rangle\rightsquigarrow\langle v_j,\varrho\rangle$. Then $\mu(p\parallel q, K_j)=\mu(s\parallel t, K_j)$, $\mu(p\parallel q, M)=\mu(s\parallel t, M)$;
  \item if $(\langle r,\varrho\rangle,\langle r_1,\varrho\rangle)\in S'$, then $r\equiv p\between q$ and $r_1\equiv s\between t$ for some $p,q,s,t\in \mathbb{SP}(pqACP^+_{QE})$ such that $(\langle p,\varrho\rangle,\langle s,\varrho\rangle)\in R_1$ and $(\langle q,\varrho\rangle,\langle t,\varrho\rangle)\in R_2$. According to \textbf{EM3} and \textbf{EM4}, $K_j=[u_j]_{R_1}\between[v_j]_{R_2}$ and $\langle p,\varrho\rangle\rightsquigarrow\langle u_j,\varrho\rangle$ and $\langle q,\varrho\rangle\rightsquigarrow\langle v_j,\varrho\rangle$. Then $\mu(p\between q, K_j)=\mu(s\between t, K_j)$, $\mu(p\between q, M)=\mu(s\between t, M)$.
\end{enumerate}
\end{proof}

\theorem[Soundness of $pqACP^+_{QE}$]{Let $p$ and $q$ be closed $pqACP^+_{QE}$ terms. If $pqACP^+_{QE}\vdash p=q$ then $p\underline{\leftrightarrow}_q q$.}\label{SoundnessQE}

\begin{proof}

It is already proven that $\underline{\leftrightarrow}_q$ is an equivalent and congruent relation on $\mathbb{PT}(pqACP^+_{QE})$, it is sufficient to examine every axiom in the axiom system (in Table \ref{AxiomsForQE}) for $pqACP^+_{QE}$ is sound.

\begin{itemize}
  \item \textbf{Axiom $PrMM4$}. For a relation $R$, suppose that $Eq(\langle (u,z)][(v,w),\varrho\rangle,\langle u\leftmerge w+v\leftmerge z+u\mid v+u\between v,\varrho\rangle)= R$ with $u=u+u$, $v=v+v$, and $u,v,z,w\in \mathbb{SP}(pqACP^+_{QE})$.

      By use of the probabilistic transition rules for operators $\leftmerge$, $\mid$, $][$ and $\between$, we get:

      $$\frac{\langle u,\varrho\rangle\rightsquigarrow \langle u',\varrho\rangle, \langle v,\varrho\rangle\rightsquigarrow \langle v',\varrho\rangle}{\langle (u,z)][(v,w),\varrho\rangle\rightsquigarrow\langle u'\leftmerge w+v'\leftmerge z+u'\mid v'+u'\between v',\varrho\rangle}$$

      $$\frac{\langle u,\varrho\rangle\rightsquigarrow \langle u',\varrho\rangle, \langle v,\varrho\rangle\rightsquigarrow \langle v',\varrho\rangle}{\langle u\leftmerge w+v\leftmerge z+u\mid v+u\between v,\varrho\rangle\rightsquigarrow\langle u'\leftmerge w+v'\leftmerge z+u'\mid v'+u'\between v',\varrho\rangle}$$

      In the above two probabilistic transition rules, the condition $u=u+u$ ensures that $\langle u,\varrho\rangle\rightsquigarrow \langle u',\varrho\rangle$, $\langle u,\varrho\rangle\rightsquigarrow \langle u'',\varrho\rangle$, $\langle u',\varrho\rangle\underline{\leftrightarrow}_q \langle u'',\varrho\rangle$. $v=v+v$ ensures the similar things.

      With the assumption $(u,z)][(v,w)=u\leftmerge w+v\leftmerge z+u\mid v+u\between v$, we get $R$ satisfies condition 1 in the definition of $\underline{\leftrightarrow}_q$
      (see Definition \ref{SPQB}). Note that, for operator $][$, there are not action transition rules.

      It is easy to check that the condition on PDF $\mu$ for $R$ also satisfies condition 4 in Definition \ref{SPQB}, we omit it.

      So, $R=\underline{\leftrightarrow}_q$, as desired.

  \item \textbf{Axiom $EM1$}. For a relation $R=R_1\cup R_2$, suppose that $Eq(\langle \alpha\between\circledS_{\alpha},\varrho\rangle,\langle \alpha,\varrho\rangle)= R_1$, and $Eq(\langle \breve{\alpha}\between\breve{\circledS_\alpha},\varrho\rangle,\langle \breve{\alpha},\varrho\rangle)= R_2$.

      By use of the probabilistic transition rules and the action transition rules for atomic action $\alpha,\circledS_\alpha$ and entanglement merge $\between$, we get:

      $$\frac{\langle \alpha,\varrho\rangle\rightsquigarrow \langle \breve{\alpha},\varrho\rangle, \langle \circledS_\alpha,\varrho\rangle\rightsquigarrow \langle \breve{\circledS_\alpha},\varrho\rangle}{\langle \alpha\between\circledS_\alpha,\varrho\rangle\rightsquigarrow\langle \breve{\alpha}\between\breve{\circledS_\alpha},\varrho\rangle}$$

      $$\frac{\langle \breve{\alpha},\varrho\rangle\xrightarrow{\alpha} \langle \surd,\varrho'\rangle,\langle \breve{\circledS_\alpha},\varrho'\rangle\rightarrow \langle \surd,\varrho'\rangle}{\langle \breve{\alpha}\between\breve{\circledS_\alpha},\varrho\rangle\xrightarrow{\alpha}\langle \surd,\varrho'\rangle}$$

      With the assumptions $\alpha\between \circledS_\alpha=\alpha$ and $\breve{\alpha}\between\breve{\circledS_\alpha}=\breve{\alpha}$, we get $R$ satisfies conditions 1, 2 and 3 in the definition of $\underline{\leftrightarrow}_q$
      (see Definition \ref{SPQB}).

      It is easy to check that the condition on PDF $\mu$ for $R$ also satisfies condition 4 in Definition \ref{SPQB}, we omit it.

      So, $R=\underline{\leftrightarrow}_q$, as desired.

  \item \textbf{Axiom $EM2$}. For a relation $R=R_1\cup R_2$, suppose that $Eq(\langle \circledS_\alpha\between\alpha,\varrho\rangle,\langle \alpha,\varrho\rangle)= R_1$, and $Eq(\langle \breve{\circledS_\alpha}\between\breve{\alpha},\varrho\rangle,\langle \breve{\alpha},\varrho\rangle)= R_2$.

      By use of the probabilistic transition rules and the action transition rules for atomic action $\alpha,\circledS_\alpha$ and entanglement merge $\between$, we get:

      $$\frac{\langle \circledS_\alpha,\varrho\rangle\rightsquigarrow \langle \breve{\circledS_\alpha},\varrho\rangle, \langle \alpha,\varrho\rangle\rightsquigarrow \langle \breve{\alpha},\varrho\rangle}{\langle \circledS_\alpha\between\alpha,\varrho\rangle\rightsquigarrow\langle \breve{\circledS_\alpha}\between\breve{\alpha},\varrho\rangle}$$

      $$\frac{\langle \breve{\alpha},\varrho\rangle\xrightarrow{\alpha} \langle \surd,\varrho'\rangle,\langle \breve{\circledS_\alpha},\varrho'\rangle\rightarrow \langle \surd,\varrho'\rangle}{\langle \breve{\circledS_\alpha}\between\breve{\alpha},\varrho\rangle\xrightarrow{\alpha}\langle \surd,\varrho'\rangle}$$

      With the assumptions $\circledS_\alpha\between \alpha=\alpha$ and $\breve{\circledS_\alpha}\between\breve{\alpha}=\breve{\alpha}$, we get $R$ satisfies conditions 1, 2 and 3 in the definition of $\underline{\leftrightarrow}_q$
      (see Definition \ref{SPQB}).

      It is easy to check that the condition on PDF $\mu$ for $R$ also satisfies condition 4 in Definition \ref{SPQB}, we omit it.

      So, $R=\underline{\leftrightarrow}_q$, as desired.

  \item \textbf{Axiom $EM3$}. For a relation $R=R_1\cup R_2$, suppose that $Eq(\langle \alpha\between\circledS_\alpha\cdot u,\varrho\rangle,\langle \alpha\cdot u,\varrho\rangle)= R_1$ with $u\in \mathbb{SP}(pqACP^+_{QE})$, and $Eq(\langle \breve{\alpha}\between\breve{\circledS_\alpha}\cdot u,\varrho\rangle,\langle \breve{\alpha}\cdot u,\varrho\rangle)= R_2$ with $u\in \mathbb{SP}(pqACP^+_{QE})$.

      By use of the probabilistic transition rules and the action transition rules for atomic action $\alpha,\circledS_\alpha$, sequential composition $\cdot$ and entanglement merge $\between$, we get:

      $$\frac{\langle \alpha,\varrho\rangle\rightsquigarrow \langle \breve{\alpha},\varrho\rangle, \langle \circledS_\alpha,\varrho\rangle\rightsquigarrow \langle \breve{\circledS_\alpha},\varrho\rangle}{\langle \alpha\between \circledS_\alpha\cdot u,\varrho\rangle\rightsquigarrow\langle \breve{\alpha}\between\breve{\circledS_\alpha}\cdot u,\varrho\rangle}$$

      $$\frac{\langle \alpha,\varrho\rangle\rightsquigarrow \langle \breve{\alpha},\varrho\rangle}{\langle \alpha\cdot u,\varrho\rangle\rightsquigarrow\langle \breve{\alpha}\cdot u,\varrho\rangle}$$

      $$\frac{\langle \breve{\alpha},\varrho\rangle\xrightarrow{\alpha} \langle \surd,\varrho'\rangle,\langle \breve{\circledS_\alpha},\varrho'\rangle\rightarrow \langle \surd,\varrho'\rangle}{\langle \breve{\alpha}\between\breve{\circledS_\alpha}\cdot u,\varrho\rangle\xrightarrow{\alpha}\langle u,\varrho'\rangle}$$

      $$\frac{\langle \breve{\alpha},\varrho\rangle\xrightarrow{\alpha} \langle \surd,\varrho'\rangle}{\langle \breve{\alpha}\cdot u,\varrho\rangle\xrightarrow{\alpha}\langle u,\varrho'\rangle}$$

      With the assumptions $\alpha\between \circledS_\alpha\cdot u=\alpha\cdot u$ and $\breve{\alpha}\between\breve{\circledS_\alpha}\cdot u=\breve{\alpha}\cdot u$, we get $R$ satisfies conditions 1, 2 and 3 in the definition of $\underline{\leftrightarrow}_q$
      (see Definition \ref{SPQB}).

      It is easy to check that the condition on PDF $\mu$ for $R$ also satisfies condition 4 in Definition \ref{SPQB}, we omit it.

      So, $R=\underline{\leftrightarrow}_q$, as desired.

  \item \textbf{Axiom $EM4$}. For a relation $R=R_1\cup R_2$, suppose that $Eq(\langle \circledS_\alpha\between\alpha\cdot u,\varrho\rangle,\langle \alpha\cdot u,\varrho\rangle)= R_1$ with $u\in \mathbb{SP}(pqACP^+_{QE})$, and $Eq(\langle \breve{\circledS_\alpha}\between\breve{\alpha}\cdot u,\varrho\rangle,\langle \breve{\alpha}\cdot u,\varrho\rangle)= R_2$ with $u\in \mathbb{SP}(pqACP^+_{QE})$.

      By use of the probabilistic transition rules and the action transition rules for atomic action $\alpha,\circledS_\alpha$, sequential composition $\cdot$ and entanglement merge $\between$, we get:

      $$\frac{\langle \alpha,\varrho\rangle\rightsquigarrow \langle \breve{\alpha},\varrho\rangle, \langle \circledS_\alpha,\varrho\rangle\rightsquigarrow \langle \breve{\circledS_\alpha},\varrho\rangle}{\langle \circledS_\alpha\between \alpha\cdot u,\varrho\rangle\rightsquigarrow\langle \breve{\circledS_\alpha}\between\breve{\alpha}\cdot u,\varrho\rangle}$$

      $$\frac{\langle \alpha,\varrho\rangle\rightsquigarrow \langle \breve{\alpha},\varrho\rangle}{\langle \alpha\cdot u,\varrho\rangle\rightsquigarrow\langle \breve{\alpha}\cdot u,\varrho\rangle}$$

      $$\frac{\langle \breve{\alpha},\varrho\rangle\xrightarrow{\alpha} \langle \surd,\varrho'\rangle,\langle \breve{\circledS_\alpha},\varrho'\rangle\rightarrow \langle \surd,\varrho'\rangle}{\langle \breve{\circledS_\alpha}\between\breve{\alpha}\cdot u,\varrho\rangle\xrightarrow{\alpha}\langle u,\varrho'\rangle}$$

      $$\frac{\langle \breve{\alpha},\varrho\rangle\xrightarrow{\alpha} \langle \surd,\varrho'\rangle}{\langle \breve{\alpha}\cdot u,\varrho\rangle\xrightarrow{\alpha}\langle u,\varrho'\rangle}$$

      With the assumptions $\circledS_\alpha\between \alpha\cdot u=\alpha\cdot u$ and $\breve{\circledS_\alpha}\between\breve{\alpha}\cdot u=\breve{\alpha}\cdot u$, we get $R$ satisfies conditions 1, 2 and 3 in the definition of $\underline{\leftrightarrow}_q$
      (see Definition \ref{SPQB}).

      It is easy to check that the condition on PDF $\mu$ for $R$ also satisfies condition 4 in Definition \ref{SPQB}, we omit it.

      So, $R=\underline{\leftrightarrow}_q$, as desired.

  \item \textbf{Axiom $EM5$}. For a relation $R=R_1\cup R_2$, suppose that $Eq(\langle \alpha\cdot u\between \circledS_\alpha,\varrho\rangle,\langle \alpha\cdot u,\varrho\rangle)= R_1$ with $u\in \mathbb{SP}(pqACP^+_{QE})$, and $Eq(\langle \breve{\alpha}\cdot u\between\breve{\circledS_\alpha},\varrho\rangle,\langle \breve{\alpha}\cdot u,\varrho\rangle)= R_2$ with $u\in \mathbb{SP}(pqACP^+_{QE})$.

      By use of the probabilistic transition rules and the action transition rules for atomic action $\alpha,\circledS_\alpha$, sequential composition $\cdot$ and entanglement merge $\between$, we get:

      $$\frac{\langle \alpha,\varrho\rangle\rightsquigarrow \langle \breve{\alpha},\varrho\rangle, \langle \circledS_\alpha,\varrho\rangle\rightsquigarrow \langle \breve{\circledS_\alpha},\varrho\rangle}{\langle \alpha\cdot u\between \circledS_\alpha,\varrho\rangle\rightsquigarrow\langle \breve{\alpha}\cdot u\between\breve{\circledS_\alpha},\varrho\rangle}$$

      $$\frac{\langle \alpha,\varrho\rangle\rightsquigarrow \langle \breve{\alpha},\varrho\rangle}{\langle \alpha\cdot u,\varrho\rangle\rightsquigarrow\langle \breve{\alpha}\cdot u,\varrho\rangle}$$

      $$\frac{\langle \breve{\alpha},\varrho\rangle\xrightarrow{\alpha} \langle \surd,\varrho'\rangle,\langle \breve{\circledS_\alpha},\varrho'\rangle\rightarrow \langle \surd,\varrho'\rangle}{\langle \breve{\alpha}\cdot u\between\breve{\circledS_\alpha},\varrho\rangle\xrightarrow{\alpha}\langle u,\varrho'\rangle}$$

      $$\frac{\langle \breve{\alpha},\varrho\rangle\xrightarrow{\alpha} \langle \surd,\varrho'\rangle}{\langle \breve{\alpha}\cdot u,\varrho\rangle\xrightarrow{\alpha}\langle u,\varrho'\rangle}$$

      With the assumptions $\alpha\cdot u\between\circledS_\alpha=\alpha\cdot u$ and $\breve{\alpha}\cdot u\between\breve{\circledS_\alpha}=\breve{\alpha}\cdot u$, we get $R$ satisfies conditions 1, 2 and 3 in the definition of $\underline{\leftrightarrow}_q$
      (see Definition \ref{SPQB}).

      It is easy to check that the condition on PDF $\mu$ for $R$ also satisfies condition 4 in Definition \ref{SPQB}, we omit it.

      So, $R=\underline{\leftrightarrow}_q$, as desired.

  \item \textbf{Axiom $EM6$}. For a relation $R=R_1\cup R_2$, suppose that $Eq(\langle \circledS_\alpha\cdot u\between \alpha,\varrho\rangle,\langle \alpha\cdot u,\varrho\rangle)= R_1$ with $u\in \mathbb{SP}(pqACP^+_{QE})$, and $Eq(\langle \breve{\circledS_\alpha}\cdot u\between\breve{\alpha},\varrho\rangle,\langle \breve{\alpha}\cdot u,\varrho\rangle)= R_2$ with $u\in \mathbb{SP}(pqACP^+_{QE})$.

      By use of the probabilistic transition rules and the action transition rules for atomic action $\alpha,\circledS_\alpha$, sequential composition $\cdot$ and entanglement merge $\between$, we get:

      $$\frac{\langle \alpha,\varrho\rangle\rightsquigarrow \langle \breve{\alpha},\varrho\rangle, \langle \circledS_\alpha,\varrho\rangle\rightsquigarrow \langle \breve{\circledS_\alpha},\varrho\rangle}{\langle \circledS_\alpha\cdot u\between \alpha,\varrho\rangle\rightsquigarrow\langle \breve{\circledS_\alpha}\cdot u\between\breve{\alpha},\varrho\rangle}$$

      $$\frac{\langle \alpha,\varrho\rangle\rightsquigarrow \langle \breve{\alpha},\varrho\rangle}{\langle \alpha\cdot u,\varrho\rangle\rightsquigarrow\langle \breve{\alpha}\cdot u,\varrho\rangle}$$

      $$\frac{\langle \breve{\alpha},\varrho\rangle\xrightarrow{\alpha} \langle \surd,\varrho'\rangle,\langle \breve{\circledS_\alpha},\varrho'\rangle\rightarrow \langle \surd,\varrho'\rangle}{\langle \breve{\circledS_\alpha}\cdot u\between\breve{\alpha},\varrho\rangle\xrightarrow{\alpha}\langle u,\varrho'\rangle}$$

      $$\frac{\langle \breve{\alpha},\varrho\rangle\xrightarrow{\alpha} \langle \surd,\varrho'\rangle}{\langle \breve{\alpha}\cdot u,\varrho\rangle\xrightarrow{\alpha}\langle u,\varrho'\rangle}$$

      With the assumptions $\circledS_\alpha\cdot u\between\alpha=\alpha\cdot u$ and $\breve{\circledS_\alpha}\cdot u\between\breve{\alpha}=\breve{\alpha}\cdot u$, we get $R$ satisfies conditions 1, 2 and 3 in the definition of $\underline{\leftrightarrow}_q$
      (see Definition \ref{SPQB}).

      It is easy to check that the condition on PDF $\mu$ for $R$ also satisfies condition 4 in Definition \ref{SPQB}, we omit it.

      So, $R=\underline{\leftrightarrow}_q$, as desired.

  \item \textbf{Axiom $EM7$}. For a relation $R=R_1\cup R_2$, suppose that $Eq(\langle \alpha\cdot u\between \circledS_\alpha\cdot v,\varrho\rangle,\langle \alpha\cdot (u\parallel v),\varrho\rangle)= R_1$ with $u,v\in \mathbb{SP}(pqACP^+_{QE})$, and $Eq(\langle \breve{\alpha}\cdot u\between\breve{\circledS_\alpha}\cdot v,\varrho\rangle,\langle \breve{\alpha}\cdot (u\parallel v),\varrho\rangle)= R_2$ with $u,v\in \mathbb{SP}(pqACP^+_{QE})$.

      By use of the probabilistic transition rules and the action transition rules for atomic action $\alpha,\circledS_\alpha$, sequential composition $\cdot$ and entanglement merge $\between$, we get:

      $$\frac{\langle \alpha,\varrho\rangle\rightsquigarrow \langle \breve{\alpha},\varrho\rangle, \langle \circledS_\alpha,\varrho\rangle\rightsquigarrow \langle \breve{\circledS_\alpha},\varrho\rangle}{\langle \alpha\cdot u\between \circledS_\alpha\cdot v,\varrho\rangle\rightsquigarrow\langle \breve{\alpha}\cdot u\between\breve{\circledS_\alpha}\cdot v,\varrho\rangle}$$

      $$\frac{\langle \alpha,\varrho\rangle\rightsquigarrow \langle \breve{\alpha},\varrho\rangle}{\langle \alpha\cdot (u\parallel v),\varrho\rangle\rightsquigarrow\langle \breve{\alpha}\cdot (u\parallel v),\varrho\rangle}$$

      $$\frac{\langle \breve{\alpha},\varrho\rangle\xrightarrow{\alpha} \langle \surd,\varrho'\rangle,\langle \breve{\circledS_\alpha},\varrho'\rangle\rightarrow \langle \surd,\varrho'\rangle}{\langle \breve{\alpha}\cdot u\between\breve{\circledS_\alpha}\cdot v,\varrho\rangle\xrightarrow{\alpha}\langle (u\parallel v),\varrho'\rangle}$$

      $$\frac{\langle \breve{\alpha},\varrho\rangle\xrightarrow{\alpha} \langle \surd,\varrho'\rangle}{\langle \breve{\alpha}\cdot (u\parallel v),\varrho\rangle\xrightarrow{\alpha}\langle (u\parallel v),\varrho'\rangle}$$

      With the assumptions $\alpha\cdot u\between \circledS_\alpha\cdot v=\alpha\cdot (u\parallel v)$ and $\breve{\alpha}\cdot u\between\breve{\circledS_\alpha}\cdot v=\breve{\alpha}\cdot (u\parallel v)$, we get $R$ satisfies conditions 1, 2 and 3 in the definition of $\underline{\leftrightarrow}_q$
      (see Definition \ref{SPQB}).

      It is easy to check that the condition on PDF $\mu$ for $R$ also satisfies condition 4 in Definition \ref{SPQB}, we omit it.

      So, $R=\underline{\leftrightarrow}_q$, as desired.

  \item \textbf{Axiom $EM8$}. For a relation $R=R_1\cup R_2$, suppose that $Eq(\langle \circledS_\alpha\cdot u\between \alpha\cdot v,\varrho\rangle,\langle \alpha\cdot (u\parallel v),\varrho\rangle)= R_1$ with $u,v\in \mathbb{SP}(pqACP^+_{QE})$, and $Eq(\langle \breve{\circledS_\alpha}\cdot u\between\breve{\alpha}\cdot v,\varrho\rangle,\langle \breve{\alpha}\cdot (u\parallel v),\varrho\rangle)= R_2$ with $u,v\in \mathbb{SP}(pqACP^+_{QE})$.

      By use of the probabilistic transition rules and the action transition rules for atomic action $\alpha,\circledS_\alpha$, sequential composition $\cdot$ and entanglement merge $\between$, we get:

      $$\frac{\langle \alpha,\varrho\rangle\rightsquigarrow \langle \breve{\alpha},\varrho\rangle, \langle \circledS_\alpha,\varrho\rangle\rightsquigarrow \langle \breve{\circledS_\alpha},\varrho\rangle}{\langle \circledS_\alpha\cdot u\between \alpha\cdot v,\varrho\rangle\rightsquigarrow\langle \breve{\circledS_\alpha}\cdot u\between\breve{\alpha}\cdot v,\varrho\rangle}$$

      $$\frac{\langle \alpha,\varrho\rangle\rightsquigarrow \langle \breve{\alpha},\varrho\rangle}{\langle \alpha\cdot (u\parallel v),\varrho\rangle\rightsquigarrow\langle \breve{\alpha}\cdot (u\parallel v),\varrho\rangle}$$

      $$\frac{\langle \breve{\alpha},\varrho\rangle\xrightarrow{\alpha} \langle \surd,\varrho'\rangle,\langle \breve{\circledS_\alpha},\varrho'\rangle\rightarrow \langle \surd,\varrho'\rangle}{\langle \breve{\circledS_\alpha}\cdot u\between\breve{\alpha}\cdot v,\varrho\rangle\xrightarrow{\alpha}\langle (u\parallel v),\varrho'\rangle}$$

      $$\frac{\langle \breve{\alpha},\varrho\rangle\xrightarrow{\alpha} \langle \surd,\varrho'\rangle}{\langle \breve{\alpha}\cdot (u\parallel v),\varrho\rangle\xrightarrow{\alpha}\langle (u\parallel v),\varrho'\rangle}$$

      With the assumptions $\circledS_\alpha\cdot u\between \alpha\cdot v=\alpha\cdot (u\parallel v)$ and $\breve{\circledS_\alpha}\cdot u\between\breve{\alpha}\cdot v=\breve{\alpha}\cdot (u\parallel v)$, we get $R$ satisfies conditions 1, 2 and 3 in the definition of $\underline{\leftrightarrow}_q$
      (see Definition \ref{SPQB}).

      It is easy to check that the condition on PDF $\mu$ for $R$ also satisfies condition 4 in Definition \ref{SPQB}, we omit it.

      So, $R=\underline{\leftrightarrow}_q$, as desired.

  \item \textbf{Axiom $PrEM1$}. For a relation $R$, suppose that $Eq(\langle (u\boxplus_{\pi}v)\between w,\varrho\rangle,\langle u\between w\boxplus_{\pi}v\between w,\varrho\rangle)= R$ with $u,v,w\in \mathbb{SP}(pqACP^+_{QE})$.

      By use of the probabilistic transition rules for probabilistic choice operator $\boxplus_{\pi}$ and entanglement merge $\between$, we get:

      $$\frac{\langle u,\varrho\rangle\rightsquigarrow \langle u',\varrho\rangle}{\langle (u\boxplus_{\pi}v)\between w,\varrho\rangle\rightsquigarrow\langle u'\between w,\varrho\rangle}$$

      $$\frac{\langle u,\varrho\rangle\rightsquigarrow \langle u',\varrho\rangle}{\langle u\between w\boxplus_{\pi}v\between w,\varrho\rangle\rightsquigarrow\langle u'\between w,\varrho\rangle}$$

      With the assumption $(u\boxplus_{\pi}v)\between w=u\between w\boxplus_{\pi}v\between w$, we get $R$ satisfies condition 1 in the definition of $\underline{\leftrightarrow}_q$
      (see Definition \ref{SPQB}). Note that, for probabilistic choice operator $\boxplus_{\pi}$, there are not action transition rules.

      It is easy to check that the condition on PDF $\mu$ for $R$ also satisfies condition 4 in Definition \ref{SPQB}, we omit it.

      So, $R=\underline{\leftrightarrow}_q$, as desired.

  \item \textbf{Axiom $PrEM2$}. For a relation $R$, suppose that $Eq(\langle u\between (v\boxplus_{\pi}w),\varrho\rangle,\langle u\between v\boxplus_{\pi}u\between w,\varrho\rangle)= R$ with $u,v,w\in \mathbb{SP}(pqACP^+_{QE})$.

      By use of the probabilistic transition rules for probabilistic choice operator $\boxplus_{\pi}$ and entanglement merge $\between$, we get:

      $$\frac{\langle v,\varrho\rangle\rightsquigarrow \langle v',\varrho\rangle}{\langle u\between (v\boxplus_{\pi}w),\varrho\rangle\rightsquigarrow\langle u\between v',\varrho\rangle}$$

      $$\frac{\langle v,\varrho\rangle\rightsquigarrow \langle v',\varrho\rangle}{\langle u\between v\boxplus_{\pi}u\between w,\varrho\rangle\rightsquigarrow\langle u\between v',\varrho\rangle}$$

      With the assumption $u\between (v\boxplus_{\pi}w)=u\between v\boxplus_{\pi}u\between w$, we get $R$ satisfies condition 1 in the definition of $\underline{\leftrightarrow}_q$
      (see Definition \ref{SPQB}). Note that, for probabilistic choice operator $\boxplus_{\pi}$, there are not action transition rules.

      It is easy to check that the condition on PDF $\mu$ for $R$ also satisfies condition 4 in Definition \ref{SPQB}, we omit it.

      So, $R=\underline{\leftrightarrow}_q$, as desired.

  \item \textbf{Axiom $PrEM3$}. For a relation $R=R_1\cup R_2$, suppose that $Eq(\langle (p+q)\between s,\varrho\rangle,\langle p\between s + q\between s,\varrho\rangle)= R_1$ with $p,q,s\in \mathbb{DP}(pqACP^+_{QE})$, and $Eq(\langle (u+v)\between w,\varrho\rangle,\langle u\between w + v\between w,\varrho\rangle)= R_2$ with $w=w+w$ and $u,v,w\in \mathbb{SP}(pqACP^+_{QE})$.

      By use of the probabilistic transition rules and the action transition rules for entanglement merge $\between$, alternative operator $+$, we get:

      $$\frac{\langle u,\varrho\rangle\rightsquigarrow \langle u',\varrho\rangle, \langle v,\varrho\rangle\rightsquigarrow \langle v',\varrho\rangle,\langle w,\varrho\rangle\rightsquigarrow \langle w',\varrho\rangle}{\langle (u + v)\between w,\varrho\rangle\rightsquigarrow\langle (u' + v')\between w',\varrho\rangle}$$

      $$\frac{\langle u,\varrho\rangle\rightsquigarrow \langle u',\varrho\rangle, \langle v,\varrho\rangle\rightsquigarrow \langle v',\varrho\rangle,\langle w,\varrho\rangle\rightsquigarrow \langle w',\varrho\rangle}{\langle u\between w+v\between w,\varrho\rangle\rightsquigarrow\langle u'\between w'+v'\between w',\varrho\rangle}$$

      In the above two probabilistic transition rules, the condition $w=w+w$ ensures that $\langle w,\varrho\rangle\rightsquigarrow \langle w',\varrho\rangle$, $\langle w,\varrho\rangle\rightsquigarrow \langle w'',\varrho\rangle$, $\langle w',\varrho\rangle\underline{\leftrightarrow}_q \langle w'',\varrho\rangle$.

      With the assumptions $(u+v)\between w=u\between w+v\between w$ and $(u'+v')\between w'=u'\between w'+v'\between w'$, we get $R_2$ satisfies condition 1 in the definition of $\underline{\leftrightarrow}_q$
      (see Definition \ref{SPQB}).

      $$\frac{\langle p,\varrho\rangle\xrightarrow{\alpha} \langle p',\varrho'\rangle,\langle s,\varrho'\rangle\rightarrow \langle s',\varrho'\rangle}{\langle (p+ q)\between s,\varrho\rangle\xrightarrow{\alpha}\langle p'\parallel s',\varrho'\rangle}$$

      $$\frac{\langle p,\varrho\rangle\xrightarrow{\alpha} \langle p',\varrho'\rangle,\langle s,\varrho'\rangle\rightarrow \langle s',\varrho'\rangle}{\langle p\between s+q\between s,\varrho\rangle\xrightarrow{\alpha}\langle p'\parallel s',\varrho'\rangle}$$

      $$\frac{\langle s,\varrho\rangle\xrightarrow{\alpha} \langle s',\varrho'\rangle,\langle p,\varrho'\rangle\rightarrow \langle p',\varrho'\rangle}{\langle (p+ q)\between s,\varrho\rangle\xrightarrow{\alpha}\langle p'\parallel s',\varrho'\rangle}$$

      $$\frac{\langle s,\varrho\rangle\xrightarrow{\alpha} \langle s',\varrho'\rangle,\langle p,\varrho'\rangle\rightarrow \langle p',\varrho'\rangle}{\langle p\between s+q\between s,\varrho\rangle\xrightarrow{\alpha}\langle p'\parallel s',\varrho'\rangle}$$

      $$\frac{\langle p,\varrho\rangle\xrightarrow{\alpha} \langle p',\varrho'\rangle,\langle s,\varrho'\rangle\rightarrow \langle \surd,\varrho'\rangle}{\langle (p+ q)\between s,\varrho\rangle\xrightarrow{\alpha}\langle p',\varrho\rangle}$$

      $$\frac{\langle p,\varrho\rangle\xrightarrow{\alpha} \langle p',\varrho'\rangle,\langle s,\varrho'\rangle\rightarrow \langle \surd,\varrho'\rangle}{\langle p\between s+q\between s,\varrho\rangle\xrightarrow{\alpha}\langle p',\varrho\rangle}$$

      $$\frac{\langle s,\varrho\rangle\xrightarrow{\alpha} \langle s',\varrho'\rangle,\langle p,\varrho'\rangle\rightarrow \langle \surd,\varrho'\rangle}{\langle (p+ q)\between s,\varrho\rangle\xrightarrow{\alpha}\langle s',\varrho\rangle}$$

      $$\frac{\langle s,\varrho\rangle\xrightarrow{\alpha} \langle s',\varrho'\rangle,\langle p,\varrho'\rangle\rightarrow \langle \surd,\varrho'\rangle}{\langle p\between s+q\between s,\varrho\rangle\xrightarrow{\alpha}\langle s',\varrho\rangle}$$

      $$\frac{\langle p,\varrho\rangle\xrightarrow{\alpha} \langle \surd,\varrho'\rangle,\langle s,\varrho'\rangle\rightarrow \langle \surd,\varrho'\rangle}{\langle (p+ q)\between s,\varrho\rangle\xrightarrow{\alpha}\langle \surd,\varrho'\rangle}$$

      $$\frac{\langle p,\varrho\rangle\xrightarrow{\alpha} \langle \surd,\varrho'\rangle,\langle s,\varrho'\rangle\rightarrow \langle \surd,\varrho'\rangle}{\langle p\between s+q\between s,\varrho\rangle\xrightarrow{\alpha}\langle \surd,\varrho'\rangle}$$

      $$\frac{\langle s,\varrho\rangle\xrightarrow{\alpha} \langle \surd,\varrho'\rangle,\langle p,\varrho'\rangle\rightarrow \langle \surd,\varrho'\rangle}{\langle (p+ q)\between s,\varrho\rangle\xrightarrow{\alpha}\langle \surd,\varrho'\rangle}$$

      $$\frac{\langle s,\varrho\rangle\xrightarrow{\alpha} \langle \surd,\varrho'\rangle,\langle p,\varrho'\rangle\rightarrow \langle \surd,\varrho'\rangle}{\langle p\between s+q\between s,\varrho\rangle\xrightarrow{\alpha}\langle \surd,\varrho'\rangle}$$

      With the assumptions $(p+q)\between s=p\between s + q\between s$, we get $R_1$ satisfies conditions 2 and 3 in the definition of $\underline{\leftrightarrow}_q$
      (see Definition \ref{SPQB}).

      It is easy to check that the condition on PDF $\mu$ for $R$ also satisfies condition 4 in Definition \ref{SPQB}, we omit it.

      So, $R=\underline{\leftrightarrow}_q$, as desired.

  \item \textbf{Axiom $PrEM4$}. For a relation $R=R_1\cup R_2$, suppose that $Eq(\langle s\between(p+q),\varrho\rangle,\langle s\between p + s\between q,\varrho\rangle)= R_1$ with $p,q,s\in \mathbb{DP}(pqACP^+_{QE})$, and $Eq(\langle w\between(u+v),\varrho\rangle,\langle w\between u + w\between v,\varrho\rangle)= R_2$ with $w=w+w$ and $u,v,w\in \mathbb{SP}(pqACP^+_{QE})$.

      By use of the probabilistic transition rules and the action transition rules for entanglement merge $\between$, alternative operator $+$, we get:

      $$\frac{\langle u,\varrho\rangle\rightsquigarrow \langle u',\varrho\rangle, \langle v,\varrho\rangle\rightsquigarrow \langle v',\varrho\rangle,\langle w,\varrho\rangle\rightsquigarrow \langle w',\varrho\rangle}{\langle w\between(u + v),\varrho\rangle\rightsquigarrow\langle w'\between(u' + v'),\varrho\rangle}$$

      $$\frac{\langle u,\varrho\rangle\rightsquigarrow \langle u',\varrho\rangle, \langle v,\varrho\rangle\rightsquigarrow \langle v',\varrho\rangle,\langle w,\varrho\rangle\rightsquigarrow \langle w',\varrho\rangle}{\langle w\between u+w\between v,\varrho\rangle\rightsquigarrow\langle w'\between u'+w'\between v',\varrho\rangle}$$

      In the above two probabilistic transition rules, the condition $w=w+w$ ensures that $\langle w,\varrho\rangle\rightsquigarrow \langle w',\varrho\rangle$, $\langle w,\varrho\rangle\rightsquigarrow \langle w'',\varrho\rangle$, $\langle w',\varrho\rangle\underline{\leftrightarrow}_q \langle w'',\varrho\rangle$.

      With the assumptions $w\between(u+v)=w\between u+w\between v$ and $w'\between(u'+v')=w'\between u'+w'\between v'$, we get $R_2$ satisfies condition 1 in the definition of $\underline{\leftrightarrow}_q$
      (see Definition \ref{SPQB}).

      $$\frac{\langle p,\varrho\rangle\xrightarrow{\alpha} \langle p',\varrho'\rangle,\langle s,\varrho'\rangle\rightarrow \langle s',\varrho'\rangle}{\langle s\between(p+ q),\varrho\rangle\xrightarrow{\alpha}\langle s'\parallel p',\varrho'\rangle}$$

      $$\frac{\langle p,\varrho\rangle\xrightarrow{\alpha} \langle p',\varrho'\rangle,\langle s,\varrho'\rangle\rightarrow \langle s',\varrho'\rangle}{\langle s\between p+s\between q,\varrho\rangle\xrightarrow{\alpha}\langle s'\parallel p',\varrho'\rangle}$$

      $$\frac{\langle s,\varrho\rangle\xrightarrow{\alpha} \langle s',\varrho'\rangle,\langle p,\varrho'\rangle\rightarrow \langle p',\varrho'\rangle}{\langle s\between(p+ q),\varrho\rangle\xrightarrow{\alpha}\langle s'\parallel p',\varrho'\rangle}$$

      $$\frac{\langle s,\varrho\rangle\xrightarrow{\alpha} \langle s',\varrho'\rangle,\langle p,\varrho'\rangle\rightarrow \langle p',\varrho'\rangle}{\langle s\between p+s\between q,\varrho\rangle\xrightarrow{\alpha}\langle s'\parallel p',\varrho'\rangle}$$

      $$\frac{\langle p,\varrho\rangle\xrightarrow{\alpha} \langle p',\varrho'\rangle,\langle s,\varrho'\rangle\rightarrow \langle \surd,\varrho'\rangle}{\langle s\between(p+ q),\varrho\rangle\xrightarrow{\alpha}\langle p',\varrho'\rangle}$$

      $$\frac{\langle p,\varrho\rangle\xrightarrow{\alpha} \langle p',\varrho\rangle,\langle s,\varrho'\rangle\rightarrow \langle \surd,\varrho'\rangle}{\langle s\between p+s\between q,\varrho\rangle\xrightarrow{\alpha}\langle p',\varrho'\rangle}$$

      $$\frac{\langle s,\varrho\rangle\xrightarrow{\alpha} \langle s',\varrho'\rangle,\langle p,\varrho'\rangle\rightarrow \langle \surd,\varrho'\rangle}{\langle s\between(p+ q),\varrho\rangle\xrightarrow{\alpha}\langle s',\varrho'\rangle}$$

      $$\frac{\langle s,\varrho\rangle\xrightarrow{\alpha} \langle s',\varrho\rangle,\langle p,\varrho'\rangle\rightarrow \langle \surd,\varrho'\rangle}{\langle s\between p+s\between q,\varrho\rangle\xrightarrow{\alpha}\langle s',\varrho'\rangle}$$

      $$\frac{\langle p,\varrho\rangle\xrightarrow{\alpha} \langle \surd,\varrho'\rangle,\langle s,\varrho'\rangle\rightarrow \langle \surd,\varrho'\rangle}{\langle s\between(p+ q),\varrho\rangle\xrightarrow{\alpha}\langle \surd,\varrho'\rangle}$$

      $$\frac{\langle p,\varrho\rangle\xrightarrow{\alpha} \langle \surd,\varrho'\rangle,\langle s,\varrho'\rangle\rightarrow \langle \surd,\varrho'\rangle}{\langle s\between p+s\between q,\varrho\rangle\xrightarrow{\alpha}\langle \surd,\varrho'\rangle}$$

      $$\frac{\langle s,\varrho\rangle\xrightarrow{\alpha} \langle \surd,\varrho'\rangle,\langle p,\varrho'\rangle\rightarrow \langle \surd,\varrho'\rangle}{\langle s\between(p+ q),\varrho\rangle\xrightarrow{\alpha}\langle \surd,\varrho'\rangle}$$

      $$\frac{\langle s,\varrho\rangle\xrightarrow{\alpha} \langle \surd,\varrho'\rangle,\langle p,\varrho'\rangle\rightarrow \langle \surd,\varrho'\rangle}{\langle s\between p+s\between q,\varrho\rangle\xrightarrow{\alpha}\langle \surd,\varrho'\rangle}$$

      With the assumptions $s\between(p+q)=s\between p + s\between q$, we get $R_1$ satisfies conditions 2 and 3 in the definition of $\underline{\leftrightarrow}_q$
      (see Definition \ref{SPQB}).

      It is easy to check that the condition on PDF $\mu$ for $R$ also satisfies condition 4 in Definition \ref{SPQB}, we omit it.

      So, $R=\underline{\leftrightarrow}_q$, as desired.

\end{itemize}
\end{proof}

\proposition[Conservativity of $\mathcal{T}_{pqACP^+_{QE}}$ with respect to $\mathcal{T}_{pqACP^+}$]{The term-deduction system $\mathcal{T}_{pqACP^+_{QE}}$ is an operationally conservative extension of the term-deduction system $\mathcal{T}_{pqACP^+}$.}\label{CQE1}

\begin{proof}
It is sufficient to verify the following conditions:

\begin{itemize}
  \item $\mathcal{T}_{pqACP^+_{QE}}$ is a term-deduction system in path format;
  \item $\mathcal{T}_{pqACP^+}\oplus\mathcal{T}_{pqACP^+_{QE}}$ is defined;
  \item There are no conclusions $\langle s,\varrho\rangle\xrightarrow{\alpha} \langle t,\varrho'\rangle$ or $\langle s,\varrho\rangle\xrightarrow{\alpha} \langle \surd,\varrho'\rangle$ of a rule in $\mathcal{T}_{pqACP^+_{QE}}$ such $s=x$ or $s=f(x_1,\cdots,x_n)$ for some operator $f$ of $pqACP^+$.
\end{itemize}

These conditions hold which can be trivially checked.
\end{proof}

\proposition{The term-deduction system $\mathcal{T}_{pqACP^+_{QE}}$ is an operationally conservative extension up to the probabilistic quantum bisimulation of the term-deduction system $\mathcal{T}_{pqACP^+}$.}\label{CQE2}

\begin{proof}
In the definition of (strong) probabilistic quantum bisimulation (Definition \ref{SPQB}), apart from the 4th clause in Definition \ref{SPQB}, probabilistic quantum bisimulation is defined in terms of predicate and relation symbols. From Proposition \ref{CQE1}, for each closed $pqACP^+$ term $s$, its term-relation-predicate diagrams in both $\mathcal{T}_{pqACP^+}$ and $\mathcal{T}_{pqACP^+_{QE}}$ are the same. The concepts of PDF $\mu$ and quantum state $\varrho$ do not disturb the notion of the probabilistic quantum bisimulation defined only in terms of predicate and relation symbols.
\end{proof}

\proposition[Conservativity of $pqACP^+_{QE}$ with respect to $pqACP^+$]{$pqACP^+_{QE}$ is an equationally conservative extension of $pqACP^+$, that is, if $t$ and $s$ are closed $pqACP^+$ terms, then $pqACP^+\vdash t=s\Leftrightarrow pqACP^+_{QE}\vdash t=s$.}\label{CQE3}

\begin{proof}
It is proven by the following three facts:

\begin{enumerate}
  \item $\mathcal{T}_{pqACP^+_{QE}}$ is an operationally conservative extension of $\mathcal{T}_{pqACP^+}$ up to probabilistic quantum bisimulation (by Proposition \ref{CQE2});
  \item $pqACP^+$ is a complete axiomatization with respect to the bisimulation model (by Theorem \ref{CompletePQACP});
  \item $\mathcal{T}_{pqACP^+_{QE}}$ with respect to the probabilistic quantum bisimulation equivalence induces a model of $pqACP^+_{QE}$ (by Theorem \ref{SoundnessQE}).
\end{enumerate}
\end{proof}

\theorem[Completeness of $pqACP^+_{QE}$]{Let $z$ and $u$ are closed $pqACP^+_{QE}$ terms, if $z\underline{\leftrightarrow}_q u$, then $pACP^+_{QE}\vdash z=u$.}

\begin{proof}
It is proven based on the following two facts:

\begin{enumerate}
  \item $pqACP^+_{QE}$ has the elimination property for $pqBPA$ (by Theorem \ref{ElimiQE});
  \item $pqACP^+_{QE}$ is an operationally conservative extension of $pqACP^+$ (by Proposition \ref{CQE3}).
\end{enumerate}
\end{proof}

\section{Unifying Quantum and Classical Computing}\label{Unifying}

We use a quantum process configurations $\langle p,\varrho\rangle$ to represent information related to the execution of a (probabilistic) quantum process, in which $p$ represents the structural properties of a quantum process and $\varrho$ expresses the quantum properties of a quantum process. We have established a whole theory about quantum processes in closed quantum systems based on probabilistic process algebra, which is called $PQRA$.

In $PQRA$, the set of actions is consisted of atomic unitary operators and projections of quantum measurements, and also the deadlock $\delta$ and the silent step $\tau$. The execution of an atomic unitary operator $\alpha$ and a projection $\beta_i$ not only influences of the structural part $p$, but also changes the state of quantum variables $\varrho$. We still use the framework of a quantum process configuration $\langle p,\varrho\rangle$ under the situation of classical computing. In classical computing, the execution of a (classical) atomic action $a$ only influence the structural part $p$, and maintain the quantum state $\varrho$ unchanged. Note that, this kind of actions are alike to quantum communicating actions in section \ref{UO}, and range over the set $C$ of quantum communicating actions. In nature, quantum communicating actions are some kind of classical actions in contrast to quantum operations, because they are unrelated to the quantum state $\varrho$. The difference of a quantum communicating action and a classical communicating action is that they exchange different contents, a classical communicating action exchange the classical data by value or by reference, while a quantum communicating action exchange the quantum variables only by reference. We extend the set $C$ of quantum communicating actions to classical atomic actions (including classical communicating actions), let classical actions $a,b\in C$.

Base on the fact that a classical action $a$ does not affect the quantum state $\varrho$, we can generalize classical probabilistic process algebra $PRA$ under the framework of quantum process configuration $\langle p, \varrho\rangle$. We only take an example of $pBPA$, while $pBPA+PR$, $pACP^+$ and $fpBPA_{\tau}$ are omitted.

We give the probabilistic transition rules under quantum configuration for $pBPA$ as follows.

$$\frac{}{\langle a,\varrho\rangle\rightsquigarrow\langle\breve{a},\varrho\rangle} \quad\quad\frac{}{\langle\delta,\varrho\rangle\rightsquigarrow\langle\breve{\delta},\varrho\rangle}$$

$$\frac{\langle x,\varrho\rangle\rightsquigarrow \langle x',\varrho\rangle}{\langle x\cdot y,\varrho\rangle\rightsquigarrow\langle x'\cdot y,\varrho\rangle}
\quad\quad\frac{\langle x,\varrho\rangle\rightsquigarrow \langle x',\varrho\rangle, \langle y,\varrho\rangle\rightsquigarrow \langle y',\varrho\rangle}{\langle x + y,\varrho\rangle\rightsquigarrow\langle x' + y',\varrho\rangle}
\quad\quad\frac{\langle x,\varrho\rangle\rightsquigarrow \langle z,\varrho\rangle}{\langle x\boxplus_{\pi} y,\varrho\rangle\rightsquigarrow \langle z,\varrho\rangle, \langle y\boxplus_{\pi} x,\varrho\rangle\rightsquigarrow\langle z,\varrho\rangle}$$

The action transition rules for $pBPA$ based on quantum configuration are as follows.

$$\frac{}{\langle\breve{a},\varrho\rangle\xrightarrow{a}\langle\surd,\varrho\rangle}
\quad\quad\frac{\langle x,\varrho\rangle\xrightarrow{a} \langle x',\varrho\rangle}{\langle x\cdot y,\varrho\rangle\xrightarrow{a}\langle x'\cdot y,\varrho\rangle}
\quad\quad\frac{\langle x,\varrho\rangle\xrightarrow{a} \langle \surd,\varrho\rangle}{\langle x\cdot y,\varrho\rangle\xrightarrow{a}\langle y,\varrho\rangle}$$

$$\frac{\langle x,\varrho\rangle\xrightarrow{a} \langle x',\varrho\rangle}{\langle x+ y,\varrho\rangle\xrightarrow{a}\langle x',\varrho\rangle,\langle y+x,\varrho\rangle\xrightarrow{a}\langle x',\varrho\rangle}
\quad\quad\frac{\langle x,\varrho\rangle\xrightarrow{a} \langle \surd,\varrho\rangle}{\langle x+ y,\varrho\rangle\xrightarrow{a}\langle \surd,\varrho\rangle,\langle y+x,\varrho\rangle\xrightarrow{a}\langle \surd,\varrho\rangle}$$

We can get the following conclusions naturally.

\theorem[Soundness of $pBPA+PR$ under quantum configuration]{Let $x$ and $y$ be $pBPA+PR$ terms. If $pBPA+PR\vdash x=y$ then $\langle x,\varrho\rangle\underline{\leftrightarrow}_q \langle y,\varrho\rangle$.}

\theorem[$AIP^-$ in $\mathbb{PT}^{(\infty)}(pBPA+PR)$ under quantum configuration]{If for all $n\geq 1$, $\langle \Pi_n(p),\varrho\rangle\underline{\leftrightarrow}_q\langle\Pi_n(q),\varrho\rangle$, then $\langle p,\varrho\rangle\underline{\leftrightarrow}_q \langle q,\varrho\rangle$.}

\theorem[Soundness of $pBPA$ under quantum configuration]{Let $x$ and $y$ be closed $pBPA$ terms. If $pBPA\vdash x=y$ then $\langle x,\varrho\rangle\underline{\leftrightarrow}_q \langle y,\varrho\rangle$.}

\theorem[Completeness of $pBPA$ under quantum configuration]{Let $z$ and $u$ are closed $pBPA$ terms, if $\langle z,\varrho\rangle\underline{\leftrightarrow}_q \langle u,\varrho\rangle$, then $pBPA\vdash z=u$.}

\theorem[Soundness of $pACP^+$ under quantum configuration]{Let $p$ and $q$ be closed $pACP^+$ terms. If $pACP^+\vdash p=q$ then $\langle p,\varrho\rangle\underline{\leftrightarrow}_q \langle q,\varrho\rangle$.}

\theorem[Completeness of $pACP^+$ under quantum configuration]{Let $z$ and $u$ are closed $pACP^+$ terms, if $\langle z,\varrho\rangle\underline{\leftrightarrow}_q \langle u,\varrho\rangle$, then $pACP^+\vdash z=u$.}

\theorem[Soundness of $fpBPA_{\tau}+PVR_1+PVR_2+\cdots$ under quantum configuration]{G/$\underline{\leftrightarrow}_{pqrb}$ is a model of $fpBPA_{\tau}+PVR_1+PVR_2+\cdots$.}

So, $PQRA$ and $PRA$ are unified under the framework of quantum process configuration $\langle p, \varrho\rangle$, that is, $PQRA$ and classical $PRA$ have the same equational logics and the same semantic models based on quantum configuration.

The unification of $PQRA$ and classical $PRA$ has an important significance, because most quantum protocols, like the famous BB84 protocol \cite{BB84} and E91 protocol \cite{E91} (their verifications are shown in section \ref{App}), are mixtures of quantum information and classical information, and quantum computing and classical computing. This unification can be used widely in verification for all quantum protocols with an assumption of closed systems.

\section{Applications}\label{App}

Quantum and classical computing in closed systems are unified with probabilistic process algebra, which have the same equational logic and the same quantum configuration based operational semantics. The unification can be used widely in verification for the behaviors of quantum and classical computing mixed systems. In this section, we show its usage in verification for the quantum communication protocols: the famous quantum teleportation \cite{QT}, the first quantum communication protocol -- BB84 \cite{BB84}, and the first quantum communication protocol using quantum entanglement -- E91 \cite{E91}.

\subsection{Verification for Quantum Teleportation Protocol}\label{VQT}

\begin{figure}
  \centering
  \includegraphics{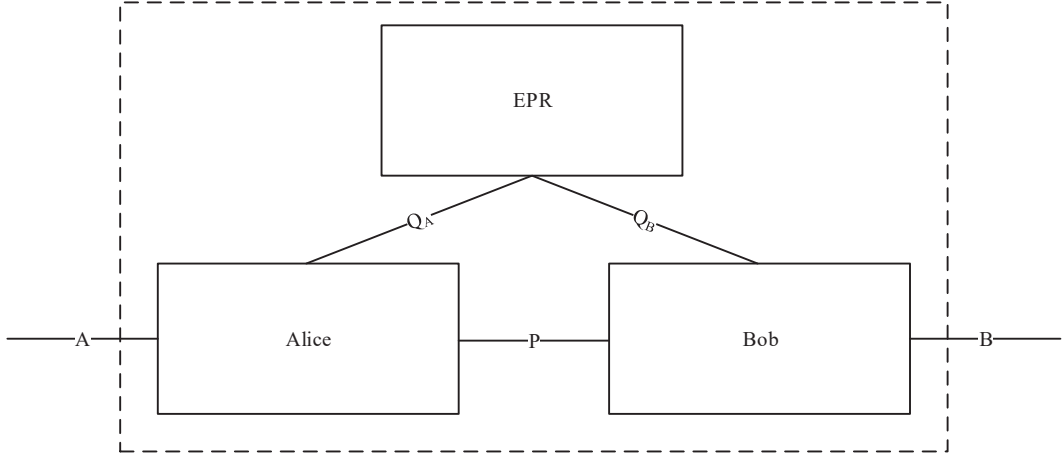}
  \caption{Quantum teleportation protocol.}
  \label{QT}
\end{figure}

Quantum teleportation \cite{QT} is a famous quantum protocol in quantum information theory to teleport an unknown quantum state by sending only classical information, provided that the sender and the receiver, Alice and Bob, shared an entangled state in advance. Firstly, we introduce the basic quantum teleportation protocol briefly, which is illustrated in Fig.\ref{QT}. In this section, we show how to process quantum entanglement in an implicit way, while in section \ref{VE91}, we show how to process quantum entanglement in an explicit way.

\begin{enumerate}
  \item EPR generates 2-qubits entangled EPR pair $q=q_1\otimes q_2$, and he sends $q_1$ to Alice through quantum channel $Q_A$ and $q_2$ to Bob through quantum channel $Q_B$;
  \item Alice receives $q_1$, after some preparations, she measures on $q_1$, and sends the measurement results $x$ to Bob through classical channel $P$;
  \item Bob receives $q_2$ from EPR, and also the classical information $x$ from Alice. According to $x$, he chooses specific Pauli transformation on $q_2$.
\end{enumerate}

We re-introduce the basic quantum teleportation protocol in an abstract way with more technical details as Fig.\ref{QT} illustrates.

Now, we assume the generation of 2-qubits $q$ through two unitary operators $Set[q]$ and $H[q]$. EPR sends $q_1$ to Alice through the quantum channel $Q_A$ by quantum communicating action $send_{Q_A}(q_1)$ and Alice receives $q_1$ through $Q_A$ by quantum communicating action $receive_{Q_A}(q_1)$. Similarly, for Bob, those are $send_{Q_B}(q_2)$ and $receive_{Q_B}(q_2)$. After Alice receives $q_1$, she does some preparations, including a unitary transformation $CNOT$ and a Hadamard transformation $H$, then Alice do measurement $M=\sum^3_{i=0}M_i$, and sends measurement results $x$ to Bob through the public classical channel $P$ by classical communicating action $send_{P}(x)$, and Bob receives $x$ through channel $P$ by classical communicating action $receive_{P}(x)$. According to $x$, Bob performs specific Pauli transformations $\sigma_x$ on $q_2$. Let Alice, Bob and EPR be a system $ABE$ and let interactions between Alice, Bob and EPR be internal actions. $ABE$ receives external input $D_i$ through channel $A$ by communicating action $receive_A(D_i)$ and sends results $D_o$ through channel $B$ by communicating action $send_B(D_o)$. Note that the entangled EPR pair $q=q_1\otimes q_2$ is within $ABE$, so quantum entanglement can be processed implicitly.

Then the state transitions of EPR can be described by PQRA as follows.

\begin{eqnarray}
&&E=Set[q]\cdot E_1\nonumber\\
&&E_1=H[q]\cdot E_2\nonumber\\
&&E_2=send_{Q_A}(q_1)\cdot E_3\nonumber\\
&&E_3=send_{Q_B}(q_2)\cdot E\nonumber
\end{eqnarray}

And the state transitions of Alice can be described by PQRA as follows.

\begin{eqnarray}
&&A=\sum_{D_i\in \Delta_i}receive_A(D_i)\cdot A_1\nonumber\\
&&A_1=receive_{Q_A}(q_1)\cdot A_2\nonumber\\
&&A_2=CNOT\cdot A_3\nonumber\\
&&A_3=H\cdot A_4\nonumber\\
&&A_4=(M_0\cdot send_P(0)\boxplus_{\frac{1}{4}}M_1\cdot send_P(1)\boxplus_{\frac{1}{4}}M_2\cdot send_P(2)\boxplus_{\frac{1}{4}}M_3\cdot send_P(3))\cdot A\nonumber
\end{eqnarray}

where $\Delta_i$ is the collection of the input data.

And the state transitions of Bob can be described by PQRA as follows.

\begin{eqnarray}
&&B=receive_{Q_B}(q_2)\cdot B_1\nonumber\\
&&B_1=(receive_P(0)\cdot\sigma_0\boxplus_{\frac{1}{4}}receive_P(1)\cdot\sigma_1\boxplus_{\frac{1}{4}}receive_P(2) \cdot\sigma_2\boxplus_{\frac{1}{4}}receive_P(3)\cdot\sigma_3)\cdot B_2\nonumber\\
&&B_2=\sum_{D_o\in\Delta_o}send_B(D_o)\cdot B\nonumber
\end{eqnarray}

where $\Delta_o$ is the collection of the output data.

The send action and receive action of the same data through the same channel can communicate each other, otherwise, a deadlock $\delta$ will be caused. We define the following communication functions.

\begin{eqnarray}
&&\gamma(send_{Q_A}(q_1),receive_{Q_A}(q_1))\triangleq c_{Q_A}(q_1)\nonumber\\
&&\gamma(send_{Q_B}(q_2),receive_{Q_B}(q_2))\triangleq c_{Q_B}(q_2)\nonumber\\
&&\gamma(send_P(0),receive_P(0))\triangleq c_P(0)\nonumber\\
&&\gamma(send_P(1),receive_P(1))\triangleq c_P(1)\nonumber\\
&&\gamma(send_P(2),receive_P(2))\triangleq c_P(2)\nonumber\\
&&\gamma(send_P(3),receive_P(3))\triangleq c_P(3)\nonumber
\end{eqnarray}

Let $A$, $B$ and $E$ in parallel, then the system $ABE$ can be represented by the following process term.

$$\tau_I(\partial_H(A\parallel B\parallel E))$$

where $H=\{send_{Q_A}(q_1), receive_{Q_A}(q_1), send_{Q_B}(q_2), receive_{Q_B}(q_2),\\
send_P(0), receive_P(0), send_P(1), receive_P(1),\\
send_P(2), receive_P(2), send_P(3), receive_P(3)\}$ and $I=\{Set[q], H[q], CNOT, H, M_0, M_1, M_2, M_3, \sigma_0, \sigma_1, \sigma_2, \sigma_3, \\ c_{Q_A}(q_1), c_{Q_B}(q_2), c_P(0), c_P(1), c_P(2), c_P(3)\}$.

Then we get the following conclusion.

\theorem{The basic quantum teleportation protocol $\tau_I(\partial_H(A\parallel B\parallel E))$ exhibits desired external behaviors.}

\begin{proof}
\begin{eqnarray}
&&\partial_H(A\parallel B \parallel E)=\sum_{D_i\in \Delta_i}receive_A(D_i)\cdot\partial_H(A_1\parallel B\parallel E)\nonumber\\
&&\partial_H(A_1\parallel B\parallel E)=Set[q]\cdot\partial_H(A_1\parallel B\parallel E_1)\nonumber\\
&&\partial_H(A_1\parallel B\parallel E_1)=H[q]\cdot\partial_H(A_1\parallel B\parallel E_2)\nonumber\\
&&\partial_H(A_1\parallel B\parallel E_2)=c_{Q_A}(q_1)\cdot \partial_H(A_2\parallel B\parallel E_3)\nonumber\\
&&\partial_H(A_2\parallel B\parallel E_3)=CNOT\cdot \partial_H(A_3\parallel B\parallel E_3)\nonumber\\
&&\partial_H(A_3\parallel B\parallel E_3)=H\cdot \partial_H(A_4\parallel B\parallel E_3)\nonumber\\
&&\partial_H(A_4\parallel B\parallel E_3)=c_{Q_B}(q_2)\cdot \partial_H(A_4\parallel B_1\parallel E)\nonumber\\
&&\partial_H(A_4\parallel B_1\parallel E)=(M_0\cdot c_P(0)\cdot\sigma_0\boxplus_{\frac{1}{4}}M_1\cdot c_P(1)\cdot\sigma_1\boxplus_{\frac{1}{4}}M_2\cdot c_P(2)\cdot\sigma_2\boxplus_{\frac{1}{4}}M_3\cdot c_P(3)\cdot\sigma_3)\cdot \partial_H(A\parallel B_2\parallel E)\nonumber\\
&&\partial_H(A\parallel B_2\parallel E)=\sum_{D_o\in\Delta_o}send_B(D_o)\cdot\partial_H(A\parallel B\parallel E)\nonumber
\end{eqnarray}

Let $\partial_H(A\parallel B\parallel E)=\langle X_1|E\rangle$, where E is the following guarded linear recursion specification:

\begin{eqnarray}
&&\{X_1=\sum_{D_i\in \Delta_i}receive_A(D_i)\cdot X_2,X_2=Set[q]\cdot X_3,X_3=H[q]\cdot X_4,\nonumber\\
&&X_4=c_{Q_A}(q_1)\cdot X_5,X_5=CNOT\cdot X_6,X_6=H\cdot X_7,X_7=c_{Q_B}(q_2)\cdot X_8,\nonumber\\
&&X_8=(M_0\cdot c_P(0)\cdot\sigma_0\boxplus_{\frac{1}{4}}M_1\cdot c_P(1)\cdot\sigma_1\boxplus_{\frac{1}{4}}M_2\cdot c_P(2)\cdot\sigma_2\boxplus_{\frac{1}{4}}M_3\cdot c_P(3)\cdot\sigma_3)\cdot X_9,\nonumber\\
&&X_9=\sum_{D_o\in\Delta_o}send_B(D_o)\cdot X_1\}\nonumber
\end{eqnarray}

Then we apply abstraction operator $\tau_I$ into $\langle X_1|E\rangle$.

\begin{eqnarray}
\tau_I(\langle X_1|E\rangle)
&=&\sum_{D_i\in \Delta_i}receive_A(D_i)\cdot\tau_I(\langle X_2|E\rangle)\nonumber\\
&=&\sum_{D_i\in \Delta_i}receive_A(D_i)\cdot\tau_I(\langle X_3|E\rangle)\nonumber\\
&=&\sum_{D_i\in \Delta_i}receive_A(D_i)\cdot\tau_I(\langle X_4|E\rangle)\nonumber\\
&=&\sum_{D_i\in \Delta_i}receive_A(D_i)\cdot\tau_I(\langle X_5|E\rangle)\nonumber\\
&=&\sum_{D_i\in \Delta_i}receive_A(D_i)\cdot\tau_I(\langle X_6|E\rangle)\nonumber\\
&=&\sum_{D_i\in \Delta_i}receive_A(D_i)\cdot\tau_I(\langle X_7|E\rangle)\nonumber\\
&=&\sum_{D_i\in \Delta_i}receive_A(D_i)\cdot\tau_I(\langle X_8|E\rangle)\nonumber\\
&=&\sum_{D_i\in \Delta_i}receive_A(D_i)\cdot\tau_I(\langle X_9|E\rangle)\nonumber\\
&=&\sum_{D_i\in \Delta_i}\sum_{D_o\in\Delta_o}receive_A(D_i)\cdot send_B(D_o)\cdot \tau_I(\langle X_1|E\rangle)\nonumber
\end{eqnarray}

We get $\tau_I(\langle X_1|E\rangle)=\sum_{D_i\in \Delta_i}\sum_{D_o\in\Delta_o}receive_A(D_i)\cdot send_B(D_o)\cdot \tau_I(\langle X_1|E\rangle)$, that is, $\tau_I(\partial_H(A\parallel B\parallel E))=\sum_{D_i\in \Delta_i}\sum_{D_o\in\Delta_o}receive_A(D_i)\cdot send_B(D_o)\cdot \tau_I(\partial_H(A\parallel B\parallel E))$. So, the basic quantum teleportation protocol $\tau_I(\partial_H(A\parallel B\parallel E))$ exhibits desired external behaviors.
\end{proof}

\subsection{Verification for BB84 Protocol}\label{VBB84}

\begin{figure}
  \centering
  \includegraphics{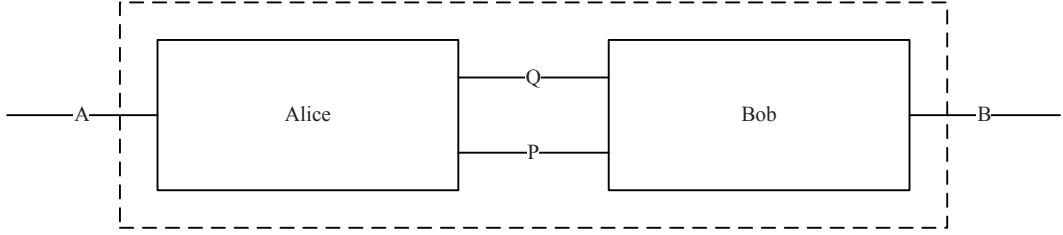}
  \caption{BB84 protocol.}
  \label{BB84}
\end{figure}

The BB84 protocol \cite{BB84} is used to create a private key between two parities, Alice and Bob. Firstly, we introduce the basic BB84 protocol briefly, which is illustrated in Fig.\ref{BB84}.

\begin{enumerate}
  \item Alice create two string of bits with size $n$ randomly, denoted as $B_a$ and $K_a$;
  \item Alice generates a string of qubits $q$ with size $n$, and the $i$th qubit in $q$ is $|x_y\rangle$, where $x$ is the $i$th bit of $B_a$ and $y$ is the $i$th bit of $K_a$;
  \item Alice sends $q$ to Bob through a quantum channel $Q$ between Alice and Bob;
  \item Bob receives $q$ and randomly generates a string of bits $B_b$ with size $n$;
  \item Bob measures each qubit of $q$ according to a basis by bits of $B_b$. And the measurement results would be $K_b$, which is also with size $n$;
  \item Bob sends his measurement bases $B_b$ to Alice through a public channel $P$;
  \item Once receiving $B_b$, Alice sends her bases $B_a$ to Bob through channel $P$, and Bob receives $B_a$;
  \item Alice and Bob determine that at which position the bit strings $B_a$ and $B_b$ are equal, and they discard the mismatched bits of $B_a$ and $B_b$. Then the remaining bits of $K_a$ and $K_b$, denoted as $K_a'$ and $K_b'$ with $K_{a,b}=K_a'=K_b'$.
\end{enumerate}

We re-introduce the basic BB84 protocol in an abstract way with more technical details as Fig.\ref{BB84} illustrates.

Now, we assume a special measurement operation $Rand[q;B_a]=\sum^{2n-1}_{i=0}Rand[q;B_a]_i$ which create a string of $n$ random bits $B_a$ from the $q$ quantum system, and the same as $Rand[q;K_a]=\sum^{2n-1}_{i=0}Rand[q;K_a]_i$, $Rand[q';B_b]=\sum^{2n-1}_{i=0}Rand[q';B_b]_i$. $M[q;K_b]=\sum^{2n-1}_{i=0}M[q;K_b]_i$ denotes the Bob's measurement on $q$. The generation of $n$ qubits $q$ through two unitary operators $Set_{K_a}[q]$ and $H_{B_a}[q]$. Alice sends $q$ to Bob through the quantum channel $Q$ by quantum communicating action $send_{Q}(q)$ and Bob receives $q$ through $Q$ by quantum communicating action $receive_{Q}(q)$. Bob sends $B_b$ to Alice through the public classical channel $P$ by classical communicating action $send_{P}(B_b)$ and Alice receives $B_b$ through channel $P$ by classical communicating action $receive_{P}(B_b)$, and the same as $send_{P}(B_a)$ and $receive_{P}(B_a)$. Alice and Bob generate the private key $K_{a,b}$ by a classical comparison action $cmp(K_{a,b},K_a,K_b,B_a,B_b)$. Let Alice and Bob be a system $AB$ and let interactions between Alice and Bob be internal actions. $AB$ receives external input $D_i$ through channel $A$ by communicating action $receive_A(D_i)$ and sends results $D_o$ through channel $B$ by communicating action $send_B(D_o)$.

Then the state transitions of Alice can be described by PQRA as follows.

\begin{eqnarray}
&&A=\sum_{D_i\in \Delta_i}receive_A(D_i)\cdot A_1\nonumber\\
&&A_1=\boxplus_{\frac{1}{2n},i=0}^{2n-1}Rand[q;B_a]_i\cdot A_2\nonumber\\
&&A_2=\boxplus_{\frac{1}{2n},i=0}^{2n-1}Rand[q;K_a]_i\cdot A_3\nonumber\\
&&A_3=Set_{K_a}[q]\cdot A_4\nonumber\\
&&A_4=H_{B_a}[q]\cdot A_5\nonumber\\
&&A_5=send_Q(q)\cdot A_6\nonumber\\
&&A_6=receive_P(B_b)\cdot A_7\nonumber\\
&&A_7=send_P(B_a)\cdot A_8\nonumber\\
&&A_8=cmp(K_{a,b},K_a,K_b,B_a,B_b)\cdot A\nonumber
\end{eqnarray}

where $\Delta_i$ is the collection of the input data.

And the state transitions of Bob can be described by PQRA as follows.

\begin{eqnarray}
&&B=receive_Q(q)\cdot B_1\nonumber\\
&&B_1=\boxplus_{\frac{1}{2n},i=0}^{2n-1}Rand[q';B_b]_i\cdot B_2\nonumber\\
&&B_2=\boxplus_{\frac{1}{2n},i=0}^{2n-1}M[q;K_b]_i\cdot B_3\nonumber\\
&&B_3=send_P(B_b)\cdot B_4\nonumber\\
&&B_4=receive_P(B_a)\cdot B_5\nonumber\\
&&B_5=cmp(K_{a,b},K_a,K_b,B_a,B_b)\cdot B_6\nonumber\\
&&B_6=\sum_{D_o\in\Delta_o}send_B(D_o)\cdot B\nonumber
\end{eqnarray}

where $\Delta_o$ is the collection of the output data.

The send action and receive action of the same data through the same channel can communicate each other, otherwise, a deadlock $\delta$ will be caused. We define the following communication functions.

\begin{eqnarray}
&&\gamma(send_Q(q),receive_Q(q))\triangleq c_Q(q)\nonumber\\
&&\gamma(send_P(B_b),receive_P(B_b))\triangleq c_P(B_b)\nonumber\\
&&\gamma(send_P(B_a),receive_P(B_a))\triangleq c_P(B_a)\nonumber
\end{eqnarray}

Let $A$ and $B$ in parallel, then the system $AB$ can be represented by the following process term.

$$\tau_I(\partial_H(A\parallel B))$$

where $H=\{send_Q(q),receive_Q(q),send_P(B_b),receive_P(B_b),send_P(B_a),receive_P(B_a)\}$ and $I=\{Rand[q;B_a]_i,\\ Rand[q;K_a]_i, Set_{K_a}[q], H_{B_a}[q], Rand[q';B_b]_i, M[q;K_b]_i, c_Q(q), c_P(B_b), c_P(B_a), cmp(K_{a,b},K_a,K_b,B_a,B_b)\}$.

Then we get the following conclusion.

\theorem{The basic BB84 protocol $\tau_I(\partial_H(A\parallel B))$ exhibits desired external behaviors.}

\begin{proof}
\begin{eqnarray}
&&\partial_H(A\parallel B)=\sum_{D_i\in \Delta_i}receive_A(D_i)\cdot\partial_H(A_1\parallel B)\nonumber\\
&&\partial_H(A_1\parallel B)=\boxplus_{\frac{1}{2n},i=0}^{2n-1}Rand[q;B_a]_i\cdot\partial_H(A_2\parallel B)\nonumber\\
&&\partial_H(A_2\parallel B)=\boxplus_{\frac{1}{2n},i=0}^{2n-1}Rand[q;K_a]_i\cdot\partial_H(A_3\parallel B)\nonumber\\
&&\partial_H(A_3\parallel B)=Set_{K_a}[q]\cdot\partial_H(A_4\parallel B)\nonumber\\
&&\partial_H(A_4\parallel B)=H_{B_a}[q]\cdot\partial_H(A_5\parallel B)\nonumber\\
&&\partial_H(A_5\parallel B)=c_Q(q)\cdot\partial_H(A_6\parallel B_1)\nonumber\\
&&\partial_H(A_6\parallel B_1)=\boxplus_{\frac{1}{2n},i=0}^{2n-1}Rand[q';B_b]_i\cdot\partial_H(A_6\parallel B_2)\nonumber\\
&&\partial_H(A_6\parallel B_2)=\boxplus_{\frac{1}{2n},i=0}^{2n-1}M[q;K_b]_i\cdot\partial_H(A_6\parallel B_3)\nonumber\\
&&\partial_H(A_6\parallel B_3)=c_P(B_b)\cdot\partial_H(A_7\parallel B_4)\nonumber\\
&&\partial_H(A_7\parallel B_4)=c_P(B_a)\cdot\partial_H(A_8\parallel B_5)\nonumber\\
&&\partial_H(A_8\parallel B_5)=cmp(K_{a,b},K_a,K_b,B_a,B_b)\cdot\partial_H(A\parallel B_5)\nonumber\\
&&\partial_H(A\parallel B_5)=cmp(K_{a,b},K_a,K_b,B_a,B_b)\cdot\partial_H(A\parallel B_6)\nonumber\\
&&\partial_H(A\parallel B_6)=\sum_{D_o\in\Delta_o}send_B(D_o)\cdot\partial_H(A\parallel B)\nonumber
\end{eqnarray}

Let $\partial_H(A\parallel B)=\langle X_1|E\rangle$, where E is the following guarded linear recursion specification:

\begin{eqnarray}
&&\{X_1=\sum_{D_i\in \Delta_i}receive_A(D_i)\cdot X_2,X_2=\boxplus_{\frac{1}{2n},i=0}^{2n-1}Rand[q;B_a]_i\cdot X_3,X_3=\boxplus_{\frac{1}{2n},i=0}^{2n-1}Rand[q;K_a]_i\cdot X_4,\nonumber\\
&&X_4=Set_{K_a}[q]\cdot X_5,X_5=H_{B_a}[q]\cdot X_6,X_6=c_Q(q)\cdot X_7,\nonumber\\
&&X_7=\boxplus_{\frac{1}{2n},i=0}^{2n-1}Rand[q';B_b]_i\cdot X_8,X_8=\boxplus_{\frac{1}{2n},i=0}^{2n-1}M[q;K_b]_i\cdot X_9,X_9=c_P(B_b)\cdot X_{10},X_{10}=c_P(B_a)\cdot X_{11},\nonumber\\
&&X_{11}=cmp(K_{a,b},K_a,K_b,B_a,B_b)\cdot X_{12},X_{12}=cmp(K_{a,b},K_a,K_b,B_a,B_b)\cdot X_{13},\nonumber\\
&&X_{13}=\sum_{D_o\in\Delta_o}send_B(D_o)\cdot X_1\}\nonumber
\end{eqnarray}

Then we apply abstraction operator $\tau_I$ into $\langle X_1|E\rangle$.

\begin{eqnarray}
\tau_I(\langle X_1|E\rangle)
&=&\sum_{D_i\in \Delta_i}receive_A(D_i)\cdot\tau_I(\langle X_2|E\rangle)\nonumber\\
&=&\sum_{D_i\in \Delta_i}receive_A(D_i)\cdot\tau_I(\langle X_3|E\rangle)\nonumber\\
&=&\sum_{D_i\in \Delta_i}receive_A(D_i)\cdot\tau_I(\langle X_4|E\rangle)\nonumber\\
&=&\sum_{D_i\in \Delta_i}receive_A(D_i)\cdot\tau_I(\langle X_5|E\rangle)\nonumber\\
&=&\sum_{D_i\in \Delta_i}receive_A(D_i)\cdot\tau_I(\langle X_6|E\rangle)\nonumber\\
&=&\sum_{D_i\in \Delta_i}receive_A(D_i)\cdot\tau_I(\langle X_7|E\rangle)\nonumber\\
&=&\sum_{D_i\in \Delta_i}receive_A(D_i)\cdot\tau_I(\langle X_8|E\rangle)\nonumber\\
&=&\sum_{D_i\in \Delta_i}receive_A(D_i)\cdot\tau_I(\langle X_9|E\rangle)\nonumber\\
&=&\sum_{D_i\in \Delta_i}receive_A(D_i)\cdot\tau_I(\langle X_{10}|E\rangle)\nonumber\\
&=&\sum_{D_i\in \Delta_i}receive_A(D_i)\cdot\tau_I(\langle X_{11}|E\rangle)\nonumber\\
&=&\sum_{D_i\in \Delta_i}receive_A(D_i)\cdot\tau_I(\langle X_{12}|E\rangle)\nonumber\\
&=&\sum_{D_i\in \Delta_i}receive_A(D_i)\cdot\tau_I(\langle X_{13}|E\rangle)\nonumber\\
&=&\sum_{D_i\in \Delta_i}\sum_{D_o\in\Delta_o}receive_A(D_i)\cdot send_B(D_o)\cdot \tau_I(\langle X_1|E\rangle)\nonumber
\end{eqnarray}

We get $\tau_I(\langle X_1|E\rangle)=\sum_{D_i\in \Delta_i}\sum_{D_o\in\Delta_o}receive_A(D_i)\cdot send_B(D_o)\cdot \tau_I(\langle X_1|E\rangle)$, that is, $\tau_I(\partial_H(A\parallel B))=\sum_{D_i\in \Delta_i}\sum_{D_o\in\Delta_o}receive_A(D_i)\cdot send_B(D_o)\cdot \tau_I(\partial_H(A\parallel B))$. So, the basic BB84 protocol $\tau_I(\partial_H(A\parallel B))$ exhibits desired external behaviors.
\end{proof}

\subsection{Verification for E91 Protocol}\label{VE91}

\begin{figure}
  \centering
  \includegraphics{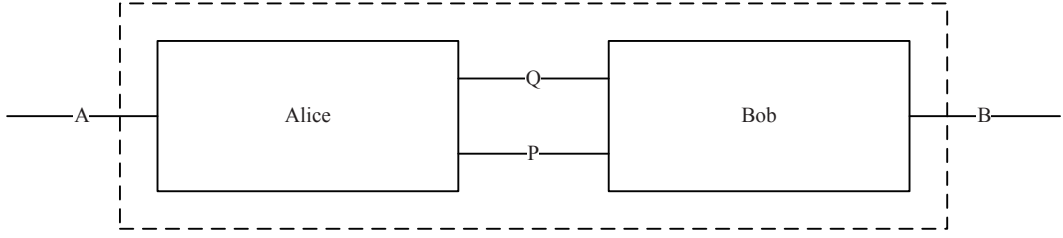}
  \caption{E91 protocol.}
  \label{E91}
\end{figure}

With support of Entanglement merge $\between$, PQRA can be used to verify quantum protocols utilizing entanglement explicitly. E91 protocol\cite{E91} is the first quantum protocol which utilizes entanglement. E91 protocol is used to create a private key between two parities, Alice and Bob. Firstly, we introduce the basic E91 protocol briefly, which is illustrated in Fig.\ref{E91}.

\begin{enumerate}
  \item Alice generates a string of EPR pairs $q$ with size $n$, i.e., $2n$ particles, and sends a string of qubits $q_b$ from each EPR pair with $n$ to Bob through a quantum channel $Q$, remains the other string of qubits $q_a$ from each pair with size $n$;
  \item Alice create two string of bits with size $n$ randomly, denoted as $B_a$ and $K_a$;
  \item Bob receives $q_b$ and randomly generates a string of bits $B_b$ with size $n$;
  \item Alice measures each qubit of $q_a$ according to a basis by bits of $B_a$. And the measurement results would be $K_a$, which is also with size $n$;
  \item Bob measures each qubit of $q_b$ according to a basis by bits of $B_b$. And the measurement results would be $K_b$, which is also with size $n$;
  \item Bob sends his measurement bases $B_b$ to Alice through a public channel $P$;
  \item Once receiving $B_b$, Alice sends her bases $B_a$ to Bob through channel $P$, and Bob receives $B_a$;
  \item Alice and Bob determine that at which position the bit strings $B_a$ and $B_b$ are equal, and they discard the mismatched bits of $B_a$ and $B_b$. Then the remaining bits of $K_a$ and $K_b$, denoted as $K_a'$ and $K_b'$ with $K_{a,b}=K_a'=K_b'$.
\end{enumerate}

We re-introduce the basic E91 protocol in an abstract way with more technical details as Fig.\ref{E91} illustrates.

Now, $M[q_a;K_a]=\sum_{i=0}^{2n-1}M[q_a;K_a]_i$ denotes the Alice's measurement operation of $q_a$, and $\circledS_{M[q_a;K_a]}=\sum_{i=0}^{2n-1}\circledS_{M[q_a;K_a]_i}$ denotes the responding shadow constant; $M[q_b;K_b]=\sum_{i=0}^{2n-1}M[q_b;K_b]_i$ denotes the Bob's measurement operation of $q_b$, and $\circledS_{M[q_b;K_b]}=\sum_{i=0}^{2n-1}\circledS_{M[q_b;K_b]_i}$ denotes the responding shadow constant. Alice sends $q_b$ to Bob through the quantum channel $Q$ by quantum communicating action $send_{Q}(q_b)$ and Bob receives $q_b$ through $Q$ by quantum communicating action $receive_{Q}(q_b)$. Bob sends $B_b$ to Alice through the public channel $P$ by classical communicating action $send_{P}(B_b)$ and Alice receives $B_b$ through channel $P$ by classical communicating action $receive_{P}(B_b)$, and the same as $send_{P}(B_a)$ and $receive_{P}(B_a)$. Alice and Bob generate the private key $K_{a,b}$ by a classical comparison action $cmp(K_{a,b},K_a,K_b,B_a,B_b)$. Let Alice and Bob be a system $AB$ and let interactions between Alice and Bob be internal actions. $AB$ receives external input $D_i$ through channel $A$ by communicating action $receive_A(D_i)$ and sends results $D_o$ through channel $B$ by communicating action $send_B(D_o)$.

Then the state transitions of Alice can be described by PQRA as follows.

\begin{eqnarray}
&&A=\sum_{D_i\in \Delta_i}receive_A(D_i)\cdot A_1\nonumber\\
&&A_1=send_Q(q_b)\cdot A_2\nonumber\\
&&A_2=\boxplus_{\frac{1}{2n},i=0}^{2n-1}M[q_a;K_a]_i\cdot A_3\nonumber\\
&&A_3=\boxplus_{\frac{1}{2n},i=0}^{2n-1}\circledS_{M[q_b;K_b]_i}\cdot A_4\nonumber\\
&&A_4=receive_P(B_b)\cdot A_5\nonumber\\
&&A_5=send_P(B_a)\cdot A_6\nonumber\\
&&A_6=cmp(K_{a,b},K_a,K_b,B_a,B_b)\cdot A\nonumber
\end{eqnarray}

where $\Delta_i$ is the collection of the input data.

And the state transitions of Bob can be described by PQRA as follows.

\begin{eqnarray}
&&B=receive_Q(q_b)\cdot B_1\nonumber\\
&&B_1=\boxplus_{\frac{1}{2n},i=0}^{2n-1}\circledS_{M[q_a;K_a]_i}\cdot B_2\nonumber\\
&&B_2=\boxplus_{\frac{1}{2n},i=0}^{2n-1}M[q_b;K_b]_i\cdot B_3\nonumber\\
&&B_3=send_P(B_b)\cdot B_4\nonumber\\
&&B_4=receive_P(B_a)\cdot B_5\nonumber\\
&&B_5=cmp(K_{a,b},K_a,K_b,B_a,B_b)\cdot B_6\nonumber\\
&&B_6=\sum_{D_o\in\Delta_o}send_B(D_o)\cdot B\nonumber
\end{eqnarray}

where $\Delta_o$ is the collection of the output data.

The send action and receive action of the same data through the same channel can communicate each other, otherwise, a deadlock $\delta$ will be caused. The quantum operation and its shadow constant pair will lead entanglement occur, otherwise, a deadlock $\delta$ will occur. We define the following communication functions.

\begin{eqnarray}
&&\gamma(send_Q(q_b),receive_Q(q_b))\triangleq c_Q(q_b)\nonumber\\
&&\gamma(send_P(B_b),receive_P(B_b))\triangleq c_P(B_b)\nonumber\\
&&\gamma(send_P(B_a),receive_P(B_a))\triangleq c_P(B_a)\nonumber
\end{eqnarray}

Let $A$ and $B$ in parallel, then the system $AB$ can be represented by the following process term.

$$\tau_I(\partial_H(A\parallel B))$$

where $H=\{send_Q(q_b),receive_Q(q_b),send_P(B_b),receive_P(B_b),send_P(B_a),receive_P(B_a),\\ M[q_a;K_a]_i, \circledS_{M[q_a;K_a]_i}, M[q_b;K_b]_i, \circledS_{M[q_b;K_b]_i}\}$ and $I=\{c_Q(q_b), c_P(B_b), c_P(B_a), M[q_a;K_a], M[q_b;K_b],\\ cmp(K_{a,b},K_a,K_b,B_a,B_b)\}$.

Then we get the following conclusion.

\theorem{The basic E91 protocol $\tau_I(\partial_H(A\parallel B))$ exhibits desired external behaviors.}

\begin{proof}
\begin{eqnarray}
&&\partial_H(A\parallel B)=\sum_{D_i\in \Delta_i}receive_A(D_i)\cdot\partial_H(A_1\parallel B)\nonumber\\
&&\partial_H(A_1\parallel B)=c_Q(q_b)\cdot\partial_H(A_2\parallel B_1)\nonumber\\
&&\partial_H(A_2\parallel B_1)=\boxplus_{\frac{1}{2n},i=0}^{2n-1}M[q_a;K_a]_i\cdot\partial_H(A_3\parallel B_2)\nonumber\\
&&\partial_H(A_3\parallel B_2)=\boxplus_{\frac{1}{2n},i=0}^{2n-1}M[q_b;K_b]_i\cdot\partial_H(A_4\parallel B_3)\nonumber\\
&&\partial_H(A_4\parallel B_3)=c_P(B_b)\cdot\partial_H(A_5\parallel B_4)\nonumber\\
&&\partial_H(A_5\parallel B_4)=c_P(B_a)\cdot\partial_H(A_6\parallel B_5)\nonumber\\
&&\partial_H(A_6\parallel B_5)=cmp(K_{a,b},K_a,K_b,B_a,B_b)\cdot\partial_H(A\parallel B_5)\nonumber\\
&&\partial_H(A\parallel B_5)=cmp(K_{a,b},K_a,K_b,B_a,B_b)\cdot\partial_H(A\parallel B_6)\nonumber\\
&&\partial_H(A\parallel B_6)=\sum_{D_o\in\Delta_o}send_B(D_o)\cdot\partial_H(A\parallel B)\nonumber
\end{eqnarray}

Let $\partial_H(A\parallel B)=\langle X_1|E\rangle$, where E is the following guarded linear recursion specification:

\begin{eqnarray}
&&\{X_1=\sum_{D_i\in \Delta_i}receive_A(D_i)\cdot X_2,X_2=c_Q(q_b)\cdot X_3,\nonumber\\
&&X_3=\boxplus_{\frac{1}{2n},i=0}^{2n-1}M[q_a;K_a]_i\cdot X_4,X_4=\boxplus_{\frac{1}{2n},i=0}^{2n-1}M[q_b;K_b]_i\cdot X_5,X_5=c_P(B_b)\cdot X_{6},X_{6}=c_P(B_a)\cdot X_{7},\nonumber\\
&&X_{7}=cmp(K_{a,b},K_a,K_b,B_a,B_b)\cdot X_{8},X_{8}=cmp(K_{a,b},K_a,K_b,B_a,B_b)\cdot X_{9},\nonumber\\
&&X_{9}=\sum_{D_o\in\Delta_o}send_B(D_o)\cdot X_1\}\nonumber
\end{eqnarray}

Then we apply abstraction operator $\tau_I$ into $\langle X_1|E\rangle$.

\begin{eqnarray}
\tau_I(\langle X_1|E\rangle)
&=&\sum_{D_i\in \Delta_i}receive_A(D_i)\cdot\tau_I(\langle X_2|E\rangle)\nonumber\\
&=&\sum_{D_i\in \Delta_i}receive_A(D_i)\cdot\tau_I(\langle X_3|E\rangle)\nonumber\\
&=&\sum_{D_i\in \Delta_i}receive_A(D_i)\cdot\tau_I(\langle X_4|E\rangle)\nonumber\\
&=&\sum_{D_i\in \Delta_i}receive_A(D_i)\cdot\tau_I(\langle X_5|E\rangle)\nonumber\\
&=&\sum_{D_i\in \Delta_i}receive_A(D_i)\cdot\tau_I(\langle X_6|E\rangle)\nonumber\\
&=&\sum_{D_i\in \Delta_i}receive_A(D_i)\cdot\tau_I(\langle X_7|E\rangle)\nonumber\\
&=&\sum_{D_i\in \Delta_i}receive_A(D_i)\cdot\tau_I(\langle X_8|E\rangle)\nonumber\\
&=&\sum_{D_i\in \Delta_i}receive_A(D_i)\cdot\tau_I(\langle X_9|E\rangle)\nonumber\\
&=&\sum_{D_i\in \Delta_i}\sum_{D_o\in\Delta_o}receive_A(D_i)\cdot send_B(D_o)\cdot \tau_I(\langle X_1|E\rangle)\nonumber
\end{eqnarray}

We get $\tau_I(\langle X_1|E\rangle)=\sum_{D_i\in \Delta_i}\sum_{D_o\in\Delta_o}receive_A(D_i)\cdot send_B(D_o)\cdot \tau_I(\langle X_1|E\rangle)$, that is, $\tau_I(\partial_H(A\parallel B))=\sum_{D_i\in \Delta_i}\sum_{D_o\in\Delta_o}receive_A(D_i)\cdot send_B(D_o)\cdot \tau_I(\partial_H(A\parallel B))$. So, the basic E91 protocol $\tau_I(\partial_H(A\parallel B))$ exhibits desired external behaviors.
\end{proof}

\section{Extensions}\label{Extensions}

One of the most fascinating characteristics is the modularity of ACP-like process algebra. Based on the concept of conservative extension, by introducing new operators or new constants, $PRA$ and $PQRA$ can have more properties. We have already seen the mechanism of extension in section \ref{QE}.

In this section, we introducing two extensions: renaming and priorities. In these extensions, an atomic action can either be an atomic unitary operator $\alpha$, a projection $\beta_i$, or be a classical atomic action $a$, since they are unified under the semantic model of quantum configuration. For simplicity, we only consider atomic unitary operators.

\subsection{Renaming}

We take an example of renaming operator $\rho_f(\alpha)$ which is used to rename the atomic actions $\alpha$ to $f(\alpha)$. $pACP^+_{RN}$ is obtained by extending $pACP^+$ with renaming operator $\rho_f$, we obtain $pqACP^+_{RN}$ make $pACP^+_{RN}$ based on quantum configuration. The axiom system for priorities is shown in Table \ref{AxiomsForR}.

\begin{center}
\begin{table}
  \begin{tabular}{@{}ll@{}}
\hline No. &Axiom\\
  $RN1$ & $\rho_f(a)=f(a)$\\
  $RN2$ & $\rho_f(\delta)=\delta$\\
  $RN3$ & $\rho_f(x+y)=\rho_f(x)+\rho_f(y)$\\
  $RN4$ & $\rho_f(x\cdot y)=\rho_f(x)\cdot\rho_f(y)$\\
  $PrRN1$ & $\rho_f(x\boxplus_{\pi} y)=\rho_f(x)\boxplus_{\pi}\rho_f(y)$\\
\end{tabular}
\caption{Axioms for renaming}
\label{AxiomsForR}
\end{table}
\end{center}

The (probabilistic and action) transition rules for renaming operator are as follows.

$$\frac{\langle x,\varrho\rangle\rightsquigarrow \langle x',\varrho\rangle}{\langle\rho_f(x),\varrho\rangle\rightsquigarrow\langle\rho_f(x'),\varrho\rangle}$$

$$\frac{\langle x,\varrho\rangle\xrightarrow{\alpha}\langle x',\varrho'\rangle}{\langle \rho_f(x),\varrho\rangle\xrightarrow{f(\alpha)}\langle\rho_f(x'),\varrho'\rangle}
\quad\quad\frac{\langle x,\varrho\rangle\xrightarrow{\alpha}\langle\surd,\varrho'\rangle}{\langle \rho_f(x),\varrho\rangle\xrightarrow{f(\alpha)}\langle\surd,\varrho'\rangle}$$

\definition[PDF $\mu$ for renaming operator]{We define PDF $\mu$ for operator $\rho_f$ as follows, and the definitions of $\mu$ for other operators are same as those in $pqACP^+$.

$$\mu(\rho_f(x),\rho_f(x'))=\mu(x,x')$$
}

\proposition{$\mu$ and $\mu^*$ are well-defined on $\mathbb{PT}(pqACP^+_{RN})$.}

\begin{proof}
It is easy to check $\mu$ is well-defined on $\mathbb{PT}(pqACP^+_{RN})$, it follows the two cases:

\begin{enumerate}
  \item Case $t\in\mathbb{SP}(pqACP^+_{RN})$ processes. For $t\equiv\rho_f(s)$, $\mu(\rho_f(s),u)=\mu(s,v)$, if $u\equiv\rho_f(v)$; otherwise, $\mu(\rho_f(s),u)=0$. $\mu(s,v)$ is defined by the inductive hypothesis, so $\mu(\rho_f(s),u)$ is defined as well.
  \item Case $t\in\mathbb{DP}(pqACP^+_{RN})$ processes. For $t\equiv\rho_f(s)$, $\mu(\rho_f(s),u)=\mu(s,v)$, if $u\equiv\rho_f(v)$; otherwise, $\mu(\rho_f(s),u)=0$. $\mu(s,v)$ is defined by the inductive hypothesis $\mu(s,v)=0$, so $\mu(\rho_f(s),u)=0$ is defined as well.
\end{enumerate}

It is easy to check cPDF $\mu^*$ is also well-defined on $\mathbb{PT}(pqACP^+_{RN})$, we omit it.
\end{proof}

\theorem[Elimination theorem of the renaming operator]{Let $p$ be a closed $pqACP^+_{RN}$ term. Then there is a closed $pqBPA$ term such that $pqACP^+_{RN}\vdash p=q$.}\label{ElimiRN}

\begin{proof}
The equational logic of $pqACP^+_{RN}$ ($pqBPA$ and $pqACP^+$) is same as that of $pACP^+_{RN}$ ($pBPA$ and $pACP^+$), and the same elimination properties. We only need to treat the new case of renaming operator $\rho_f$.

Assume that $p\equiv \rho_f(p_1)$ for certain closed $pqACP^+_{RN}$ term $p_1$. By the induction hypothesis there is a closed $pqBPA$ term $q_1$ such that $pqACP^+_{RN}\vdash p_1=q_1$. Then by the elimination theorem of $pqBPA$ (see in \cite{PPA}), there is a basic term $r_1$ such that $pqBPA\vdash q_1=r_1$. Then, $pqACP^+_{RN}\vdash p_1=r_1$. By induction on the structure of basic term $r_1$, we prove that there is a basic (closed) term $r$ such that $pqACP^+_{RN}\vdash\rho_f(r_1)=r$, and if $r_1\in\mathcal{B}_+$ then $r\in\mathcal{B}_+$.

\begin{enumerate}
  \item Case $r_1\equiv \alpha,\alpha\in \{A\cup B\}_\delta$. Then $pqACP^+_{RN}\vdash\rho_f(r_1)=\rho_f(\alpha)=f(\alpha)$ and $f(\alpha)$ is a basic $pqBPA$ term and $f(\alpha)\in\mathcal{B}_+$;
  \item Case $r_1\equiv \delta$. Then $pqACP^+_{RN}\vdash\rho_f(r_1)=\rho_f(\delta)=\delta$ and $\delta$ is a basic $pqBPA$ term and $\delta\in\mathcal{B}_+$;
  \item Case $r_1=\alpha\cdot r_1',\alpha\in \{A\cup B\}_\delta$ and basic term $r_1'$. Then $pqACP^+_{RN}\vdash\rho_f(r_1)=\rho_f(\alpha\cdot r_1')=\rho_f(\alpha)\cdot\rho_f(r_1')=f(\alpha)\cdot\rho_f(r_1')$. By induction there is a basic term $s'$ such that $pqACP^+_{RN}\vdash\rho_f(r_1')=s'$. So, $f(\alpha)\cdot s'$ is a basic term and $f(\alpha)\cdot s'\in\mathcal{B}_+$;
  \item Case $r_1=r_1'+r_1''$ for basic $\mathcal{B}_+$ terms $r_1'$ and $r_1''$. Then $pqACP^+_{RN}\vdash\rho_f(r_1)=\rho_f(r_1'+r_1'')=\rho_f(r_1')+\rho_f(r_1'')$. By induction there are a basic term $s'$ and a basic term $s''$ such that $pqACP^+_{RN}\vdash\rho_f(r_1')=s'$ and $pqACP^+_{RN}\vdash\rho_f(r_1'')=s''$. So, $s'+s''$ is a basic term and $s'+s''\in\mathcal{B}_+$;
  \item Case $r_1=r_1'\boxplus_{\pi}r_1''$ for basic $\mathcal{B}_+$ terms $r_1'$ and $r_1''$. Then $pqACP^+_{RN}\vdash\rho_f(r_1)=\rho_f(r_1'\boxplus_{\pi}r_1'')=\rho_f(r_1')\boxplus_{\pi} \rho_f(r_1'')$. By induction there are a basic term $s'$ and a basic term $s''$ such that $pqACP^+_{RN}\vdash\rho_f(r_1')=s'$ and $pqACP^+_{RN}\vdash\rho_f(r_1'')=s''$. So, $s'\boxplus_{\pi}s''$ is a basic term and $s'\boxplus_{\pi}s''\in\mathcal{B}_+$.
\end{enumerate}
\end{proof}

\theorem[Congruence theorem of $pqACP^+_{RN}$]{$\underline{\leftrightarrow}_q$ is a congruence relation on $\mathbb{PT}(pqACP^+_{RN})$ with respect to the $+$, $\cdot$, $\boxplus_{\pi}$, $\leftmerge$, $\mid$, $\parallel$, $][$, $\partial_H$ and $\rho_f$ operators.}

\begin{proof}
The part of the proof for the operators of $pqACP^+$ is the same as the proof of the congruence of $pqACP^+$. Here we only give the rest of the proof which concerns the operator $\rho_f$.

Let $x$ and $y$ be $\mathbb{PT}(pqACP^+_{RN})$ processes such that $\langle x,\varrho\rangle\underline{\leftrightarrow}_q \langle y,\varrho\rangle$. So, there exists probabilistic bisimulations $R_1$ such that $(\langle x,\varrho\rangle,\langle y,\varrho\rangle)\in R_1$. We define a relation $R$ in the following way:

$$R=Eq(S\cup D\cup R_1)$$

where $S=\{(\langle \rho_f(p),\varrho\rangle,\langle \rho_f(q),\varrho\rangle):p,q\in\mathbb{SP}(pqACP^+_{RN}),(\langle p,\varrho\rangle,\langle q,\varrho\rangle)\in R_1\}$,

$D=\{(\langle \rho_f(u),\varrho\rangle,\langle \rho_f(v),\varrho\rangle):u,v\in\mathbb{DP}(pqACP^+_{RN}),(\langle u,\varrho\rangle,\langle v,\varrho\rangle)\in R_1\}$.

\begin{itemize}
  \item \textbf{RN1}: $S$ and $D$ are equivalence relation, and $S$, $R_1$ contain pairs of static processes relevant to $R$;
  \item \textbf{RN2}: if $(\langle \rho_f(p),\varrho\rangle,\langle \rho_f(q),\varrho\rangle)\in S$ and $K\in \mathbb{DP}(pqACP^+_{RN})/D$, then $\langle \rho_f(p),\varrho\rangle\rightsquigarrow\langle K,\varrho\rangle$ iff $\langle \rho_f(q),\varrho\rangle\rightsquigarrow\langle K,\varrho\rangle$;
  \item \textbf{RN3}: if $\langle \rho_f(p),\varrho\rangle\rightsquigarrow\langle K,\varrho\rangle$ for $K\in \mathbb{DP}(pqACP^+_{RN})/D$, then $K=[\rho_f(u)]_{D}$ for some $u$ such that $\langle p,\varrho\rangle\rightsquigarrow \langle u,\varrho\rangle$. From the definition of $D$, $K=\rho_f([u]_{R_1})$;
  \item \textbf{RN4}: since $R_1$, $D$ are all subsets of $R$ and they are equivalence relations themselves, if $M\in\mathbb{DP}(pqACP^+_{RN})/R$, then $M=\bigcup_{i\in I}M_i$, $M=\bigcup_{j\in J}K_j$ for some non-empty index set $I,J$ and some equivalence classes $M_i(i\in I)$, $K_j(j\in J)$ of $R_1$, and $D$ respectively.
\end{itemize}

Now suppose that $(\langle r,\varrho\rangle,\langle r_1,\varrho\rangle)\in R$ for some $r,r_1\in\mathbb{SP}(pqACP^+_{RN})$ and $M\in\mathbb{DP}(pqACP^+_{RN})/R$, then:
\begin{enumerate}
  \item if $(\langle r,\varrho\rangle,\langle r_1,\varrho\rangle)\in R_1$, then the result follows from \textbf{RN4};
  \item if $(\langle r,\varrho\rangle,\langle r_1,\varrho\rangle)\in S$, then $r\equiv \rho_f(p)$ and $r_1\equiv \rho_f(q)$ for some $p,q\in \mathbb{SP}(pqACP^+_{RN})$ such that $(\langle p,\varrho\rangle,\langle s,\varrho\rangle)\in R_1$. According to \textbf{RN3} and \textbf{RN4}, $K_j=\rho_f([u_j]_{R_1})$ and $\langle p,\varrho\rangle\rightsquigarrow\langle u_j,\varrho\rangle$. Then $\mu(\rho_f(p), K_j)=\mu(\rho_f(q), K_j)$, $\mu(\rho_f(p), M)=\mu(\rho_f(q), M)$.
\end{enumerate}
\end{proof}

\theorem[Soundness of $pqACP^+_{RN}$]{Let $p$ and $q$ be closed $pqACP^+_{RN}$ terms. If $pqACP^+_{RN}\vdash p=q$ then $\langle p,\varrho\rangle\underline{\leftrightarrow}_q \langle q,\varrho\rangle$.}\label{SoundnessRN}

\begin{proof}
It is already proven that $\underline{\leftrightarrow}_q$ is an equivalent and congruent relation on $\mathbb{PT}(pqACP^+_{RN})$, it is sufficient to examine every axiom in the axiom system (in Table \ref{AxiomsForR}) for $pqACP^+_{RN}$ is sound.

\begin{itemize}
  \item \textbf{Axiom $RN1$}. For a relation $R=R_1\cup R_2$, suppose that $Eq(\langle \rho_f(\alpha),\varrho\rangle,\langle f(\alpha),\varrho\rangle)= R_1$, and $Eq(\langle \rho_f(\breve{\alpha}),\varrho\rangle,\langle f(\breve{\alpha}),\varrho\rangle)= R_2$.

      By use of the probabilistic transition rules and the action transition rules for atomic unitary operator and renaming operator $\rho_f$, we get:

      $$\frac{\langle \alpha,\varrho\rangle\rightsquigarrow \langle \breve{\alpha},\varrho\rangle}{\langle \rho_f(\alpha),\varrho\rangle\rightsquigarrow\langle \rho_f(\breve{\alpha}),\varrho\rangle}$$

      $$\frac{\langle \alpha,\varrho\rangle\rightsquigarrow \langle \breve{\alpha},\varrho\rangle}{\langle f(\alpha),\varrho\rangle\rightsquigarrow\langle f(\breve{\alpha}),\varrho\rangle}$$

      With the assumptions $\rho_f(\alpha)=f(\alpha)$ and $\rho_f(\breve{\alpha})=f(\breve{\alpha})$, we get $R_1$ satisfies condition 1 in the definition of $\underline{\leftrightarrow}_q$
      (see Definition \ref{SPQB}).

      $$\frac{\langle \breve{\alpha},\varrho\rangle\xrightarrow{\alpha} \langle \surd,\varrho'\rangle}{\langle \rho_f(\breve{\alpha}),\varrho\rangle\xrightarrow{f(\alpha)}\langle \surd,\varrho'\rangle}$$

      $$\frac{\langle \breve{\alpha},\varrho\rangle\xrightarrow{\alpha} \langle \surd,\varrho'\rangle}{\langle f(\breve{\alpha}),\varrho\rangle\xrightarrow{f(\alpha)}\langle \surd,\varrho'\rangle}$$

      With the assumption $\rho_f(\breve{\alpha})=f(\breve{\alpha})$, we get $R_2$ satisfies conditions 2 and 3 in the definition of $\underline{\leftrightarrow}_q$
      (see Definition \ref{SPQB}).

      It is easy to check that the condition on PDF $\mu$ for $R$ also satisfies condition 4 in Definition \ref{SPQB}, we omit it.

      So, $R=\underline{\leftrightarrow}_q$, as desired.

  \item \textbf{Axiom $RN2$}. For a relation $R=R_1\cup R_2$, suppose that $Eq(\langle \rho_f(\delta),\varrho\rangle,\langle \delta,\varrho\rangle)= R_1$, and $Eq(\langle \rho_f(\breve{\delta}),\varrho\rangle,\langle \breve{\delta},\varrho\rangle)= R_2$.

      By use of the probabilistic transition rules and the action transition rules for atomic unitary operator and renaming operator $\rho_f$, we get:

      $$\frac{\langle \delta,\varrho\rangle\rightsquigarrow \langle \breve{\delta},\varrho\rangle}{\langle \rho_f(\delta),\varrho\rangle\rightsquigarrow\langle \rho_f(\breve{\delta}),\varrho\rangle}$$

      There are not action transition rules for the constant $\breve{\delta}$, which mean that $\breve{\delta}$ leads to inaction processes, accompany with the action transition rules of operator $\rho_f$, $\rho_f(\breve{\delta})\nrightarrow$ and $\breve{\delta}\nrightarrow$ all lead to inaction processes.

      With the assumptions $\rho_f(\delta)=\delta$ and $\rho_f(\breve{\delta})=\breve{\delta}$, we get $R$ satisfies conditions 1, 2 and 3 in the definition of $\underline{\leftrightarrow}_q$
      (see Definition \ref{SPQB}).

      It is easy to check that the condition on PDF $\mu$ for $R$ also satisfies condition 4 in Definition \ref{SPQB}, we omit it.

      So, $R=\underline{\leftrightarrow}_q$, as desired.

  \item \textbf{Axiom $RN3$}. For a relation $R=Eq(\langle \rho_f(u+v),\varrho\rangle,\langle \rho_f(u)+\rho_f(v),\varrho\rangle)\cup Eq(\langle \rho_f(p+q),\varrho\rangle,\langle \rho_f(p)+\rho_f(q),\varrho\rangle)$, with $u,v\in\mathbb{SP}(pqACP^+_{RN})$ and $p,q\in\mathbb{DP}(pqACP^+_{RN})$.

      By use of the probabilistic transition rules and the action transition rules for alternative operator $+$ and renaming operator $\rho_f$, we get:

      $$\frac{\langle u,\varrho\rangle\rightsquigarrow \langle u',\varrho\rangle,\langle v,\varrho\rangle\rightsquigarrow \langle v',\varrho\rangle}{\langle \rho_f(u+v),\varrho\rangle\rightsquigarrow\langle \rho_f(u'+v'),\varrho\rangle}$$

      $$\frac{\langle u,\varrho\rangle\rightsquigarrow \langle u',\varrho\rangle,\langle v,\varrho\rangle\rightsquigarrow \langle v',\varrho\rangle}{\langle \rho_f(u)+\rho_f(v),\varrho\rangle\rightsquigarrow\langle \rho_f(u')+\rho_f(v'),\varrho\rangle}$$

      With the assumptions $\rho_f(u+v)=\rho_f(u)+\rho_f(v)$ and $\rho_f(u'+v')=\rho_f(u')+\rho_f(v')$, we get $R$ satisfies condition 1 in the definition of $\underline{\leftrightarrow}_q$
      (see Definition \ref{SPQB}).

      $$\frac{\langle p,\varrho\rangle\xrightarrow{\alpha} \langle p',\varrho'\rangle}{\langle \rho_f(p+q),\varrho\rangle\xrightarrow{\alpha}\langle \rho_f(p'),\varrho'\rangle}$$

      $$\frac{\langle p,\varrho\rangle\xrightarrow{\alpha} \langle p',\varrho'\rangle}{\langle \rho_f(p)+\rho_f(q),\varrho\rangle\xrightarrow{\alpha}\langle \rho_f(p'),\varrho'\rangle}$$

      $$\frac{\langle p,\varrho\rangle\xrightarrow{\alpha} \langle \surd,\varrho'\rangle}{\langle \rho_f(p+q),\varrho\rangle\xrightarrow{\alpha}\langle \surd,\varrho'\rangle}$$

      $$\frac{\langle p,\varrho\rangle\xrightarrow{\alpha} \langle \surd,\varrho'\rangle}{\langle \rho_f(p)+\rho_f(q),\varrho\rangle\xrightarrow{\alpha}\langle \surd,\varrho'\rangle}$$

      With the assumption $\rho_f(p+q)=\rho_f(p)+\rho_f(q)$, we get $R$ satisfies conditions 2 and 3 in the definition of $\underline{\leftrightarrow}_q$
      (see Definition \ref{SPQB}).

      It is easy to check that the condition on PDF $\mu$ for $R$ also satisfies condition 4 in Definition \ref{SPQB}, we omit it.

      So, $R=\underline{\leftrightarrow}_q$, as desired.

  \item \textbf{Axiom $RN4$}. For a relation $R=Eq(\langle \rho_f(u\cdot v),\varrho\rangle,\langle \rho_f(u)\cdot\rho_f(v),\varrho\rangle)\cup Eq(\langle \rho_f(p\cdot v),\varrho\rangle,\langle \rho_f(p)\cdot\rho_f(v),\varrho\rangle)$, with $u,v\in\mathbb{SP}(pqACP^+_{RN})$ and $p\in\mathbb{DP}(pqACP^+_{RN})$.

      By use of the probabilistic transition rules and the action transition rules for sequential composition $\cdot$ and renaming operator $\rho_f$, we get:

      $$\frac{\langle u,\varrho\rangle\rightsquigarrow \langle u',\varrho\rangle}{\langle \rho_f(u\cdot v),\varrho\rangle\rightsquigarrow\langle \rho_f(u'\cdot v),\varrho\rangle}$$

      $$\frac{\langle u,\varrho\rangle\rightsquigarrow \langle u',\varrho\rangle}{\langle \rho_f(u)\cdot\rho_f(v),\varrho\rangle\rightsquigarrow\langle \rho_f(u')\cdot\rho_f(v),\varrho\rangle}$$

      With the assumptions $\rho_f(u\cdot v)=\rho_f(u)\cdot\rho_f(v)$ and $\rho_f(u'\cdot v)=\rho_f(u')\cdot\rho_f(v)$, we get $R$ satisfies condition 1 in the definition of $\underline{\leftrightarrow}_q$
      (see Definition \ref{SPQB}).

      $$\frac{\langle p,\varrho\rangle\xrightarrow{\alpha} \langle p',\varrho'\rangle}{\langle \rho_f(p\cdot v),\varrho\rangle\xrightarrow{\alpha}\langle \rho_f(p'\cdot v),\varrho'\rangle}$$

      $$\frac{\langle p,\varrho\rangle\xrightarrow{\alpha} \langle p',\varrho'\rangle}{\langle \rho_f(p)\cdot\rho_f(v),\varrho\rangle\xrightarrow{\alpha}\langle \rho_f(p')\cdot\rho_f(v),\varrho'\rangle}$$

      $$\frac{\langle p,\varrho\rangle\xrightarrow{\alpha} \langle \surd,\varrho'\rangle}{\langle \rho_f(p\cdot v),\varrho\rangle\xrightarrow{\alpha}\langle \rho_f(v),\varrho'\rangle}$$

      $$\frac{\langle p,\varrho\rangle\xrightarrow{\alpha} \langle \surd,\varrho'\rangle}{\langle \rho_f(p)\cdot\rho_f(v),\varrho\rangle\xrightarrow{\alpha}\langle \rho_f(v),\varrho'\rangle}$$

      With the assumption $\rho_f(p\cdot v)=\rho_f(p)\cdot\rho_f(v)$ and $\rho_f(p'\cdot v)=\rho_f(p')\cdot\rho_f(v)$, we get $R$ satisfies conditions 2 and 3 in the definition of $\underline{\leftrightarrow}_q$
      (see Definition \ref{SPQB}).

      It is easy to check that the condition on PDF $\mu$ for $R$ also satisfies condition 4 in Definition \ref{SPQB}, we omit it.

      So, $R=\underline{\leftrightarrow}_q$, as desired.

  \item \textbf{Axiom $PrRN1$}. For a relation $R=Eq(\langle \rho_f(u\boxplus_{\pi}v),\varrho\rangle,\langle \rho_f(u)\boxplus_{\pi}\rho_f(v),\varrho\rangle)$, with $u,v\in\mathbb{SP}(pqACP^+_{RN})$.

      By use of the probabilistic transition rules for probabilistic choice operator $\boxplus_{\pi}$ and renaming operator $\rho_f$, we get:

      $$\frac{\langle u,\varrho\rangle\rightsquigarrow \langle u',\varrho\rangle}{\langle \rho_f(u\boxplus_{\pi}v),\varrho\rangle\rightsquigarrow\langle \rho_f(u'),\varrho\rangle}$$

      $$\frac{\langle u,\varrho\rangle\rightsquigarrow \langle u',\varrho\rangle}{\langle \rho_f(u)\boxplus_{\pi}\rho_f(v),\varrho\rangle\rightsquigarrow\langle \rho_f(u'),\varrho\rangle}$$

      With the assumption $\rho_f(u\boxplus_{\pi}v)=\rho_f(u)\boxplus_{\pi}\rho_f(v)$, we get $R$ satisfies condition 1 in the definition of $\underline{\leftrightarrow}_q$
      (see Definition \ref{SPQB}).

      Note that, there are not action transition rules for probabilistic choice operator $\boxplus_{\pi}$.

      It is easy to check that the condition on PDF $\mu$ for $R$ also satisfies condition 4 in Definition \ref{SPQB}, we omit it.

      So, $R=\underline{\leftrightarrow}_q$, as desired.

\end{itemize}
\end{proof}

\proposition[Conservativity of $\mathcal{T}_{pqACP^+_{RN}}$ with respect to $\mathcal{T}_{pqACP^+}$]{The term-deduction system $\mathcal{T}_{pqACP^+_{RN}}$ is an operationally conservative extension of the term-deduction system $\mathcal{T}_{pqACP^+}$.}\label{CRN1}

\begin{proof}
It is sufficient to verify the following conditions:

\begin{itemize}
  \item $\mathcal{T}_{pqACP^+_{RN}}$ is a term-deduction system in path format;
  \item $\mathcal{T}_{pqACP^+}\oplus\mathcal{T}_{pqACP^+_{RN}}$ is defined;
  \item There are no conclusions $\langle s,\varrho\rangle\xrightarrow{\alpha} \langle t,\varrho'\rangle$ or $\langle s,\varrho\rangle\xrightarrow{\alpha} \langle\surd,\varrho'\rangle$ of a rule in $\mathcal{T}_{pqACP^+_{RN}}$ such $s=x$ or $s=f(x_1,\cdots,x_n)$ for some operator $f$ of $pqACP^+$.
\end{itemize}

These conditions hold which can be trivially checked.
\end{proof}

\proposition{The term-deduction system $\mathcal{T}_{pqACP^+_{RN}}$ is an operationally conservative extension up to the probabilistic quantum bisimulation of the term-deduction system $\mathcal{T}_{pqACP^+}$.}\label{CRN2}

\begin{proof}
In the definition of (strong) probabilistic quantum bisimulation (Definition \ref{SPQB}), apart from the 4th clause in Definition \ref{SPQB}, probabilistic quantum bisimulation is defined in terms of predicate and relation symbols. From Proposition \ref{CRN1}, for each closed $pqACP^+$ term $s$, its term-relation-predicate diagrams in both $\mathcal{T}_{pqACP^+}$ and $\mathcal{T}_{pqACP^+_{RN}}$ are the same. The concepts of PDF $\mu$ and quantum state $\varrho$ do not disturb the notion of the probabilistic quantum bisimulation defined only in terms of predicate and relation symbols.
\end{proof}

\proposition[Conservativity of $pqACP^+_{RN}$ with respect to $pqACP^+$]{$pqACP^+_{RN}$ is an equationally conservative extension of $pqACP^+$, that is, if $t$ and $s$ are closed $pqACP^+$ terms, then $pqACP^+\vdash t=s\Leftrightarrow pqACP^+_{RN}\vdash t=s$.}\label{CRN3}

\begin{proof}
It is proven by the following three facts:

\begin{enumerate}
  \item $\mathcal{T}_{pqACP^+_{RN}}$ is an operationally conservative extension of $\mathcal{T}_{pqACP^+}$ up to probabilistic quantum bisimulation (by Proposition \ref{CRN2});
  \item $pqACP^+$ is a complete axiomatization with respect to the bisimulation model (by Theorem \ref{CompletePQACP});
  \item $\mathcal{T}_{pqACP^+_{RN}}$ with respect to the probabilistic quantum bisimulation equivalence induces a model of $pqACP^+_{RN}$ (by Theorem \ref{SoundnessRN}).
\end{enumerate}
\end{proof}

\theorem[Completeness of $pqACP^+_{RN}$]{Let $z$ and $u$ are closed $pqACP^+_{RN}$ terms, if $z\underline{\leftrightarrow}_q u$, then $pACP^+_{RN}\vdash z=u$.}

\begin{proof}
It is proven based on the following two facts:

\begin{enumerate}
  \item $pqACP^+_{RN}$ has the elimination property for $pqBPA$ (by Theorem \ref{ElimiRN});
  \item $pqACP^+_{RN}$ is an operationally conservative extension of $pqACP^+$ (by Proposition \ref{CRN3}).
\end{enumerate}
\end{proof}

\subsection{Priorities}

In \cite{PPA}, $pACP^+_\Theta$ is obtained by extending $pACP^+$ with two operators: the priority operator $\Theta$ and the auxiliary unless operator $\triangleleft$, to express priorities as a mechanism for process interruption. There is a partial order $<$ on the set of atomic actions. Corresponding to $pACP^+_\Theta$, we obtain $pqACP^+_\Theta$ make $pACP^+_\Theta$ based on quantum configuration. The axiom system for priorities is shown in Table \ref{AxiomsForP}.

\begin{center}
\begin{table}
  \begin{tabular}{@{}ll@{}}
\hline No. &Axiom\\
  $TH1$ & $\Theta(a)=a$\\
  $TH2$ & $\Theta(x\cdot y)=\Theta(x)\cdot\Theta(y)$\\
  $PrTH4$ & $\Theta(x\boxplus_{\pi} y)=\Theta(x)\boxplus_{\pi}\Theta(y)$\\
  $DyTH3$ & $x=x+x,y=y+y\Rightarrow\Theta(x+ y)=\Theta(x)\triangleleft y+\Theta(y)\triangleleft x$\\
  $P1$ & $a\triangleleft b=a\quad \textrm{ if }\neg(a<b)$\\
  $P2$ & $a\triangleleft b=\delta\quad \textrm{ if }a<b$\\
  $P3$ & $x\triangleleft (y\cdot z)=x\triangleleft y$\\
  $P4$ & $x\triangleleft (y+ z)=(x\triangleleft y)\triangleleft z$\\
  $P5$ & $x\cdot y\triangleleft z=(x\triangleleft z)\cdot y$\\
  $P6$ & $(x+y)\triangleleft z=(x\triangleleft z)+(y\triangleleft z)$\\
\end{tabular}
\caption{Axioms for priorities}
\label{AxiomsForP}
\end{table}
\end{center}

The probabilistic transition rules for priorities are as follows.

$$\frac{\langle x,\varrho\rangle\rightsquigarrow \langle x',\varrho\rangle}{\langle\Theta(x),\varrho\rangle\rightsquigarrow\langle\Theta(x'),\varrho\rangle}
\quad\quad\frac{\langle x,\varrho\rangle\rightsquigarrow \langle x',\varrho\rangle}{\langle x\triangleleft y,\varrho\rangle\rightsquigarrow \langle x'\triangleleft y,\varrho\rangle}$$

The action transition rules for priorities are as follows.

$$\frac{\langle x,\varrho\rangle\xrightarrow{\alpha} \langle x',\varrho'\rangle,\textrm{ and for all }b>a.\langle x,\varrho\rangle\nrightarrow^b}{\langle\Theta(x),\varrho\rangle\xrightarrow{\alpha}\langle\Theta(x'),\varrho'\rangle}
\quad\quad\frac{\langle x,\varrho\rangle\xrightarrow{\alpha} \langle \surd,\varrho'\rangle,\textrm{ and for all }b>a.\langle x,\varrho\rangle\nrightarrow^b}{\langle\Theta(x),\varrho\rangle\xrightarrow{\alpha}\langle\surd,\varrho'\rangle}$$

$$\frac{\langle x,\varrho\rangle\xrightarrow{\alpha} \langle x',\varrho'\rangle,\textrm{ and for all }b>a.\langle x,\varrho\rangle\nrightarrow^b}{\langle x\triangleleft y,\varrho\rangle\xrightarrow{\alpha}\langle x',\varrho'\rangle}
\quad\quad\frac{\langle x,\varrho\rangle\xrightarrow{\alpha} \langle \surd,\varrho'\rangle,\textrm{ and for all }b>a.\langle x,\varrho\rangle\nrightarrow^b}{\langle x\triangleleft y,\varrho\rangle\xrightarrow{\alpha}\langle\surd,\varrho'\rangle}$$

\theorem[Soundness of $pqACP^+_\Theta$]{Let $p$ and $q$ be closed $pqACP^+_\Theta$ terms. If $pqACP^+_\Theta\vdash p=q$ then $\langle p,\varrho\rangle\underline{\leftrightarrow}_q \langle q,\varrho\rangle$.}

\begin{proof}
It is already proven that $\underline{\leftrightarrow}_q$ is an equivalent and congruent relation on $\mathbb{PT}(pqACP^+_\Theta)$ (see details in \cite{PPA}), and $pACP^+_\Theta\vdash x=y$ then $x\underline{\leftrightarrow} y$ (that is, the conditions on PDF $\mu$ are same in $\underline{\leftrightarrow}_q$ and $\underline{\leftrightarrow}$, see in section \ref{PPA} and \cite{PPA} for details), we only need to prove that the quantum information $\varrho$ related parts are also sound.

It is sufficient to examine every axiom in the axiom system (in Table \ref{AxiomsForP}) for $pqACP^+_\Theta$ is sound.

\begin{itemize}
  \item \textbf{Axiom $TH1$}. For a relation $R=R_1\cup R_2$, suppose that $Eq(\langle \Theta(\alpha),\varrho\rangle,\langle \alpha,\varrho\rangle)= R_1$, and $Eq(\langle \Theta(\breve{\alpha}),\varrho\rangle,\langle \breve{\alpha},\varrho\rangle)= R_2$.

      By use of the probabilistic transition rules and the action transition rules for atomic unitary operator and priority operator $\Theta$, we get:

      $$\frac{\langle \alpha,\varrho\rangle\rightsquigarrow \langle \breve{\alpha},\varrho\rangle}{\langle \Theta(\alpha),\varrho\rangle\rightsquigarrow\langle \Theta(\breve{\alpha}),\varrho\rangle}$$

      $$\frac{\langle \alpha,\varrho\rangle\rightsquigarrow \langle \breve{\alpha},\varrho\rangle}{\langle \alpha,\varrho\rangle\rightsquigarrow\langle \breve{\alpha},\varrho\rangle}$$

      With the assumptions $\Theta(\alpha)=\alpha$ and $\Theta(\breve{\alpha})=\breve{\alpha}$, we get $R_1$ satisfies condition 1 in the definition of $\underline{\leftrightarrow}_q$
      (see Definition \ref{SPQB}).

      $$\frac{\langle \breve{\alpha},\varrho\rangle\xrightarrow{\alpha} \langle \surd,\varrho'\rangle}{\langle \Theta(\breve{\alpha}),\varrho\rangle\xrightarrow{\alpha}\langle \surd,\varrho'\rangle}$$

      $$\frac{\langle \breve{\alpha},\varrho\rangle\xrightarrow{\alpha} \langle \surd,\varrho'\rangle}{\langle \breve{\alpha},\varrho\rangle\xrightarrow{\alpha}\langle \surd,\varrho'\rangle}$$

      With the assumption $\Theta(\breve{\alpha})=\breve{\alpha}$, we get $R_2$ satisfies conditions 2 and 3 in the definition of $\underline{\leftrightarrow}_q$
      (see Definition \ref{SPQB}).

      So, $R=\underline{\leftrightarrow}_q$, as desired.

  \item \textbf{Axiom $TH2$}. For a relation $R=Eq(\langle \Theta(u\cdot v),\varrho\rangle,\langle \Theta(u)\cdot\Theta(v),\varrho\rangle)\cup Eq(\langle \Theta(p\cdot v),\varrho\rangle,\langle \Theta(p)\cdot\Theta(v),\varrho\rangle)$, with $u,v\in\mathbb{SP}(pqACP^+_{\Theta})$ and $p\in\mathbb{DP}(pqACP^+_{\Theta})$.

      By use of the probabilistic transition rules and the action transition rules for sequential composition $\cdot$ and priority operator $\Theta$, we get:

      $$\frac{\langle u,\varrho\rangle\rightsquigarrow \langle u',\varrho\rangle}{\langle \Theta(u\cdot v),\varrho\rangle\rightsquigarrow\langle \Theta(u'\cdot v),\varrho\rangle}$$

      $$\frac{\langle u,\varrho\rangle\rightsquigarrow \langle u',\varrho\rangle}{\langle \Theta(u)\cdot\Theta(v),\varrho\rangle\rightsquigarrow\langle \Theta(u')\cdot\Theta(v),\varrho\rangle}$$

      With the assumptions $\Theta(u\cdot v)=\Theta(u)\cdot\Theta(v)$ and $\Theta(u'\cdot v)=\Theta(u')\cdot\Theta(v)$, we get $R$ satisfies condition 1 in the definition of $\underline{\leftrightarrow}_q$
      (see Definition \ref{SPQB}).

      $$\frac{\langle p,\varrho\rangle\xrightarrow{\alpha} \langle p',\varrho'\rangle}{\langle \Theta(p\cdot v),\varrho\rangle\xrightarrow{\alpha}\langle \Theta(p'\cdot v),\varrho'\rangle}$$

      $$\frac{\langle p,\varrho\rangle\xrightarrow{\alpha} \langle p',\varrho'\rangle}{\langle \Theta(p)\cdot\Theta(v),\varrho\rangle\xrightarrow{\alpha}\langle \Theta(p')\cdot\Theta(v),\varrho'\rangle}$$

      $$\frac{\langle p,\varrho\rangle\xrightarrow{\alpha} \langle \surd,\varrho'\rangle}{\langle \Theta(p\cdot v),\varrho\rangle\xrightarrow{\alpha}\langle \Theta(v),\varrho'\rangle}$$

      $$\frac{\langle p,\varrho\rangle\xrightarrow{\alpha} \langle \surd,\varrho'\rangle}{\langle \Theta(p)\cdot\Theta(v),\varrho\rangle\xrightarrow{\alpha}\langle \Theta(v),\varrho'\rangle}$$

      With the assumption $\Theta(p\cdot v)=\Theta(p)\cdot\Theta(v)$ and $\Theta(p'\cdot v)=\Theta(p')\cdot\Theta(v)$, we get $R$ satisfies conditions 2 and 3 in the definition of $\underline{\leftrightarrow}_q$
      (see Definition \ref{SPQB}).

      So, $R=\underline{\leftrightarrow}_q$, as desired.

  \item \textbf{Axiom $PrTH4$}. For a relation $R=Eq(\langle \Theta(u\boxplus_{\pi}v),\varrho\rangle,\langle \Theta(u)\boxplus_{\pi}\Theta(v),\varrho\rangle)$, with $u,v\in\mathbb{SP}(pqACP^+_{\Theta})$.

      By use of the probabilistic transition rules for probabilistic choice operator $\boxplus_{\pi}$ and priority operator $\Theta$, we get:

      $$\frac{\langle u,\varrho\rangle\rightsquigarrow \langle u',\varrho\rangle}{\langle \Theta(u\boxplus_{\pi}v),\varrho\rangle\rightsquigarrow\langle \Theta(u'),\varrho\rangle}$$

      $$\frac{\langle u,\varrho\rangle\rightsquigarrow \langle u',\varrho\rangle}{\langle \Theta(u)\boxplus_{\pi}\Theta(v),\varrho\rangle\rightsquigarrow\langle \Theta(u'),\varrho\rangle}$$

      With the assumption $\Theta(u\boxplus_{\pi}v)=\Theta(u)\boxplus_{\pi}\Theta(v)$, we get $R$ satisfies condition 1 in the definition of $\underline{\leftrightarrow}_q$
      (see Definition \ref{SPQB}).

      Note that, there are not action transition rules for probabilistic choice operator $\boxplus_{\pi}$.

      So, $R=\underline{\leftrightarrow}_q$, as desired.

  \item \textbf{Axiom $DyTH3$}. For a relation $R=Eq(\langle \Theta(u+v),\varrho\rangle,\langle \Theta(u)\triangleleft v+\Theta(v)\triangleleft u,\varrho\rangle)\cup Eq(\langle \Theta(p+q),\varrho\rangle,\langle \Theta(p)\triangleleft q+\Theta(q)\triangleleft p,\varrho\rangle)$, with $u,v\in\mathbb{SP}(pqACP^+_{\Theta})$ and $p,q\in\mathbb{DP}(pqACP^+_{\Theta})$.

      By use of the probabilistic transition rules and the action transition rules for alternative operator $+$, priority operator $\Theta$ and unless operator $\triangleleft$, we get:

      $$\frac{\langle u,\varrho\rangle\rightsquigarrow \langle u',\varrho\rangle,\langle v,\varrho\rangle\rightsquigarrow \langle v',\varrho\rangle}{\langle \Theta(u+v),\varrho\rangle\rightsquigarrow\langle \Theta(u'+v'),\varrho\rangle}$$

      $$\frac{\langle u,\varrho\rangle\rightsquigarrow \langle u',\varrho\rangle,\langle v,\varrho\rangle\rightsquigarrow \langle v',\varrho\rangle}{\langle \Theta(u)\triangleleft v+\Theta(v)\triangleleft u,\varrho\rangle\rightsquigarrow\langle \Theta(u')\triangleleft v'+\Theta(v')\triangleleft u',\varrho\rangle}$$

      With the assumptions $\Theta(u+v)=\Theta(u)\triangleleft v+\Theta(v)\triangleleft u$ and $\Theta(u'+v')=\Theta(u')\triangleleft v'+\Theta(v')\triangleleft u'$, we get $R$ satisfies condition 1 in the definition of $\underline{\leftrightarrow}_q$
      (see Definition \ref{SPQB}).

      $$\frac{\langle p,\varrho\rangle\xrightarrow{\alpha} \langle p',\varrho'\rangle}{\langle \Theta(p+q),\varrho\rangle\xrightarrow{\alpha}\langle \Theta(p'),\varrho'\rangle}$$

      $$\frac{\langle p,\varrho\rangle\xrightarrow{\alpha} \langle p',\varrho'\rangle}{\langle \Theta(p)\triangleleft q+\Theta(q)\triangleleft p,\varrho\rangle\xrightarrow{\alpha}\langle \Theta(p'),\varrho'\rangle}$$

      $$\frac{\langle p,\varrho\rangle\xrightarrow{\alpha} \langle \surd,\varrho'\rangle}{\langle \Theta(p+q),\varrho\rangle\xrightarrow{\alpha}\langle \surd,\varrho'\rangle}$$

      $$\frac{\langle p,\varrho\rangle\xrightarrow{\alpha} \langle \surd,\varrho'\rangle}{\langle \Theta(p)\triangleleft q+\Theta(q)\triangleleft p,\varrho\rangle\xrightarrow{\alpha}\langle \surd,\varrho'\rangle}$$

      With the assumption $\Theta(p+q)=\Theta(p)\triangleleft q+\Theta(q)\triangleleft p$, we get $R$ satisfies conditions 2 and 3 in the definition of $\underline{\leftrightarrow}_q$
      (see Definition \ref{SPQB}).

      So, $R=\underline{\leftrightarrow}_q$, as desired.

  \item \textbf{Axiom $P1$}. For a relation $R=R_1\cup R_2$, suppose that $Eq(\langle \alpha\triangleleft \beta,\varrho\rangle,\langle \alpha,\varrho\rangle)= R_1$ with $\neg(\alpha<\beta)$, and $Eq(\langle \breve{\alpha}\triangleleft \beta,\varrho\rangle,\langle \breve{\alpha},\varrho\rangle)= R_2$ with $\neg(\breve{\alpha}<\beta)$.

      By use of the probabilistic transition rules and the action transition rules for atomic unitary operator and unless operator $\triangleleft$, we get:

      $$\frac{\langle \alpha,\varrho\rangle\rightsquigarrow \langle \breve{\alpha},\varrho\rangle}{\langle \alpha\triangleleft\beta,\varrho\rangle\rightsquigarrow\langle \breve{\alpha}\triangleleft\beta,\varrho\rangle}$$

      $$\frac{\langle \alpha,\varrho\rangle\rightsquigarrow \langle \breve{\alpha},\varrho\rangle}{\langle \alpha,\varrho\rangle\rightsquigarrow\langle \breve{\alpha},\varrho\rangle}$$

      With the assumptions $\alpha\triangleleft \beta=\alpha,\neg(\alpha<\beta)$ and $\breve{\alpha}\triangleleft\beta=\breve{\alpha},\neg(\alpha<\beta)$, we get $R_1$ satisfies condition 1 in the definition of $\underline{\leftrightarrow}_q$
      (see Definition \ref{SPQB}).

      $$\frac{\langle \breve{\alpha},\varrho\rangle\xrightarrow{\alpha} \langle \surd,\varrho'\rangle}{\langle \breve{\alpha}\triangleleft\beta,\varrho\rangle\xrightarrow{\alpha}\langle \surd,\varrho'\rangle}$$

      $$\frac{\langle \breve{\alpha},\varrho\rangle\xrightarrow{\alpha} \langle \surd,\varrho'\rangle}{\langle \breve{\alpha},\varrho\rangle\xrightarrow{\alpha}\langle \surd,\varrho'\rangle}$$

      With the assumption $\breve{\alpha}\triangleleft \beta=\breve{\alpha},\neg(\alpha<\beta)$, we get $R_2$ satisfies conditions 2 and 3 in the definition of $\underline{\leftrightarrow}_q$
      (see Definition \ref{SPQB}).

      So, $R=\underline{\leftrightarrow}_q$, as desired.

  \item \textbf{Axiom $P2$}. For a relation $R=Eq(\langle \alpha\triangleleft \beta,\varrho\rangle,\langle \delta,\varrho\rangle)$ with $\alpha<\beta$.

      By use of the probabilistic transition rules and the action transition rules for atomic unitary operator and unless operator $\triangleleft$, we get:

      $$\frac{\langle \delta,\varrho\rangle\rightsquigarrow \langle \breve{\delta},\varrho\rangle}{\langle \alpha\triangleleft\beta,\varrho\rangle\rightsquigarrow\langle \breve{\delta},\varrho\rangle}$$

      There are not action transition rules for the constant $\breve{\delta}$, which mean that $\breve{\delta}$ leads to inaction processes, accompany with the action transition rules of operator $\triangleleft$, $\alpha\triangleleft\beta,(\alpha<\beta)\nrightarrow$ and $\breve{\delta}\nrightarrow$ all lead to inaction processes.

      With the assumption $\alpha\triangleleft\beta=\delta$, we get $R$ satisfies conditions 1, 2 and 3 in the definition of $\underline{\leftrightarrow}_q$
      (see Definition \ref{SPQB}).

      So, $R=\underline{\leftrightarrow}_q$, as desired.

  \item \textbf{Axiom $P3$}. For a relation $R=R_1\cup R_2$, suppose that $Eq(\langle u\triangleleft(v\cdot w),\varrho\rangle,\langle u\triangleleft v,\varrho\rangle)= R_1$ with $u,v,w\in\mathbb{SP}(pqACP^+_{\Theta})$, and $Eq(\langle p\triangleleft(q\cdot w),\varrho\rangle,\langle p\triangleleft q,\varrho\rangle)= R_2$ with $p\in\mathbb{DP}(pqACP^+_{\Theta})$, $q,w\in\mathbb{SP}(pqACP^+_{\Theta})$.

      By use of the probabilistic transition rules and the action transition rules for atomic unitary operator $\alpha$, sequential composition $\cdot$ and unless operator $\triangleleft$, we get:

      $$\frac{\langle u,\varrho\rangle\rightsquigarrow \langle u',\varrho\rangle}{\langle u\triangleleft(v\cdot w),\varrho\rangle\rightsquigarrow\langle u'\triangleleft(v\cdot w),\varrho\rangle}$$

      $$\frac{\langle u,\varrho\rangle\rightsquigarrow \langle u',\varrho\rangle}{\langle u\triangleleft v,\varrho\rangle\rightsquigarrow\langle u'\triangleleft v,\varrho\rangle}$$

      With the assumptions $u\triangleleft(v\cdot w)=u\triangleleft v$ and $u'\triangleleft(v\cdot w)=u'\triangleleft v$, we get $R_1$ satisfies condition 1 in the definition of $\underline{\leftrightarrow}_q$
      (see Definition \ref{SPQB}).

      $$\frac{\langle p,\varrho\rangle\xrightarrow{\alpha} \langle p',\varrho'\rangle}{\langle p\triangleleft(q\cdot w),\varrho\rangle\xrightarrow{\alpha}\langle p',\varrho'\rangle}$$

      $$\frac{\langle p,\varrho\rangle\xrightarrow{\alpha} \langle p',\varrho'\rangle}{\langle p\triangleleft q,\varrho\rangle\xrightarrow{\alpha}\langle p',\varrho'\rangle}$$

      $$\frac{\langle p,\varrho\rangle\xrightarrow{\alpha} \langle \surd,\varrho'\rangle}{\langle p\triangleleft(q\cdot w),\varrho\rangle\xrightarrow{\alpha}\langle \surd,\varrho'\rangle}$$

      $$\frac{\langle p,\varrho\rangle\xrightarrow{\alpha} \langle \surd,\varrho'\rangle}{\langle p\triangleleft q,\varrho\rangle\xrightarrow{\alpha}\langle \surd,\varrho'\rangle}$$

      With the assumption $p\triangleleft(q\cdot w)=p\triangleleft q$, we get $R_2$ satisfies conditions 2 and 3 in the definition of $\underline{\leftrightarrow}_q$
      (see Definition \ref{SPQB}).

      So, $R=\underline{\leftrightarrow}_q$, as desired.

  \item \textbf{Axiom $P4$}. For a relation $R=R_1\cup R_2$, suppose that $Eq(\langle u\triangleleft(v+ w),\varrho\rangle,\langle u\triangleleft v\triangleleft w,\varrho\rangle)= R_1$ with $u,v,w\in\mathbb{SP}(pqACP^+_{\Theta})$, and $Eq(\langle p\triangleleft(q+ s),\varrho\rangle,\langle p\triangleleft q\triangleleft s,\varrho\rangle)= R_2$ with $p\in\mathbb{DP}(pqACP^+_{\Theta})$, $q,s\in\mathbb{SP}(pqACP^+_{\Theta})$.

      By use of the probabilistic transition rules and the action transition rules for atomic unitary operator $\alpha$, alternative composition $+$ and unless operator $\triangleleft$, we get:

      $$\frac{\langle u,\varrho\rangle\rightsquigarrow \langle u',\varrho\rangle}{\langle u\triangleleft(v+ w),\varrho\rangle\rightsquigarrow\langle u'\triangleleft(v+ w),\varrho\rangle}$$

      $$\frac{\langle u,\varrho\rangle\rightsquigarrow \langle u',\varrho\rangle}{\langle u\triangleleft v\triangleleft w,\varrho\rangle\rightsquigarrow\langle u'\triangleleft v\triangleleft w,\varrho\rangle}$$

      With the assumption $u\triangleleft(v+ w)=u\triangleleft v\triangleleft w$ and $u'\triangleleft(v+ w)=u'\triangleleft v\triangleleft w$, we get $R_1$ satisfies condition 1 in the definition of $\underline{\leftrightarrow}_q$
      (see Definition \ref{SPQB}).

      $$\frac{\langle p,\varrho\rangle\xrightarrow{\alpha} \langle p',\varrho'\rangle}{\langle p\triangleleft(q+ s),\varrho\rangle\xrightarrow{\alpha}\langle p',\varrho'\rangle}$$

      $$\frac{\langle p,\varrho\rangle\xrightarrow{\alpha} \langle p',\varrho'\rangle}{\langle p\triangleleft q\triangleleft s,\varrho\rangle\xrightarrow{\alpha}\langle p',\varrho'\rangle}$$

      $$\frac{\langle p,\varrho\rangle\xrightarrow{\alpha} \langle \surd,\varrho'\rangle}{\langle p\triangleleft(q+ s),\varrho\rangle\xrightarrow{\alpha}\langle \surd,\varrho'\rangle}$$

      $$\frac{\langle p,\varrho\rangle\xrightarrow{\alpha} \langle \surd,\varrho'\rangle}{\langle p\triangleleft q\triangleleft s,\varrho\rangle\xrightarrow{\alpha}\langle \surd,\varrho'\rangle}$$

      With the assumption $p\triangleleft(q+ s)=p\triangleleft q\triangleleft s$, we get $R_2$ satisfies conditions 2 and 3 in the definition of $\underline{\leftrightarrow}_q$
      (see Definition \ref{SPQB}).

      So, $R=\underline{\leftrightarrow}_q$, as desired.

  \item \textbf{Axiom $P5$}. For a relation $R=R_1\cup R_2$, suppose that $Eq(\langle (u\cdot v)\triangleleft w,\varrho\rangle,\langle (u\triangleleft w)\cdot v,\varrho\rangle)= R_1$ with $u,v,w\in\mathbb{SP}(pqACP^+_{\Theta})$, and $Eq(\langle (p\cdot q)\triangleleft s,\varrho\rangle,\langle (p\triangleleft s)\cdot q,\varrho\rangle)= R_2$ with $p\in\mathbb{DP}(pqACP^+_{\Theta})$, $s,q\in\mathbb{SP}(pqACP^+_{\Theta})$.

      By use of the probabilistic transition rules and the action transition rules for atomic unitary operator $\alpha$, sequential composition $\cdot$ and unless operator $\triangleleft$, we get:

      $$\frac{\langle u,\varrho\rangle\rightsquigarrow \langle u',\varrho\rangle}{\langle (u\cdot v)\triangleleft w,\varrho\rangle\rightsquigarrow\langle (u'\cdot v)\triangleleft w,\varrho\rangle}$$

      $$\frac{\langle u,\varrho\rangle\rightsquigarrow \langle u',\varrho\rangle}{\langle (u\triangleleft w)\cdot v,\varrho\rangle\rightsquigarrow\langle (u'\triangleleft w)\cdot v,\varrho\rangle}$$

      With the assumptions $(u\cdot v)\triangleleft w=(u\triangleleft w)\cdot v$ and $(u'\cdot v)\triangleleft w=(u'\triangleleft w)\cdot v$, we get $R_1$ satisfies condition 1 in the definition of $\underline{\leftrightarrow}_q$
      (see Definition \ref{SPQB}).

      $$\frac{\langle p,\varrho\rangle\xrightarrow{\alpha} \langle p',\varrho'\rangle}{\langle (p\cdot q)\triangleleft s),\varrho\rangle\xrightarrow{\alpha}\langle p'\cdot q,\varrho'\rangle}$$

      $$\frac{\langle p,\varrho\rangle\xrightarrow{\alpha} \langle p',\varrho'\rangle}{\langle (p\triangleleft s)\cdot q,\varrho\rangle\xrightarrow{\alpha}\langle p'\cdot q,\varrho'\rangle}$$

      $$\frac{\langle p,\varrho\rangle\xrightarrow{\alpha} \langle \surd,\varrho'\rangle}{\langle (p\cdot q)\triangleleft s),\varrho\rangle\xrightarrow{\alpha}\langle q,\varrho'\rangle}$$

      $$\frac{\langle p,\varrho\rangle\xrightarrow{\alpha} \langle \surd,\varrho'\rangle}{\langle (p\triangleleft s)\cdot q,\varrho\rangle\xrightarrow{\alpha}\langle q,\varrho'\rangle}$$

      With the assumption $(p\cdot q)\triangleleft s=(p\triangleleft s)\cdot q$, we get $R_2$ satisfies conditions 2 and 3 in the definition of $\underline{\leftrightarrow}_q$
      (see Definition \ref{SPQB}).

      So, $R=\underline{\leftrightarrow}_q$, as desired.

  \item \textbf{Axiom $P6$}. For a relation $R=R_1\cup R_2$, suppose that $Eq(\langle (p+q)\triangleleft s,\varrho\rangle,\langle p\triangleleft s + q\triangleleft s,\varrho\rangle)= R_1$ with $p,q\in \mathbb{DP}(pqACP^+_{\Theta})$ and $s\in \mathbb{SP}(pqACP^+_{\Theta})$, and $Eq(\langle (u+v)\triangleleft w,\varrho\rangle,\langle u\triangleleft w + v\triangleleft w,\varrho\rangle)= R_2$ with $u,v,w\in \mathbb{SP}(pqACP^+_{\Theta})$.

      By use of the probabilistic transition rules and the action transition rules for unless operator $\triangleleft$, alternative operator $+$, we get:

      $$\frac{\langle u,\varrho\rangle\rightsquigarrow \langle u',\varrho\rangle, \langle v,\varrho\rangle\rightsquigarrow \langle v',\varrho\rangle}{\langle (u + v)\triangleleft w,\varrho\rangle\rightsquigarrow\langle (u' + v')\triangleleft w,\varrho\rangle}$$

      $$\frac{\langle u,\varrho\rangle\rightsquigarrow \langle u',\varrho\rangle, \langle v,\varrho\rangle\rightsquigarrow \langle v',\varrho\rangle}{\langle u\triangleleft w+v\triangleleft w,\varrho\rangle\rightsquigarrow\langle u'\triangleleft w+v'\triangleleft w,\varrho\rangle}$$

      With the assumptions $(u+v)\triangleleft w=u\triangleleft w+v\triangleleft w$ and $(u'+v')\triangleleft w=u'\triangleleft w+v'\triangleleft w$, we get $R_2$ satisfies condition 1 in the definition of $\underline{\leftrightarrow}_q$
      (see Definition \ref{SPQB}).

      $$\frac{\langle p,\varrho\rangle\xrightarrow{\alpha} \langle p',\varrho'\rangle}{\langle (p+ q)\triangleleft s,\varrho\rangle\xrightarrow{\alpha}\langle p',\varrho'\rangle}$$

      $$\frac{\langle p,\varrho\rangle\xrightarrow{\alpha} \langle p',\varrho'\rangle}{\langle p\triangleleft s+q\triangleleft s,\varrho\rangle\xrightarrow{\alpha}\langle p',\varrho'\rangle}$$

      $$\frac{\langle p,\varrho\rangle\xrightarrow{\alpha} \langle \surd,\varrho'\rangle}{\langle (p+ q)\triangleleft s,\varrho\rangle\xrightarrow{\alpha}\langle \surd,\varrho'\rangle}$$

      $$\frac{\langle p,\varrho\rangle\xrightarrow{\alpha} \langle \surd,\varrho'\rangle}{\langle p\triangleleft s+q\triangleleft s,\varrho\rangle\xrightarrow{\alpha}\langle \surd,\varrho'\rangle}$$

      With the assumptions $(p+q)\triangleleft s=p\triangleleft s + q\triangleleft s$, we get $R_1$ satisfies conditions 2 and 3 in the definition of $\underline{\leftrightarrow}_q$
      (see Definition \ref{SPQB}).

      So, $R=\underline{\leftrightarrow}_q$, as desired.

\end{itemize}
\end{proof}

\theorem[Completeness of $pqACP^+_\Theta$]{Let $z$ and $u$ are closed $pqACP^+_\Theta$ terms, if $\langle z,\varrho\rangle\underline{\leftrightarrow}_q \langle u,\varrho\rangle$, then $pqACP^+_\Theta\vdash z=u$.}

\begin{proof}
It is based on the following three facts: let $z$ and $u$ are closed $pqACP^+_\Theta$ terms,

\begin{enumerate}
  \item $\langle z,\varrho\rangle\underline{\leftrightarrow}_q \langle u,\varrho\rangle \Rightarrow z\underline{\leftrightarrow} u$ by Proposition \ref{PQBPB};
  \item $z\underline{\leftrightarrow} u \Rightarrow pACP^+_\Theta\vdash z=u$ (see in \cite{PPA});
  \item the term systems of $pACP^+_\Theta$ and $pqACP^+_\Theta$ are same, including the same operators and the same axiom systems, but different atomic action meanings (semantics).
\end{enumerate}

So, we get $\langle z,\varrho\rangle\underline{\leftrightarrow}_q \langle u,\varrho\rangle \Rightarrow pqACP^+_\Theta\vdash z=u$, as desired.
\end{proof}

\section{Conclusions}\label{Conclusions}

With an assumption of closed quantum systems, we unify quantum and classical computing under the framework of probabilistic process algebra, in which, quantum and classical computing have the same equational logic and the same structured operational semantics based on quantum configuration. Besides quantum computing mechanism in closed systems -- unitary operator, quantum measurement and quantum entanglement, many computation properties are also involved, including sequential composition, alternative composition, probabilistic choice, recursion (projection), parallelism and communication, and abstraction. As a ACP-like process algebra, it inherits the advantages of ACP, such as modularity, which makes it can be extended easily and elegantly. Our work can be used widely in verification for quantum and classical computing mixed systems, such as most quantum communication protocol.

Note that, our work strongly relies on the progress of ACP-like probabilistic process algebra. For ACP-like probabilistic process algebra still can not solve a mixture of alternative composition, probabilistic choice and abstraction. This shortcoming is also one of our future directions.

\label{lastpage}

\end{document}